\documentclass[11pt]{article}
\usepackage{amsmath,amssymb}
\usepackage{amsthm}
\usepackage{bbm}
\usepackage{bm}
\usepackage{centernot}
\usepackage{enumitem}
\usepackage{framed}
\usepackage{graphicx}
\usepackage{indentfirst}
\usepackage{interval}
\usepackage{mathtools}
\usepackage{multibib}
\usepackage{multirow}
\usepackage[round,longnamesfirst]{natbib}
\usepackage{rotating}
\usepackage{setspace}
\usepackage{subcaption}
\usepackage{threeparttable}
\usepackage{titlesec}
\usepackage{titling}
\usepackage{authblk}
\usepackage{url}

\DeclareMathOperator*{\argmin}{\arg\!\min}

\mathtoolsset{showonlyrefs}
\newtheorem{thm}{Theorem}

\newtheorem{asm}{Assumption}
\newtheorem{rem}{Remark}

\renewenvironment{proof}[1][\proofname]{{\bfseries #1. }}{\qed}
\newtheorem{lemapp}{Lemma}

\newcites{app}{References to Appendices A and B}

\title{Testing for Stationary or Persistent Coefficient Randomness in Predictive Regressions}

\author[1]{Mikihito Nishi\footnote{I am greatly indebted to Eiji Kurozumi, my advisor, for discussions and his help, support and encouragement. I also thank the participants in a seminar at Hitotsubashi University and a conference at Hiroshima University for their helpful comments and suggestions. All errors are mine. Address correspondence to: Graduate School of Economics, Hitotsubashi University, 2-1 Naka, Kunitachi, Tokyo 186-8601, Japan; e-mail: ed225007@g.hit-u.ac.jp}}

\affil[1]{Graduate School of Economics, Hitotsubashi University}

\date{\today}

\begin{document}

\baselineskip= 6mm

    \begin{titlingpage}

        \maketitle

        \begin{abstract}
		This study considers tests for coefficient randomness in predictive regressions. Our focus is on how tests for coefficient randomness are influenced by the persistence of random coefficient. We show that when the random coefficient is stationary, or I(0), \citeauthor{nyblomTestingConstancyParameters1989}'s (\citeyear{nyblomTestingConstancyParameters1989}) LM test loses its optimality (in terms of power), which is established against the alternative of integrated, or I(1), random coefficient. We demonstrate this by constructing a test that is more powerful than the LM test when the random coefficient is stationary, although the test is dominated in terms of power by the LM test when the random coefficient is integrated. The power comparison is made under the sequence of local alternatives that approaches the null hypothesis at different rates depending on the persistence of the random coefficient and which test is considered. We revisit an earlier empirical research and apply the tests considered in this study to the U.S. stock returns data. The result mostly reverses the earlier finding.
        \end{abstract}

\medskip

\noindent


\medskip

\noindent


\end{titlingpage}

\section{Introduction}

In the applied finance and economics literature, stock return predictability has been widely discussed. Predictability analysis often involves regressing stock returns on the lag of another financial/economic variable and investigating whether the coefficient on the variable is significantly different from zero (a significantly non-zero coefficient amounts to evidence for predictability). Despite a large body of literature on this topic, researchers have reported mixed evidence of predictability. Recently, several researchers have suspected this is due to the time-varying nature of predictability (i.e., the coefficient on the predictor may be nonzero in some periods of time but may be zero in other periods) and tried to estimate its time path \citep{guidolinTimeVaryingStock2013, devpuraStockReturnPredictability2018, rahmanPredictivePowerDividend2019, hammamiUnderstandingTimevaryingShorthorizon2020}. In such studies, the coefficient at each time point is estimated in a rolling-window fashion, that is, by using data over limited sample periods, and authors plot the time path of its estimate, assessing whether the coefficient is time-varying and identifying the period when predictability exists (if any). 

The authors, however, conduct this analysis without statistically confirming the assumption that the coefficient on the predictor is time-varying. They often argue it is time-varying just because its estimated trajectory is seemingly not constant over time. However, the coefficient estimates can be seemingly time-varying even if the coefficient is, in fact, constant over the full sample. This is because the coefficient at each time point is estimated by using a subsample, which tends to be small \citep[for example,][use subsamples of 60 months to produce the sequence of coefficient estimates]{rahmanPredictivePowerDividend2019}, and thus the standard errors of the coefficient estimates tend to be large. This implies that even if the coefficient is actually constant over time, the time variations of its estimates can be so large that only a graphical inspection erroneously supports the hypothesis of time-varying predictability.

A more decent approach is to statistically test for the constancy of the coefficient before estimating it or its trajectory. If the null of constant coefficient is rejected, the rolling-window estimation may be applied to identify the period when predictability exists (otherwise, the usual method based on the full sample should be used to test whether the coefficient is zero).

Testing procedures for coefficient constancy in predictive regressions have recently been discussed by \citet{georgievTestingParameterInstability2018}. They propose several bootstrap-based tests under the assumption that the predictor is highly persistent, and find that a test based on \citeauthor{nyblomTestingConstancyParameters1989}'s (\citeyear{nyblomTestingConstancyParameters1989}) LM test performs well when the coefficient on the predictor experiences an abrupt break or follows a near unit root AR(1) process. This result is due to the fact that \citeauthor{nyblomTestingConstancyParameters1989}'s (\citeyear{nyblomTestingConstancyParameters1989}) LM test is designed to be the locally most powerful against martingale coefficient processes, which accommodate random walk and processes with abrupt breaks. However, in empirical applications, the coefficient process may not be a martingale, and it is unclear whether the LM test retains good power in such a case.

Models with non-martingale random coefficients are widely seen in the literature. \citet{swamyLinearPredictionEstimation1980} propose regression models with stationary random coefficients, gives them a justification from statistical and economic perspectives, and discuss prediction and estimation methods. \citet{linTestingParameterConstancy1999} derive the LM test for coefficient constancy against the alternative of stationary random coefficient, and \citet{fuSpecificationTestsTimevarying2023} propose a test to distinguish between stationary and persistent variations in parameters. Stationary-coefficient models are also employed in empirical studies. For instance, \citet{chiangForecastingTreasuryBill1991} estimate a model where spot interest rate is regressed on forward interest rate, and coefficients are assumed to be possibly stationary stochastic processes. \citet{leeRevisitingDemandMoney2013} estimate the money demand function with stationary random coefficients.

When stationary (or $\mathrm{I}(0)$) random coefficient is assumed as the alternative hypothesis, we can naturally test the null of constant coefficient by using a Wald-type test based on a linear model derived from the original predictive regression. This testing procedure is essentially the same as the one employed by \citet{nishiTestingCoefficientRandomness2023}, who considers AR(1) models with i.i.d. random coefficient and shows that the Wald test performs well in such a case. In this article, we show that the Wald test performs better than Nyblom's LM test in predictive regressions with stationary (rather than i.i.d.) random coefficient. The Wald test is, however, less powerful than the LM test when the coefficient process follows a near unit root (or $\mathrm{I}(1)$) process.\footnote{One can also consider intermediate cases where the coefficient process is of $\mathrm{I}(d)$, where $d\in (0,1)$, but we do not consider such cases and will say that the coefficient process is stationary if it is $\mathrm{I}(0)$.}

The power comparison between the LM and Wald tests is made under the sequence of local alternatives. As shown later, the appropriate localizing rate of the alternative hypotheses depends on the persistence of the random coefficient and which test (the LM or Wald) is considered. When the random coefficient is $\mathrm{I}(0)$, the LM test has a nontrivial asymptotic power against the sequence of local alternatives approaching the null of zero randomness at the rate of $T^{-1/2}$, while the appropriate rate of the local alternatives for the Wald test is $T^{-3/4}$, where $T$ is the sample size. This implies that the Wald test can detect smaller variations in the random coefficient than the LM test when the random coefficient is $\mathrm{I}(0)$. In contrast, when the random coefficient is $\mathrm{I}(1)$, the LM test has nontrivial asymptotic power against the sequence of local alternatives of order $T^{-3/2}$, but the Wald test against that of order $T^{-5/4}$. Thus, the LM test detects smaller coefficient variations than the Wald test when the random coefficient is $\mathrm{I}(1)$.

Because one test does not dominate the other in terms of power, we consider combining the LM and Wald tests to obtain new tests that have good power properties irrespective of the persistence of the random coefficient. Our proposal is simply to take the sum and product of the two statistics. We will investigate the power properties of four tests in total.

The asymptotic null distributions of the Wald and LM test statistics are found to depend on nuisance parameters under our setting (and so do the combinations of these two tests). To implement these tests without the knowledge about the nuisance parameters, we base them on subsampling, in which critical values are approximated by using a number of test statistics calculated from subsamples \citep{andrewsHybridSizeCorrectedSubsampling2009}.\footnote{Another way to circumvent the nuisance-parameter problem is to use bootstrap as \citet{georgievTestingParameterInstability2018} do. Because different approaches require different mathematical assumptions and discussions, we focus on subsampling. The reason why we use subsampling instead of bootstrap is explained in Remark 3 in Section 2.} We discuss the validity of the subsampling approach in our setting by some theoretical and numerical analyses.

Finally, we revisit \citeauthor{devpuraStockReturnPredictability2018}'s (\citeyear{devpuraStockReturnPredictability2018}) study. They determine whether predictability is time-varying
based on a sequence of p-values associated with the rolling-window coefficient estimates. They do this, however, without any adjustment for multiple testing, resulting in uncontrolled size. In this article, we investigate the coefficient constancy by applying the tests considered in this article. Our result mostly reverses \citeauthor{devpuraStockReturnPredictability2018}'s (\citeyear{devpuraStockReturnPredictability2018}) conclusion.

The remainder of this article is structured as follows. In Section 2, we define the Wald test statistic assuming that the random coefficient is an $\mathrm{I}(0)$ process. We analyze and compare the asymptotic behavior of the Wald and LM tests under this setting. In Section 3, these two tests are compared under the assumption that the coefficient process is a near unit root, and two combination tests of the Wald and LM tests are proposed. Section 4 proposes to base these tests on subsampling. In Section 5, the model is extended to the heteroskedastic case, and we define modified test statistics. Section 6 conducts Monte Carlo simulations to investigate finite-sample behavior of the tests, and Section 7 gives a real data application. Section 8 concludes the article. All mathematical proofs are relegated to Appendix A.

\section{The Case of $\mathrm{I}(0)$ Random Coefficient}

\subsection{Model and assumptions}
In this section, we consider a predictive regression where the coefficient on the persistent predictor is expressed by the sum of a constant $\beta$ and stationary random variable $s_{\beta,t}$:
\begin{align}
	\begin{split}
		y_t &= \alpha_y + (\beta + \omega_\beta s_{\beta,t})x_{t-1} + \varepsilon_{y,t}, \quad t=1,\ldots,T, \\
		x_t &= \alpha_x + s_{x,t}, \\
        s_{x,t} &= (1+c_xT^{-1})s_{x,t-1} + \varepsilon_{x,t},
	\end{split}
	 \label{model:pr_sta}
\end{align}
where $\mathrm{E}[s_{\beta,t}]=0$, $\mathrm{Var}(s_{\beta,t})=1$, and $s_{x,0}=O_p(1)$. Here, the predictor, $x_{t}$, is modeled as a persistent local-to-unity AR(1) process, which is a commonly employed assumption in the literature \citep[e.g.,][]{campbellEfficientTestsStock2006}. Note that $\beta$ and $\omega_\beta^2$ denote the mean and variance of the coefficient on $x_{t-1}$, respectively. Therefore, coefficient constancy is equivalent to $\omega_\beta^2=0$ in our setting. To simplify the discussion in what follows, we set $\alpha_y=\alpha_x=s_{x,0}=0$, resulting in
\begin{align}
	y_t = (\beta + \omega_\beta s_{\beta,t})x_{t-1} + \varepsilon_{y,t}, \quad t=1,\ldots,T, \label{model:pr_sta_set0}
\end{align}
with $x_t = (1+c_xT^{-1})x_{t-1} + \varepsilon_{x,t} \ (x_0=0)$.

\begin{rem}
	\citet{georgievTestingParameterInstability2018} consider cases where the intercept term as well as the slope parameter may vary over time. If the intercept term is the sum of constant $\alpha_y$ and stationary random variable $\omega_\alpha s_{\alpha,t}$, then the random part is absorbed into the disturbance term, and thus $\omega_\alpha$ is not identified. In empirical studies, however, researchers are often interested only in the time variation of the slope parameter. Therefore, we may focus on models where only the slope parameter possibly exhibits time variation. Interested readers are referred to \citet{georgievTestingParameterInstability2018} for marginal and joint tests for time variation in the intercept and slope (when they are nonstationary).
\end{rem}

We suppose the following assumptions hold.

\begin{asm}
	
	\begin{itemize}
		\item[(i)] $\varepsilon_{y,t}/\sigma_y = \gamma(\varepsilon_{x,t}/\sigma_x) + v_t$, where $\sigma_i^2 \coloneqq \mathrm{Var}(\varepsilon_{i,t})$, $i=x,y$, $\mathrm{E}[\varepsilon_{x,t}]=\mathrm{E}[v_t]=\mathrm{E}[\varepsilon_{x,t}v_t]=0$ and $\gamma \in (-1,1)$ is the correlation coefficient between $\varepsilon_{y,t}$ and $\varepsilon_{x,t}$.
		\item[(ii)] $(\varepsilon_{x,t},v_t,s_{\beta,t})$ is stationary and $\alpha$-mixing with mixing coefficients $\alpha(m) = O(m^{-\frac{r}{r-2}})$ for some $r>2$.
		\item[(iii)] $\mathrm{E}[|\varepsilon_{x,t}|^{4(r-1)+\delta}] + \mathrm{E}[|v_t|^{4(r-1)+\delta}] + \mathrm{E}[|s_{\beta,t}|^{4(r-1)+\delta}] < \infty$ for some $\delta>0$.
		\item[(iv)] $(\varepsilon_{x,t},v_t,\eta_t)$, where $\eta_t \coloneqq \varepsilon_{y,t}^2 - \sigma_y^2$, is a martingale difference sequence (m.d.s.) with respect to some filtration $\mathcal{F}_t$ to which $(\varepsilon_{x,t}, v_t, s_{\beta,t})$ is adapted.
	\end{itemize}
	\label{asm:stationary_case_wald}
\end{asm}

In Assumption \ref{asm:stationary_case_wald}(i), the disturbance $\varepsilon_{y,t}$ of predictive regression \eqref{model:pr_sta_set0} is correlated with the innovation $\varepsilon_{x,t}$ that drives $x_{t}$, with correlation coefficient $\gamma$. Assumption \ref{asm:stationary_case_wald}(ii) specifies that the innovation process is stationary and weakly serially dependent, while all the innovations except $s_{\beta,t}$ are restricted to be serially uncorrelated by Assumption \ref{asm:stationary_case_wald}(iv). Assumption \ref{asm:stationary_case_wald}(iv) also implies conditional homoskedasticity of $\varepsilon_{y,t}$, i.e., $\mathrm{E}[\varepsilon_{y,t}^2|\mathcal{F}_{t-1}] = \sigma_{y}^2$. The case where $\varepsilon_{y,t}$ is conditionally heteroskedastic is considered in Section 5.
 
Define $\sigma_\eta^2 \coloneqq \mathrm{E}[\eta_t^2]$ and $\sigma_{\beta,lr}^2 \coloneqq \mathrm{E}[s_{\beta,1}^2] + 2\sum_{i=1}^{\infty}\mathrm{E}[s_{\beta,1}s_{\beta,1+i}]$. Define also the partial sum process $(W_{x,T}, W_{y,T}, W_{\eta,T},W_{\beta,lr,T})'$ on $[0,1]$ by $W_{i,T}(r) \coloneqq T^{-1/2}\sigma_i^{-1}\sum_{t=1}^{\lfloor Tr\rfloor}\varepsilon_{i,t}$, $i=x,y$, $W_{\eta,T}(r) \coloneqq T^{-1/2}\sigma_\eta^{-1}\sum_{t=1}^{\lfloor Tr\rfloor}\eta_t$ and $W_{\beta,lr,T}(r) \coloneqq T^{-1/2}\sigma_{\beta,lr}^{-1}\sum_{t=1}^{\lfloor Tr\rfloor}s_{\beta,t}$. Then, it follows from the functional central limit theorem (FCLT) that under Assumption \ref{asm:stationary_case_wald}, as $T \to \infty$,
\begin{align}
		(W_{x,T}, W_{y,T}, W_{\eta,T}, W_{\beta,lr,T})'
	\Rightarrow 
		(W_{x}, W_{y}, W_{\eta}, W_{\beta,lr})',
\end{align}
in the Skorokhod space $D_4[0,1]$, where $(W_{x},W_{y},W_{\eta},W_{\beta,lr})'$ is a vector Brownian motion. Furthermore, by the standard continuous mapping argument, we also have $J_{x,T}(r) \coloneqq T^{-1/2}\sigma_x^{-1}x_{\lfloor Tr \rfloor} \Rightarrow J_x(r)$, where $J_x$ is an Ornstein-Uhlenbeck (OU) process satisfying $dJ_x(r) = c_xJ_x(r)dr + dW_x(r)$.

In some analysis, we rely on the following assumption instead of Assumption \ref{asm:stationary_case_wald}:

\begin{asm}
	Assumption \ref{asm:stationary_case_wald} holds with (ii) and (iii) replaced by the following conditions:
	\begin{itemize}
		\item[(ii')] $(\varepsilon_{x,t},v_t,s_{\beta,t})$ is stationary and $\alpha$-mixing with mixing coefficients $\alpha(m) = O(m^{-\frac{r}{r-2}})$ for some $2<r<2.4$.
		\item[(iii')] $\mathrm{E}[|\varepsilon_{x,t}|^{3r}] + \mathrm{E}[|v_t|^{3r}] + \mathrm{E}[|s_{\beta,t}|^{3r}] < \infty$.
	\end{itemize} \label{asm:stationary_case_lm}
\end{asm}

\subsection{Wald and LM tests and asymptotic distributions}

In the sequel, we let $\sigma_{y\beta} \coloneqq \mathrm{Cov}(\varepsilon_{y,t},s_{\beta,t})$.

Our aim here is to construct a Wald-type test for coefficient constancy in \eqref{model:pr_sta_set0}, or $\omega_\beta^2=0$. To make our point clear, suppose for now $\beta$ is known. Define $z_t(\beta) \coloneqq y_t-\beta x_{t-1} = \omega_\beta s_{\beta,t}x_{t-1} + \varepsilon_{y,t}$, and obtain
\begin{align}
	z_t^2(\beta) = \sigma_y^2 + \delta x_{t-1} + \omega_\beta^2 x_{t-1}^2 + \xi_t, \label{eqn:linearized}
\end{align}
where $\delta \coloneqq 2\omega_\beta\sigma_{y\beta}$ and $\xi_t \coloneqq (\varepsilon_{y,t}^2-\sigma_y^2) + 2\omega_\beta x_{t-1}(s_{\beta,t}\varepsilon_{y,t} - \sigma_{y\beta}) + \omega_\beta^2 x_{t-1}^2(s_{\beta,t}^2 - 1)$. Viewing \eqref{eqn:linearized} as a linear model with $\xi_t$ playing the role of the disturbance, \citet{nishiTestingCoefficientRandomness2023} proposes to test $\mathrm{H_0}:\omega_\beta^2=0$ by the Wald test for $(\delta, \omega_\beta^2) = (0,0)$, since $\omega_\beta^2=0$ if and only if $(\delta,\omega_\beta^2) = (0,0)$. Moreover, the LM test of \citet{linTestingParameterConstancy1999}, which is designed to detect stationary random coefficient, can be derived as the (squared) t-test for $\omega_\beta^2=0$ under the restriction $\delta=0$ (see Appendix B). Therefore, it is natural in our setting to employ model \eqref{eqn:linearized} as the basis for constructing a test statistic. To define the Wald test explicitly, we first regress the variables on a constant, obtaining
\begin{align}
	\widetilde{z_t^2}(\beta) = \delta \tilde{x}_{1,t-1} + \omega_\beta^2 \tilde{x}_{2,t-1} + \tilde{\xi}_t, \label{eqn:linearized_demean}
\end{align}
where $\widetilde{z_t^2}(\beta) \coloneqq z_t^2(\beta) - T^{-1}\sum_{t=1}^{T}z_t^2(\beta)$, $\tilde{x}_{1,t-1} \coloneqq x_{t-1} - T^{-1}\sum_{t=1}^Tx_{t-1}$, $\tilde{x}_{2,t-1} \coloneqq x_{t-1}^2 - T^{-1}\sum_{t=1}^{T}x_{t-1}^2$, and $\tilde{\xi}_t \coloneqq \xi_t - T^{-1}\sum_{t=1}^T\xi_t$. Letting $Z_2(\beta) \coloneqq (\widetilde{z_1^2}(\beta),\ldots,\widetilde{z_T^2}(\beta))'$, $X_1 \coloneqq (\tilde{x}_{1,0},\ldots,\tilde{x}_{1,T-1})'$, $X_2 \coloneqq (\tilde{x}_{2,0},\ldots,\tilde{x}_{2,T-1})'$, and $\Xi \coloneqq (\tilde{\xi}_1,\ldots,\tilde{\xi}_t)'$, we arrive at
\begin{align}
	Z_2(\beta) = X\theta + \Xi, \label{eqn:linearized_matrix}
\end{align}
where $X\coloneqq (X_1,X_2)$ and $\theta\coloneqq (\delta,\omega_\beta^2)'$. Then, the Wald test statistic is
\begin{align}
	\mathrm{W}_T(\beta) \coloneqq \hat{\sigma}_{\tilde{\xi}}^{-2}(\beta)\hat{\theta}_T(\beta)'(X'X)\hat{\theta}_T(\beta),
\end{align}
where $\hat{\theta}_T(\beta)$ and $\hat{\sigma}_{\tilde{\xi}}^2(\beta)$ are the OLS coefficient and variance estimators from \eqref{eqn:linearized_matrix}, respectively: $\hat{\theta}_T(\beta) = (X'X)^{-1}X'Z_2(\beta)$ and $\hat{\sigma}_{\tilde{\xi}}^2(\beta) = T^{-1}Z_2(\beta)'M_XZ_2(\beta)$ with $M_X \coloneqq I - X(X'X)^{-1}X'$.

The Wald test defined above uses the true value of $\beta$. Because $\beta$ is usually unknown in applications, we use $\mathrm{W}_T(\hat{\beta})$, where $\hat{\beta}$ is the OLS estimator of $\beta$: $\hat{\beta} =\sum_{t=1}^{T}y_t x_{t-1}/\sum_{t=1}^{T}x_{t-1}^2$. The use of $\hat{\beta}$ makes the Wald test invariant to the value of $\beta$ and feasible without any knowledge about $\beta$ (in particular, about whether it is zero or not).

We also consider \citeauthor{nyblomTestingConstancyParameters1989}'s (\citeyear{nyblomTestingConstancyParameters1989}) LM test, which is defined by
\begin{align}
	\mathrm{LM}_T \coloneqq \frac{1}{T\hat{\sigma}_T^2\sum_{t=1}^{T}x_{t-1}^2}\sum_{i=1}^{T}\biggl(\sum_{t=1}^{i}x_{t-1}z_t(\hat{\beta})\biggr)^2,
\end{align}
where $\hat{\sigma}_T^2 \coloneqq T^{-1}\sum_{t=1}^{T}z_t^2(\hat{\beta})$.\footnote{\citet{georgievTestingParameterInstability2018} also investigate the performance of the $\mathrm{Sup}F$ test proposed by \citet{andrewsTestsParameterInstability1993}. Although this test is known to be comparable in terms of power with Nyblom's LM test when parameters are $\mathrm{I}(1)$, it is not designed to detect (non)stationary randomness in parameters. Therefore, Nyblom's LM test seems more appropriate as the test to be compared with our Wald test, as argued by \citet{hansenTestsParameterInstability2002}.}

\begin{rem}
		When $(\alpha_y,\alpha_x,s_{x,0}) \neq (0,0,0)$ in \eqref{model:pr_sta}, $y_t$ and $x_{t-1}$ are replaced with their demeaned counterparts in \eqref{eqn:linearized} and the definitions of $z_t(\beta)$ and $\hat{\beta}$.
		\label{rem:demeaning}
\end{rem}

\begin{thm}
	\begin{itemize}
		\item[(i)] Suppose Assumption \ref{asm:stationary_case_wald} holds. If $\omega_\beta^2 = g^2T^{-3/2}$ and $\mathrm{Corr}(\varepsilon_{y,t},s_{\beta,t}) = \sigma_{y\beta}/\sigma_y = mT^{-1/4}$, then we have, under model \eqref{model:pr_sta_set0},
		\begin{align}
			\mathrm{W}_T(\hat{\beta}) \Rightarrow &\Bigg\{
			\begin{pmatrix}
				\int_{0}^{1}\tilde{J}_{x,1}(r)dW_\eta(r) \\ \int_{0}^{1}\tilde{J}_{x,2}(r)dW_\eta(r) 
			\end{pmatrix}
			+ \frac{\sigma_x}{\sigma_\eta} \begin{pmatrix} g^2\sigma_x\int_{0}^{1}\tilde{J}_{x,1}(r)\tilde{J}_{x,2}(r)dr+2gm\sigma_y\int_{0}^{1}\tilde{J}_{x,1}(r)^2dr \\  g^2\sigma_x\int_{0}^{1}\tilde{J}_{x,2}(r)^2dr+2gm\sigma_y\int_{0}^{1}\tilde{J}_{x,1}(r)\tilde{J}_{x,2}(r)dr
			\end{pmatrix}\Biggr\}' \notag \\ 
			&\times \begin{pmatrix}
				\int_{0}^{1}\tilde{J}_{x,1}(r)^2dr & \int_{0}^{1}\tilde{J}_{x,1}(r)\tilde{J}_{x,2}(r)dr \\ \int_{0}^{1}\tilde{J}_{x,2}(r)\tilde{J}_{x,1}(r)dr & \int_{0}^{1}\tilde{J}_{x,2}(r)^2dr
			\end{pmatrix}^{-1} \notag \\
			&\times \Bigg\{
			\begin{pmatrix}
				\int_{0}^{1}\tilde{J}_{x,1}(r)dW_\eta(r) \\ \int_{0}^{1}\tilde{J}_{x,2}(r)dW_\eta(r) 
			\end{pmatrix}
			+ \frac{\sigma_x}{\sigma_\eta} \begin{pmatrix} g^2\sigma_x\int_{0}^{1}\tilde{J}_{x,1}(r)\tilde{J}_{x,2}(r)dr+2gm\sigma_y\int_{0}^{1}\tilde{J}_{x,1}(r)^2dr \\  g^2\sigma_x\int_{0}^{1}\tilde{J}_{x,2}(r)^2dr+2gm\sigma_y\int_{0}^{1}\tilde{J}_{x,1}(r)\tilde{J}_{x,2}(r)dr
			\end{pmatrix}\Biggr\}, \label{asym_dist:wald_sta}
		\end{align}
		where $\tilde{J}_{x,1}(r) \coloneqq J_x(r) - \int_{0}^{1}J_x(s)ds$ and
		$\tilde{J}_{x,2}(r) \coloneqq J_x(r)^2 - \int_{0}^{1}J_x(s)^2ds$.\footnote{The localization $\mathrm{Corr}(\varepsilon_{y,t},s_{\beta,t}) = mT^{-1/4}$ is just a technical condition to prevent test statistics from diverging and is not of interest in itself; see the proof of Theorem \ref{thm:asym_dists_sta} and \citet{nishiTestingCoefficientRandomness2023}.}
		
		\item[(ii)] Suppose Assumption \ref{asm:stationary_case_lm} holds. If $\omega_\beta^2 = g^2T^{-1}$ and $\mathrm{Corr}(\varepsilon_{y,t},s_{\beta,t}) = mT^{-1/4}$, then we have, under model \eqref{model:pr_sta_set0},
		\begin{align}
			\mathrm{LM}_T &\Rightarrow \frac{\int_{0}^{1}\bigl[\int_{0}^{s}J_x(r)dA(r) + g(\sigma_x\sigma_{\beta,lr}/\sigma_y)\int_{0}^{s}J_x(r)dB(r)\bigr]^2ds}{\bigl(1+g^2(\sigma_x^2/\sigma_y^2)\int_{0}^{1}J_x(r)^2dr\bigr)\int_{0}^{1}J_x(r)^2dr}, \\
			\intertext{where}
			A(r) &\coloneqq W_y(r) - \frac{\int_{0}^{1}J_x(s)dW_y(s)}{\int_{0}^{1}J_x(s)^2ds}\int_{0}^{r}J_x(s)ds \\
			\intertext{and}
			B(r) \coloneqq \Biggl(\int_{0}^{r}J_x(s)dW_{\beta,lr}(s) -& \frac{\int_{0}^{1}J_x(s)^2dW_{\beta,lr}(s)}{\int_{0}^{1}J_x(s)^2ds}\int_{0}^{r}J_x(s)ds\Biggr) + \frac{\Lambda_{x\beta}}{\sigma_x\sigma_{\beta,lr}}\Biggl(r-\frac{2\int_{0}^{1}J_x(s)ds}{\int_{0}^{1}J_x(s)^2ds}\int_{0}^{r}J_x(s)ds\Biggr)
		\end{align}
		with $\Lambda_{x\beta} \coloneqq \sum_{i=1}^{\infty}\mathrm{E}[\varepsilon_{x,1}s_{\beta,1+i}]$. 
		
	\end{itemize}
	\label{thm:asym_dists_sta}
\end{thm}

There are two points worth mentioning in Theorem \ref{thm:asym_dists_sta}. First, the asymptotic distributions of the Wald and LM test statistics are derived under different local alternatives. The Wald test statistic has the limiting alternative distribution when $\omega_\beta=gT^{-3/4}$, while the LM test does when $\omega_\beta=gT^{-1/2}$. Because the former specification captures a sequence of local alternatives closer to the null of $\omega_\beta=0$, it implies the Wald test can detect smaller variations in random coefficient than the LM test, as long as the random coefficient is $\mathrm{I}(0)$. This is intuitively because the Wald test is constructed under the assumption that the random coefficient follows a stationary process, while the derivation of the LM test is based on the assumption that the coefficient process is nonstationary. This finding is in line with the results of \citet{linTestingParameterConstancy1999}. They verify through simulation that their test, which is designed to detect stationary random coefficient, outperforms \citeauthor{nyblomTestingConstancyParameters1989}'s LM test when the random coefficient is a stationary AR process. Theorem \ref{thm:asym_dists_sta} gives a theoretical explanation for their simulation results.\footnote{\citeauthor{linTestingParameterConstancy1999}'s (\citeyear{linTestingParameterConstancy1999}) LM test has the asymptotic alternative distribution when $\omega_\beta=gT^{-3/4}$, as our Wald test does. See Appendix B for details.}

The second point is that the asymptotic null distributions ($g=0$) of the Wald and LM test statistics are dependent on nuisance parameters. Specifically, the Wald test depends on the correlation between $W_x$ and $W_\eta$, and the LM test on the correlation between $W_x$ and $W_y$. Moreover, these tests depend on $c_x$, the persistence parameter of $x_t$. Therefore, information on the nuisance parameters is required to perform these tests. In Section 4, we propose to base them on subsampling to solve this problem.

\begin{rem}
	\citet{georgievTestingParameterInstability2018} propose to include $\Delta x_t$ in the right hand side of \eqref{model:pr_sta_set0} to make the LM test independent of the correlation between $W_x$ and $W_y$, which renders the use of bootstrap valid in their setting. As for the Wald test, however, a Bonferroni approach is needed to make the test free from the correlation between $W_x$ and $W_\eta$ \citep{nishiTestingCoefficientRandomness2023}. Because the joint use of the bootstrap and Bonferroni methods is impractical, we will rely on subsampling, which does not require the elimination of the correlations.
\end{rem}

\section{The Case of $\mathrm{I}(1)$ Random Coefficient}

In this section, we consider a predictive regression model where the random part of the coefficient is near $\mathrm{I}(1)$:
\begin{align}
	\begin{split}
		y_t &= (\beta + \omega_\beta s_{\beta,t})x_{t-1} + \varepsilon_{y,t}, \quad t=1,\ldots,T, \\
		s_{\beta,t} &= (1+c_\beta T^{-1})s_{\beta,t-1} + \varepsilon_{\beta,t} \ (s_0=0),
	\end{split}
	 \label{model:pr_nonsta_set0}
\end{align}
where $x_t = (1+c_xT^{-1})x_{t-1} + \varepsilon_{x,t} \ (x_0=0)$. We impose the following condition on $\varepsilon_{\beta,t}$.

\begin{asm}
	$\varepsilon_{\beta,t}$ is an m.d.s. and $\mathrm{Var}(\varepsilon_{\beta,t})=1$. \label{asm:varepsilon_beta}
\end{asm}

\citet{georgievTestingParameterInstability2018} prove that, under model \eqref{model:pr_nonsta_set0}, the LM test statistic has the asymptotic alternative distribution when $\omega_\beta = gT^{-3/2}$. The next theorem establishes that the Wald test statistic has the asymptotic alternative distribution if $\omega_\beta = gT^{-5/4}$. To state this precisely, we need the following notation: let $J_\beta(r)$ be an OU process satisfying $dJ_\beta(r) = c_\beta J_\beta(r)dr + dW_\beta(r)$, where $W_\beta(\cdot)$ is the standard Brownian motion to which $W_{\beta,T}(\cdot) \coloneqq T^{-1/2}\sum_{t=1}^{\lfloor T\cdot \rfloor}\varepsilon_{\beta,t}$ weakly converges.

\begin{thm}
	Suppose Assumption \ref{asm:stationary_case_wald} with $s_{\beta,t}$ replaced by $\varepsilon_{\beta,t}$ and Assumption \ref{asm:varepsilon_beta} hold. If $\omega_\beta = gT^{-5/4}$, then we have, under model \eqref{model:pr_nonsta_set0},
	\begin{align}
		\mathrm{W}_T(\hat{\beta}) \Rightarrow &\Bigg\{
		\begin{pmatrix}
			\int_{0}^{1}\tilde{J}_{x,1}(r)dW_\eta(r) \\ \int_{0}^{1}\tilde{J}_{x,2}(r)dW_\eta(r) 
		\end{pmatrix}
		+ \frac{g^2\sigma_x^2}{\sigma_\eta} \begin{pmatrix} \int_{0}^{1}\tilde{J}_{x,1}(r)J_x(r)^2Q(r)^2dr \\  \int_{0}^{1}\tilde{J}_{x,2}(r)J_x(r)^2Q(r)^2dr
		\end{pmatrix}\Biggr\}' \notag \\ 
		&\times \begin{pmatrix}
			\int_{0}^{1}\tilde{J}_{x,1}(r)^2dr & \int_{0}^{1}\tilde{J}_{x,1}(r)\tilde{J}_{x,2}(r)dr \\ \int_{0}^{1}\tilde{J}_{x,2}(r)\tilde{J}_{x,1}(r)dr & \int_{0}^{1}\tilde{J}_{x,2}(r)^2dr
		\end{pmatrix}^{-1} \notag \\
	    &\times \Bigg\{
		\begin{pmatrix}
			\int_{0}^{1}\tilde{J}_{x,1}(r)dW_\eta(r) \\ \int_{0}^{1}\tilde{J}_{x,2}(r)dW_\eta(r) 
		\end{pmatrix}
		+ \frac{g^2\sigma_x^2}{\sigma_\eta} \begin{pmatrix} \int_{0}^{1}\tilde{J}_{x,1}(r)J_x(r)^2Q(r)^2dr \\  \int_{0}^{1}\tilde{J}_{x,2}(r)J_x(r)^2Q(r)^2dr
		\end{pmatrix}\Biggr\}, \label{asym_dist:wald_nonsta}
	\end{align}
	where $Q(r) \coloneqq J_\beta(r) - (\int_{0}^{1}J_x(r)^2dr)^{-1}\int_{0}^{1}J_x(r)^2J_\beta(r)dr$.
	\label{thm:wald_nonsta}
\end{thm}

Based on Theorems \ref{thm:asym_dists_sta} and \ref{thm:wald_nonsta} and the result of \citet{georgievTestingParameterInstability2018}, we observe that the convergence rate of $\omega_\beta$ under which the tests have nontrivial asymptotic power depends on the nature of random coefficient $s_{\beta,t}$. We summarize this in Table \ref{tab:convergence_rates}. When the random part $s_{\beta,t}$ of the coefficient is $\mathrm{I}(0)$, the Wald test has a nontrivial power for $\omega_\beta=gT^{-3/4}$, which describes a sequence of local alternatives closer to the null of $\omega_\beta=0$ than $\omega_\beta=gT^{-1/2}$ for the LM case. However, in the case of $\mathrm{I}(1)$ $s_{\beta,t}$, the convergence rate of $\omega_\beta$ associated with the nontrivial power of the LM test is $gT^{-3/2}$, which converges to the null of $\omega_\beta=0$ faster than $gT^{-5/4}$ for the Wald test. This implies that the Wald test dominates the LM test in terms of power when $s_{\beta,t}$ is $\mathrm{I}(0)$, while the LM test is more powerful than the Wald test when $s_{\beta,t}$ is $\mathrm{I}(1)$.

\begin{rem}
	Nyblom's LM test is also the locally most powerful test against a structural break. Suppose that $s_{\beta,t} = 1\{t>\lfloor\tau_0T \rfloor\}$, where $1\{\cdot\}$ is the indicator function, and $\tau_0 \in(0,1)$. Then, it can be shown that our Wald test has a nontrivial asymptotic power when $\omega_\beta=gT^{-3/4}$, whereas Nyblom's LM test statistic has a local alternative distribution under $\omega_\beta=gT^{-1}$, as shown by \citet{georgievTestingParameterInstability2018}. This implies that Nyblom's LM test is more powerful than our Wald test when parameter instability is due to a structural break.\footnote{Note that our analysis is about the local-alternative, or small-variation case. As pointed out by \citet{perronUsefulnessLackThereof2016}, different results might be obtained for the tests studied in this article, if we consider large variations in parameters and allow for lagged dependent variables as regressors and serial correlation in the error term. We exclude such a case under model \eqref{model:pr_sta} and Assumptions \ref{asm:stationary_case_wald} and \ref{asm:stationary_case_lm}.} Because the asymptotic distribution of the Wald statistic can be derived in a similar way to that in the case of $\mathrm{I}(1)$ $s_{\beta,t}$, the detail is omitted.
\end{rem}

Given the above observation, one ideal testing procedure is to use the Wald test when $\mathrm{I}(0)$ random coefficient is suspected and use the LM test when $\mathrm{I}(1)$ random coefficient is probable. However, the persistence of the coefficient (if it is random) is often unknown, and thus it will be difficult to rely on such a procedure in practice. To detect coefficient randomness efficiently whether it is $\mathrm{I}(0)$ or $\mathrm{I}(1)$, we propose two combination tests of the Wald and LM, one being defined as the sum of them and the other as the product. Specifically, we consider the following test statistics:
\begin{align}
	\mathrm{S}_T \coloneqq \mathrm{W}_T(\hat{\beta}) + \mathrm{LM}_T \quad \text{and} \quad \mathrm{P}_T \coloneqq \mathrm{W}_T(\hat{\beta}) \times \mathrm{LM}_T.
\end{align}
We shall call $\mathrm{S}_T$ the Sum test statistic and $\mathrm{P}_T$ the Product test statistic. The motivation of these two tests is simple. When $s_{\beta,t}$ is $\mathrm{I}(0)$ and $\omega_\beta=gT^{-3/4}$, $\mathrm{W}_T(\hat{\beta})$, $\mathrm{S}_T$ and $\mathrm{P}_T$ weakly converge to alternative distributions, but $\mathrm{LM}_T$ to the null distribution. In contrast, if $s_{\beta,t}$ is $\mathrm{I}(1)$ and $\omega_\beta=gT^{-3/2}$, $\mathrm{LM}_T(\hat{\beta})$, $\mathrm{S}_T$ and $\mathrm{P}_T$ have nontrivial asymptotic power but $\mathrm{W}_T(\hat{\beta})$ does not. In other words, the Sum and Product tests are more influenced by the more powerful of the Wald and LM tests, and thus they are expected to perform well in both the $\mathrm{I}(0)$ and $\mathrm{I}(1)$ $s_{\beta,t}$ cases.\footnote{A more preferable way to combine the LM and Wald tests will be to weight those test statistics based on some information on the persistence of the random coefficient. For example, the weighting function might be a function of the test statistic proposed by \citet{fuSpecificationTestsTimevarying2023}, which is aimed at distinguishing between stationary and persistent parameters. However, their statistic cannot be directly used in our setting because their analysis is based on the assumption that regressor $x_t$ is weakly stationary.} The finite-sample performances of these tests are examined in Section 6.

\section{Subsampling Approach}

\subsection{Basics}
As observed in previous sections, tests we have considered are dependent on nuisance parameters. In the literature, the subsampling approach has been a popular solution to nuisance parameter problems \citep{romanoSubsamplingIntervalsAutoregressive2001,choiSubsamplingVectorAutoregressive2005,andrewsHybridSizeCorrectedSubsampling2009}. The method allows us to approximate the sampling distribution of a statistic by calculating a number of the statistics based on subsets of the full sample. The idea behind this strategy is that, in large samples, the sampling distribution of subsample statistics will be close to that of the full sample statistic, and hence the empirical distribution of a number of subsample statistics can approximate it.

To fix ideas, take the LM test as an example. To implement the subsampling method, one has to set the subsample size $b < T$, the number of data used to calculate each subsample statistic. Then, the first subsample statistic is calculated using the subsample ranging from $t=1$ to $t=b$, the second one is based on $t=2,\ldots,b+1$, and continue until the last statistic based on $t=T-b+1,\ldots,T$ is calculated. Letting $\mathrm{LM}_{j,b}$ denote the subsample LM statistic based on $b$ data ranging from $t=j$, the subsample test statistics can be written as
\begin{align}
	\mathrm{LM}_{j,b} = \frac{1}{b\hat{\sigma}^2_{j,b}\sum_{t=j}^{j+b-1}x_{t-1}^2}\sum_{i=j}^{j+b-1}\Bigl(\sum_{t=j}^{i}x_{t-1}z_t(\hat{\beta}_{j,b})\Bigr)^2,\quad j = 1,2,\ldots,T-b+1,
\end{align}
where $\hat{\beta}_{j,b} \coloneqq \sum_{t=j}^{j+b-1}y_tx_{t-1}/\sum_{t=j}^{j+b-1}x^2_{t-1}$ is the $\beta$ estimate based on $t=j,\ldots,j+b-1$ and $\hat{\sigma}^2_{j,b} \coloneqq b^{-1}\sum_{t=j}^{j+b-1}z^2_t(\hat{\beta}_{j,b})$. If the significance level is $\alpha$, then the critical value (denoted by $c_{\mathrm{LM}}(1-\alpha)$) is defined by the $1-\alpha$ quantile of the empirical distribution function
\begin{align}
	L_{b,T}(x) \coloneqq (T-b+1)^{-1}\sum_{j=1}^{T-b+1}1\{LM_{j,b} \leq x\},
\end{align}
where $1\{\cdot\}$ is the indicator function. The null is rejected if the full sample test statistic $\mathrm{LM}_T$ exceeds the critical value. The Wald, Sum and Product tests can be based on the subsampling approach in the same way. 

\subsection{The choice of $b$}

When implementing the subsampling method, one has to choose the subsample size $b$. One necessary condition is  $1/b + b/T\to 0$ as $T\to \infty$ \citep{romanoSubsamplingIntervalsAutoregressive2001}. One reasonable way to choose $b$ is to compute the empirical distribution $L_{b,T}$ for a number of $b$ in some set $B$ and select $b$ that leads to the most `appropriate' $L_{b,T}$ in some sense. In the literature, several ways to define $B$ and the appropriateness criterion have been proposed. We here follow \citet{jachSubsamplingInferenceMean2012} and set, for some $q \in (0,1)$, $B = \{b_j \coloneqq \lfloor q^jT \rfloor|j=j_{\mathrm{max}},j_{\mathrm{max}}-1,\ldots,j_{\mathrm{min}}\}$, with $j_{\mathrm{min}}<j_{\mathrm{max}} \ (j_{\mathrm{min}},j_{\mathrm{max}} \in \mathbb{N})$, and select $b$ such that $b^{\mathrm{sel}} \coloneqq \argmin_{b_j \in B} \  \mathrm{KS}(L_{b_j,T},L_{b_{j+1},T})$, where $\mathrm{KS}(x,y)$ denotes the Kolmogorov-Smirnov (KS) distance between distributions $x$ and $y$.

The idea behind this selection rule is that $L_{b,T}$, viewed as a function of $b$, should vary only slightly in an appropriate range of $b$, and hence values of $b$ leading to small KS distances should be selected. After preliminary experiments, we set $(q,j_{\mathrm{min}},j_{\mathrm{max}}) = (0.85,\lfloor \log_{0.85}(0.27) \rfloor,\lfloor \log_{0.85}(5/T^{0.9}) \rfloor)$ for the LM test and $(q,j_{\mathrm{min}},j_{\mathrm{max}}) = (0.95,\lfloor \log_{0.95}(0.3) \rfloor,\lfloor \log_{0.95}(15/T^{0.9}) \rfloor)$ for the Wald, Sum and Product tests, so that we consider $b$'s up to about 30\% of the sample size $T$.

\begin{rem}
	One drawback of the above approach is that it does not guarantee $b^{\mathrm{sel}}/T \to 0$ as $T\to \infty$, since $\max B/T \to q^{j_{\mathrm{min}}} >0$. To check whether this condition holds, we will investigate the behavior of $b^{\mathrm{sel}}/T$ as $T$ increases in the Monte Carlo simulation conducted in Section 6.
	\label{ref:subsample_size_condition}
\end{rem}

\subsection{Validity of subsampling}

\citet{andrewsHybridSizeCorrectedSubsampling2009} consider subsampling-based inference in nonregular models, including regression models with nearly integrated regressors, and provide sufficient conditions for subsampling-based tests to work. Although it will take a complicated analysis to check whether their sufficient conditions hold in our setting, the (in)validity of the subsampling approach can be conjectured by checking the monotonicity in $c_x$ of the critical value of the asymptotic null distributions. Here, we follow Andrews and Guggenberger's argument to discuss the validity of the subsampling approach in our setting.

For the sake of exposition, let $S_T$ denote a generic test statistic calculated using the full sample and $S_{j,b}, \ j =1,\ldots,T-b+1,$ denote its subsampling counterparts, which are functions of predictor $x_t = (1+c_x/T)x_{t-1} + \varepsilon_{x,t}$. Suppose that, under the null hypothesis, $S_T$ has an asymptotic distribution $P_{c_x}$, dependent on $c_x$. The aim of subsampling is to approximate the critical value from $P_{c_x}$ by the empirical distribution of $S_{j,b}$. However, due to the local-to-unity nature of the predictor, subsample statistics $S_{j,b}$ weakly converge to $P_0$, the asymptotic distribution of $S_T$ obtained under $c_x=0$. This is because $x_t = (1+c_x/T)x_{t-1} + \varepsilon_{x,t} = (1+c_x(b/T)/b)x_{t-1} + \varepsilon_{x,t} = (1+o(1)/b)x_{t-1} + \varepsilon_{x,t}$ (since $b/T \to 0$ as $T\to \infty$), so that under the asymptotics of $b\to \infty$ (rather than $T \to \infty$) $S_{j,b}$'s weakly converge to $P_0$. Therefore, the validity of subsampling is determined by the relation between the critical values from $P_{c_x}$ (the weak limit of $S_T$) and $P_0$ (the weak limit of $S_{j,b}$). 

\subsubsection{The LM and Wald tests}
\begin{figure}
	\centering
	\begin{subfigure}{0.8\textwidth}
		\centering
		\includegraphics[width=\textwidth]{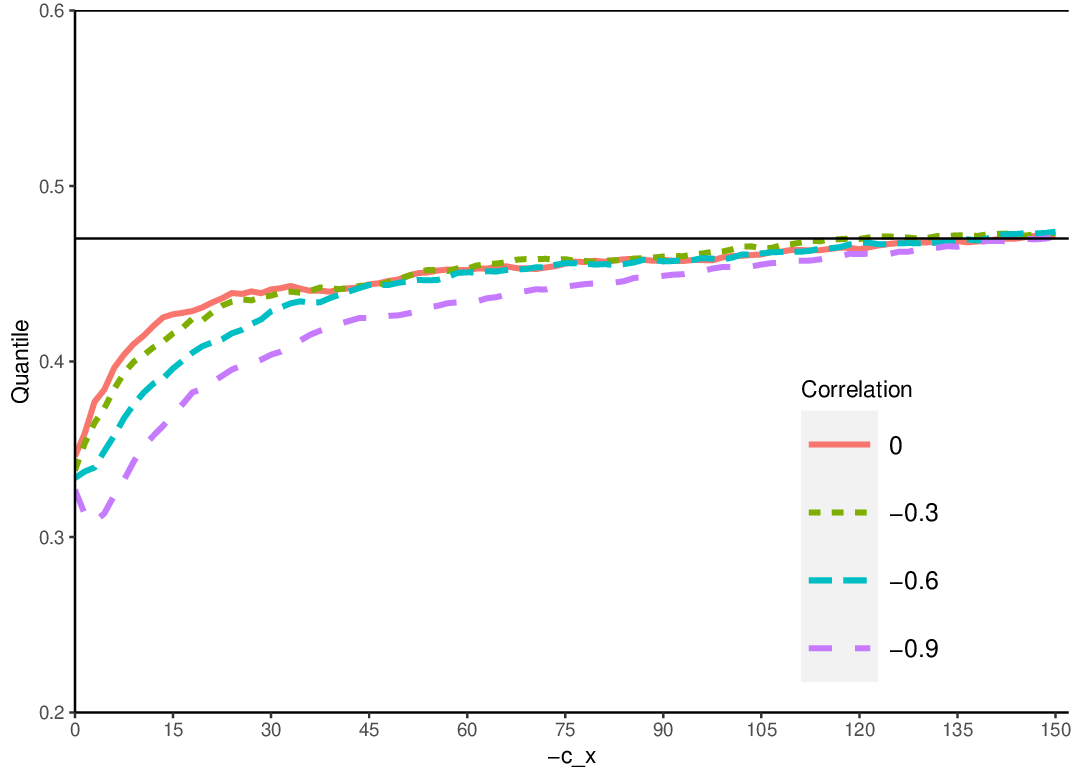}
		\caption{Asymptotic 0.95 quantiles}
		\label{fig:quantiles_asym_LM}
		\begin{minipage}{12cm}
			\footnotesize
			\centering
			{(The horizontal line indicates 0.47, the critical value under $c_x=-\infty$.) \par}
		\end{minipage}
	\end{subfigure}
	
	\begin{subfigure}{0.8\textwidth}
		\centering
		\includegraphics[width=\textwidth]{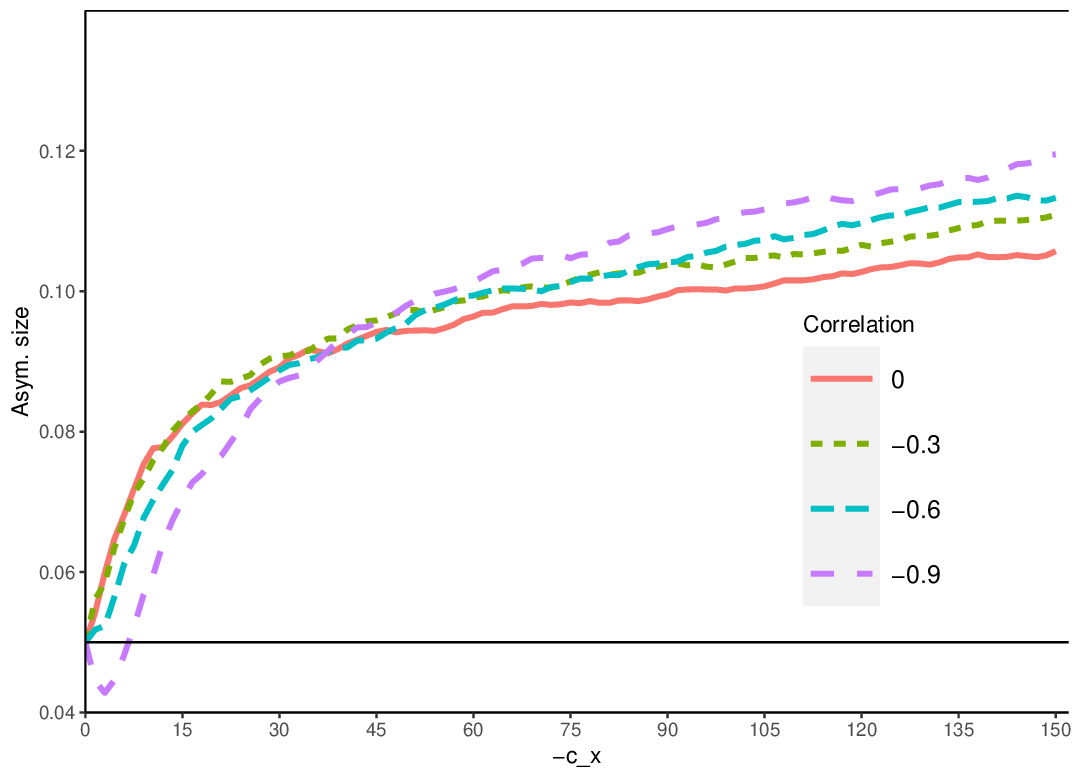}
		\caption{Asymptotic sizes under the 0.05 nominal level}
		\label{fig:sizes_asym_LM}
	\end{subfigure}
	\caption{Asymptotic behavior of the LM test}
\end{figure}

\begin{figure}
	\centering
	\begin{subfigure}{0.8\textwidth}
		\centering
		\includegraphics[width=\textwidth]{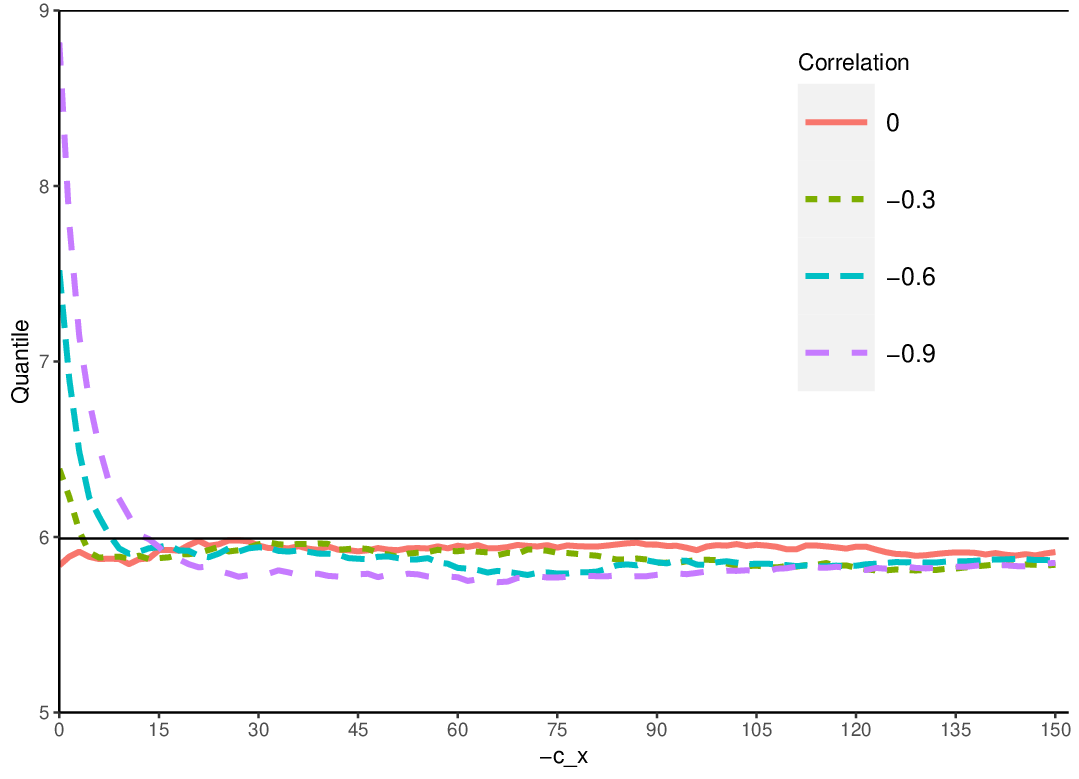}
		\caption{Asymptotic 0.95 quantiles}
		\label{fig:quantiles_asym_wald}
		\begin{minipage}{12cm}
			\footnotesize
			\centering
			{(The horizontal line indicates 5.99, the critical value under $c_x=-\infty$.) \par}
		\end{minipage}
	\end{subfigure}
	\begin{subfigure}{0.8\textwidth}
		\centering
		\includegraphics[width=\textwidth]{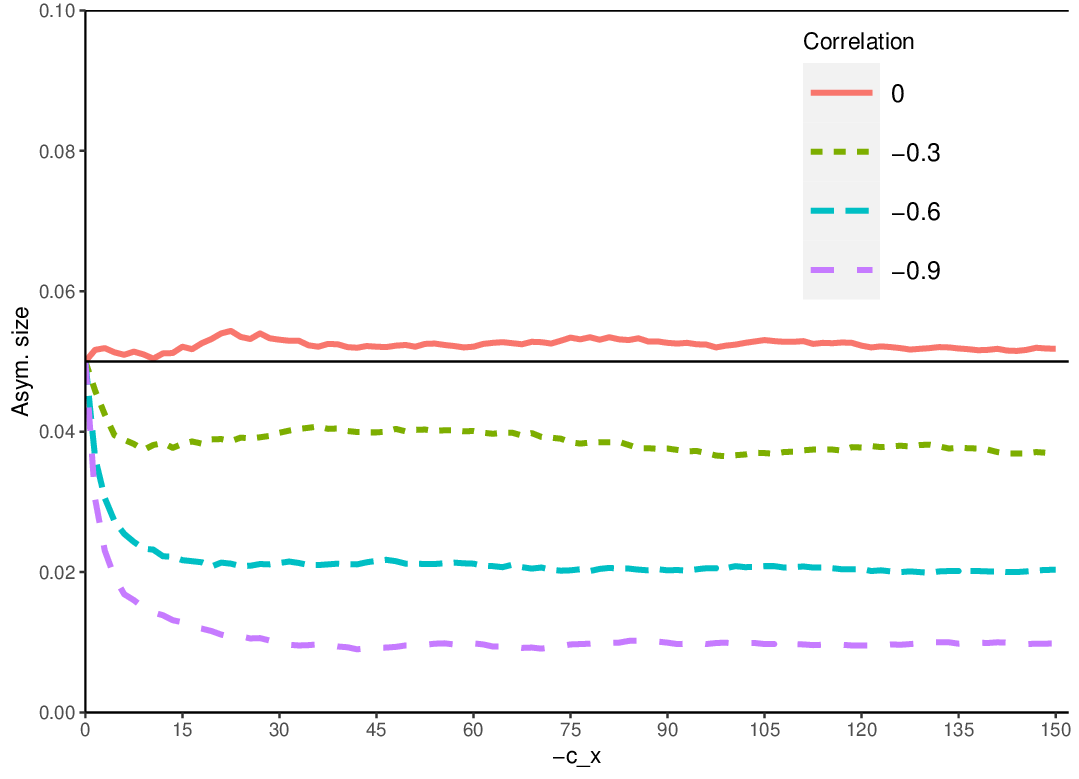}
		\caption{Asymptotic sizes under the 0.05 nominal level}
		\label{fig:sizes_asym_wald}
	\end{subfigure}
	\caption{Asymptotic behavior of the Wald test}
 \label{fig:subsampling_asym_wald}
\end{figure}

Figures \ref{fig:quantiles_asym_LM} and \ref{fig:quantiles_asym_wald} depict the 0.95 quantiles of the asymptotic distributions of the LM and Wald statistics, respectively, as a function of $c_x$. The quantiles of the LM statistic are computed for $\mathrm{Corr}(\varepsilon_{x,t},\varepsilon_{y,t}) = 0,-0.3,-0.6,-0.9$, and the quantiles of the Wald statistic are computed for $\mathrm{Corr}(\varepsilon_{x,t},\eta_t) = 0,-0.3,-0.6,-0.9$.\footnote{Quantiles are insensitive to the sign of the correlations, and thus we consider negative correlations only.} Computations are based on 20000 simulations where $(W_x,J_x,W_y,W_\eta)$ are approximated by normalized sums of 1000 steps. For the LM test, the asymptotic 0.95 quantile is minimized at $c_x=0$, and as $c_x$ tends to $-\infty$, it increases and converges to 0.47, the asymptotic 0.95 quantile obtained when $x_t$ is $\mathrm{I}(0)$ \citep{hansenTestingParameterInstability1992}, for any value of $\mathrm{Corr}(\varepsilon_{x,t},\varepsilon_{y,t})$. Note that the critical value taken from the empirical distribution of the subsample LM statistics asymptotically corresponds to that obtained under $c_x=0$. Therefore, if the true value of $c_x$ is negative, the critical value based on the subsample LM statistics is smaller than the 0.95 quantile of the asymptotic distribution of the full-sample LM statistic for any $\mathrm{Corr}(\varepsilon_{x,t},\varepsilon_{y,t})$, resulting in over-rejection. To quantify the degree of the over-rejection, we show in Figure \ref{fig:sizes_asym_LM} the asymptotic sizes of the subsampling LM test with significance level 0.05. It indicates that the LM test will be oversized for almost all the values of $c_x$, irrespective of the value of $\mathrm{Corr}(\varepsilon_{x,t},\varepsilon_{y,t})$.

To solve over-rejection problems associated with subsampling tests, \citet{andrewsHybridSizeCorrectedSubsampling2009} propose a hybrid test, where the referenced critical value is simply the maximum of the subsampling critical value and the critical value obtained under $c_x = -\infty$. In the case of the LM test with 0.05 significance level, the hybrid critical value is $c^*_{\mathrm{LM}}(0.95) \coloneqq \mathrm{max}\{c_{\mathrm{LM}}(0.95), 0.47\}$, where $c_{\mathrm{LM}}(0.95)$ is the subsampling critical value. Since the asymptotic 0.95 quantile of the full sample LM statistic under the null hypothesis is below 0.47 irrespective of the true value of $c_x$, this makes the LM test a conservative one. In what follows, we will use this hybrid critical value for the LM test.

For the Wald test, the asymptotic 0.95 quantile (Figure \ref{fig:quantiles_asym_wald}) is maximized at $-c_x=0$ irrespective of the value of $\mathrm{Corr}(\varepsilon_{x,t},\eta_t)$ and is decreasing in $-c_x$. This implies that the subsampling Wald test will asymptotically have size equal to or less than 0.05 (see Figure \ref{fig:sizes_asym_wald}). Strictly speaking, the quantile is not maximized at $-c_x=0$ when $\mathrm{Corr}(\varepsilon_{x,t},\eta_t)=0$, and thus the Wald test is slightly oversized in this case. However, we can show that the asymptotic null distribution of the Wald statistic is exactly the chi-square distribution with two degrees of freedom when $\mathrm{Corr}(\varepsilon_{x,t},\eta_t)=0$, irrespective of the value of $c_x$. This is because $W_x$ and $W_\eta$ are independent in this case, and thus $J_x$ and $W_\eta$ also are. This implies that the asymptotic null distribution of $W_T(\hat{\beta})$ given in Theorem \ref{thm:asym_dists_sta} (with $g=0$), if conditional on $W_x$, is chi square with two degrees of freedom, and so is it unconditionally. Therefore, the theoretical asymptotic 0.95 quantile of the Wald statistic is 5.99 for any value of $c_x$, and the type 1 error is exactly 0.05 for all $c_x$. The slight discrepancy between the theoretical result and the result observed in Figure \ref{fig:subsampling_asym_wald} is only due to the approximation error. In summary, the Wald test, if based on subsampling, is expected to be valid in that its size is equal to or below the nominal level.

\subsubsection{The Sum and Product tests}

\begin{sidewaysfigure}
    	\centering
	\begin{subfigure}{0.47\textwidth}
		\centering
		\includegraphics[width=\textwidth]{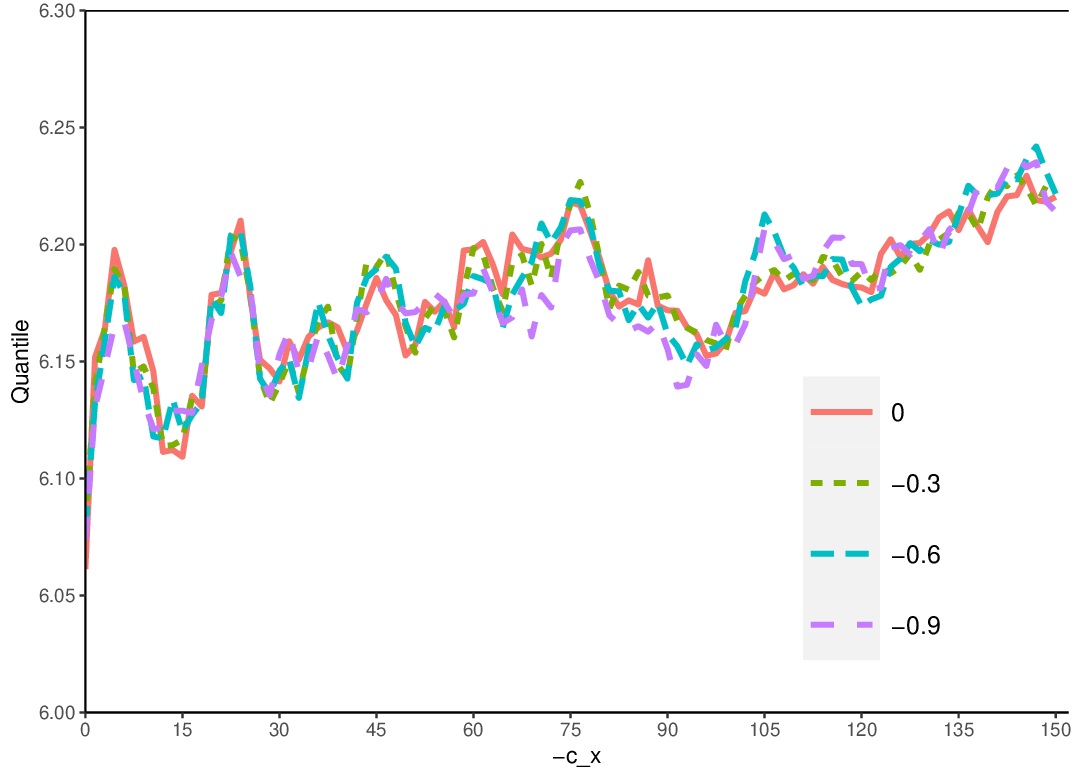}
		\caption{Asymptotic 0.95 quantiles ($\mathrm{Corr}(\varepsilon_{x,t}, \eta_t)=0$)}
		\label{fig:quantiles0_asym_sum}
	\end{subfigure}
	\begin{subfigure}{0.47\textwidth}
		\centering
		\includegraphics[width=\textwidth]{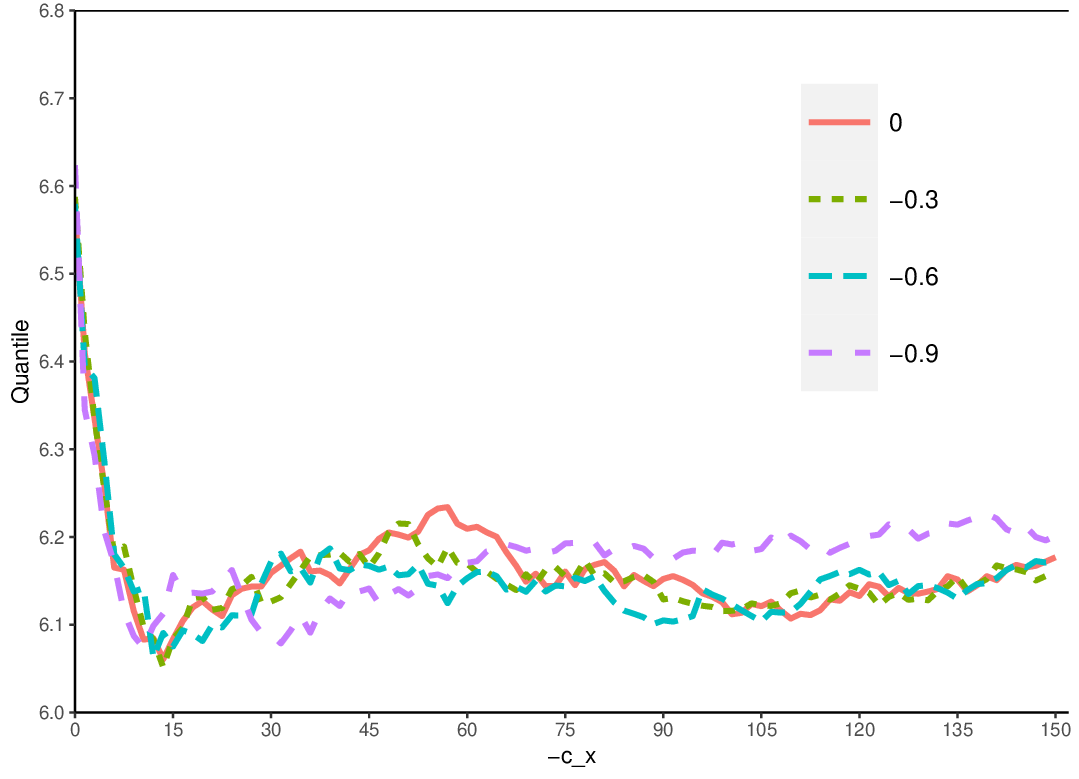}
		\caption{Asymptotic 0.95 quantiles ($\mathrm{Corr}(\varepsilon_{x,t}, \eta_t)=-0.3$)}
		\label{fig:quantiles03_asym_sum}
	\end{subfigure}
 \begin{subfigure}{0.47\textwidth}
		\centering
		\includegraphics[width=\textwidth]{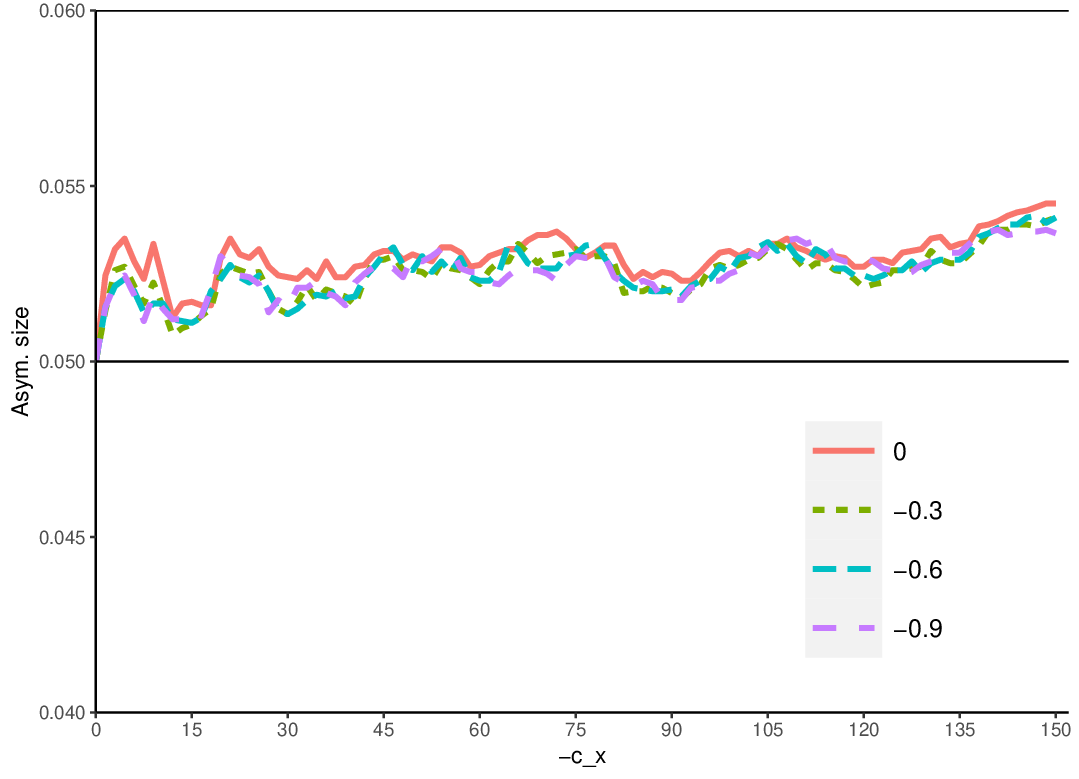}
		\caption{Asymptotic sizes under the 0.05 nominal level ($\mathrm{Corr}(\varepsilon_{x,t}, \eta_t)=0$)}
		\label{fig:size0_asym_sum}
	\end{subfigure}
	\begin{subfigure}{0.47\textwidth}
		\centering
		\includegraphics[width=\textwidth]{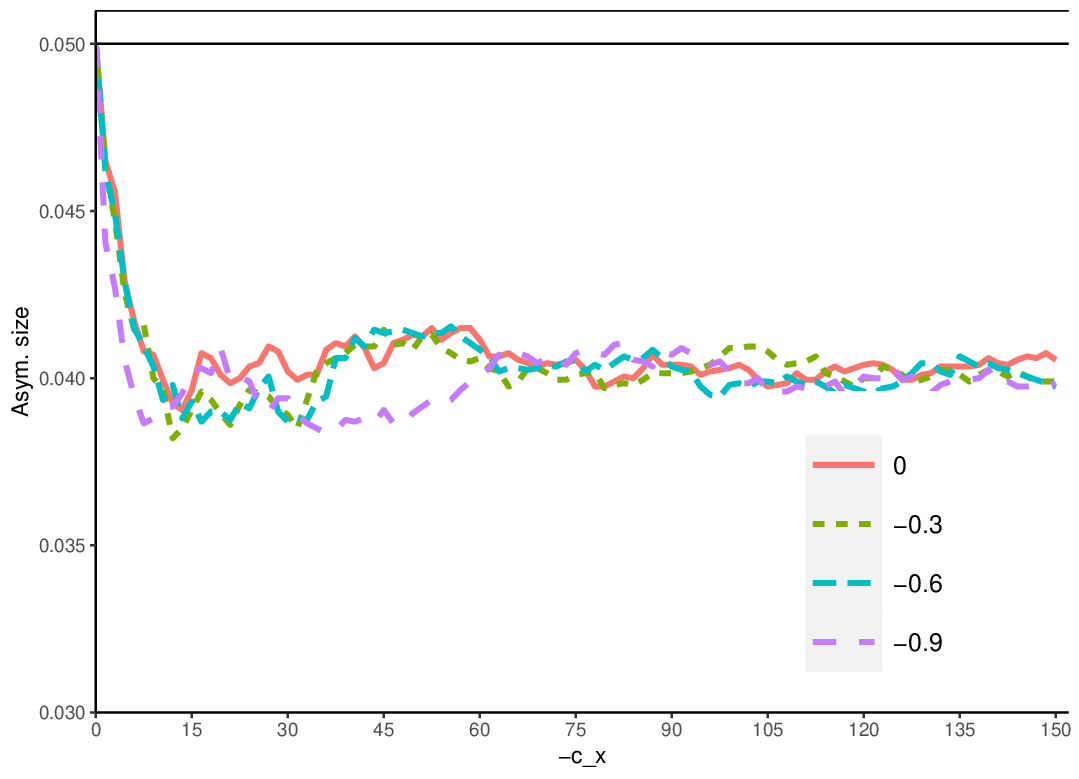}
		\caption{Asymptotic sizes under the 0.05 nominal level ($\mathrm{Corr}(\varepsilon_{x,t}, \eta_t)=-0.3$)}
		\label{fig:size03_asym_sum}
	\end{subfigure}
	\caption{Asymptotic behavior of the Sum test}
    \label{fig:sum_asym}
\end{sidewaysfigure}

\begin{sidewaysfigure}
    	\centering
	\begin{subfigure}{0.47\textwidth}
		\centering
		\includegraphics[width=\textwidth]{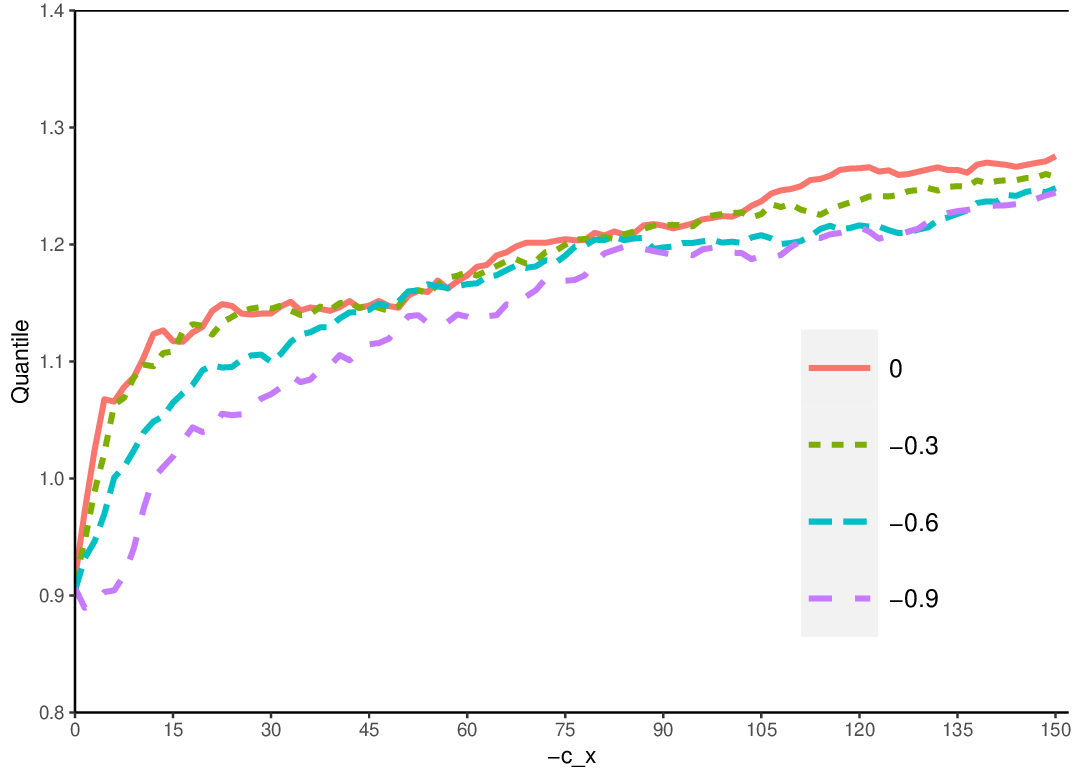}
		\caption{Asymptotic 0.95 quantiles ($\mathrm{Corr}(\varepsilon_{x,t}, \eta_t)=0$)}
		\label{fig:quantiles0_asym_prod}
	\end{subfigure}
	\begin{subfigure}{0.47\textwidth}
		\centering
		\includegraphics[width=\textwidth]{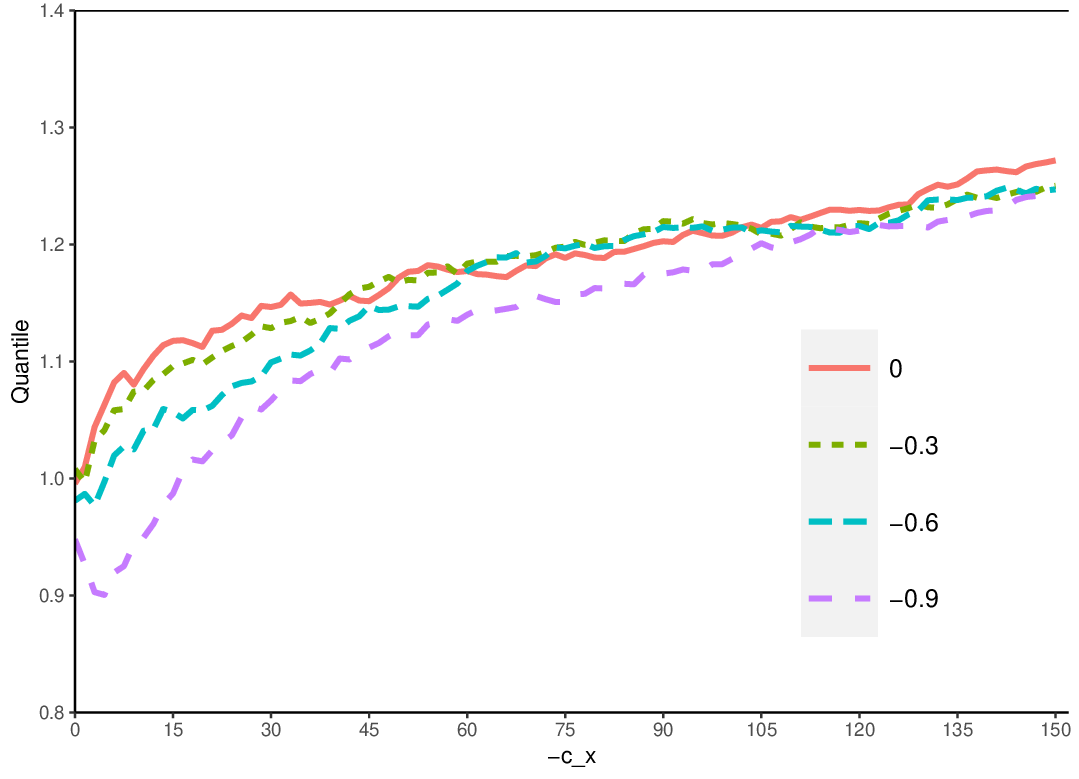}
		\caption{Asymptotic 0.95 quantiles ($\mathrm{Corr}(\varepsilon_{x,t}, \eta_t)=-0.3$)}
		\label{fig:quantiles03_asym_prod}
	\end{subfigure}
 \begin{subfigure}{0.47\textwidth}
		\centering
		\includegraphics[width=\textwidth]{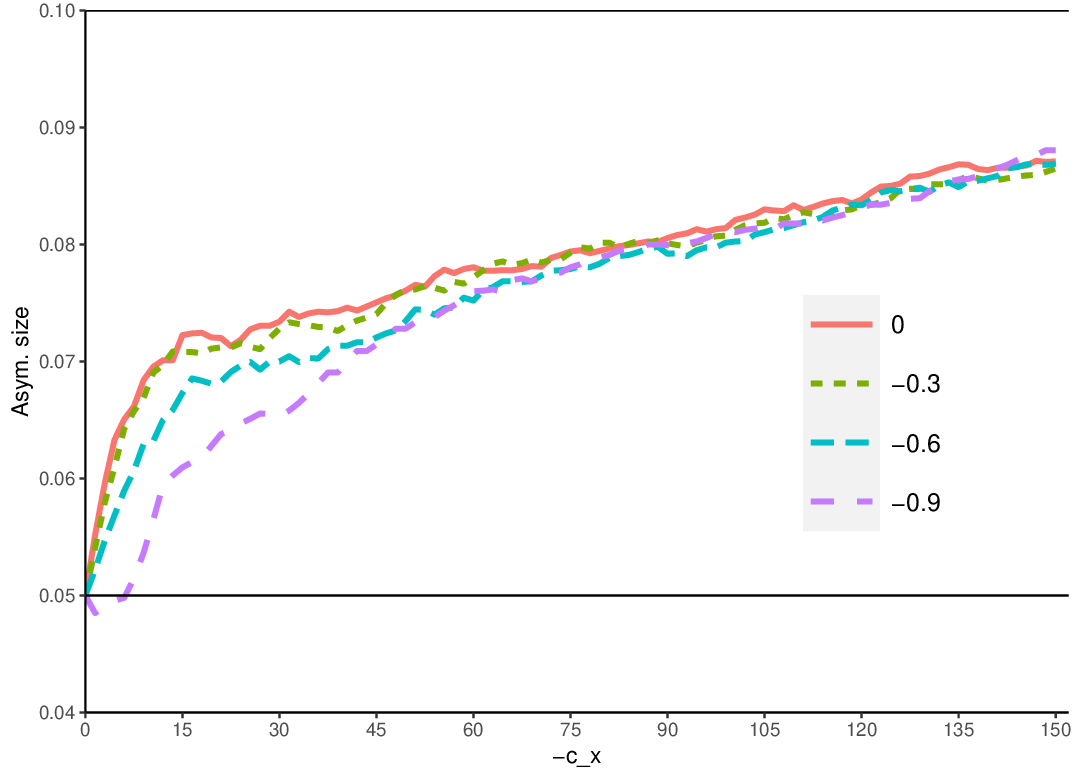}
		\caption{Asymptotic sizes under the 0.05 nominal level ($\mathrm{Corr}(\varepsilon_{x,t}, \eta_t)=0$)}
		\label{fig:size0_asym_prod}
	\end{subfigure}
	\begin{subfigure}{0.47\textwidth}
		\centering
		\includegraphics[width=\textwidth]{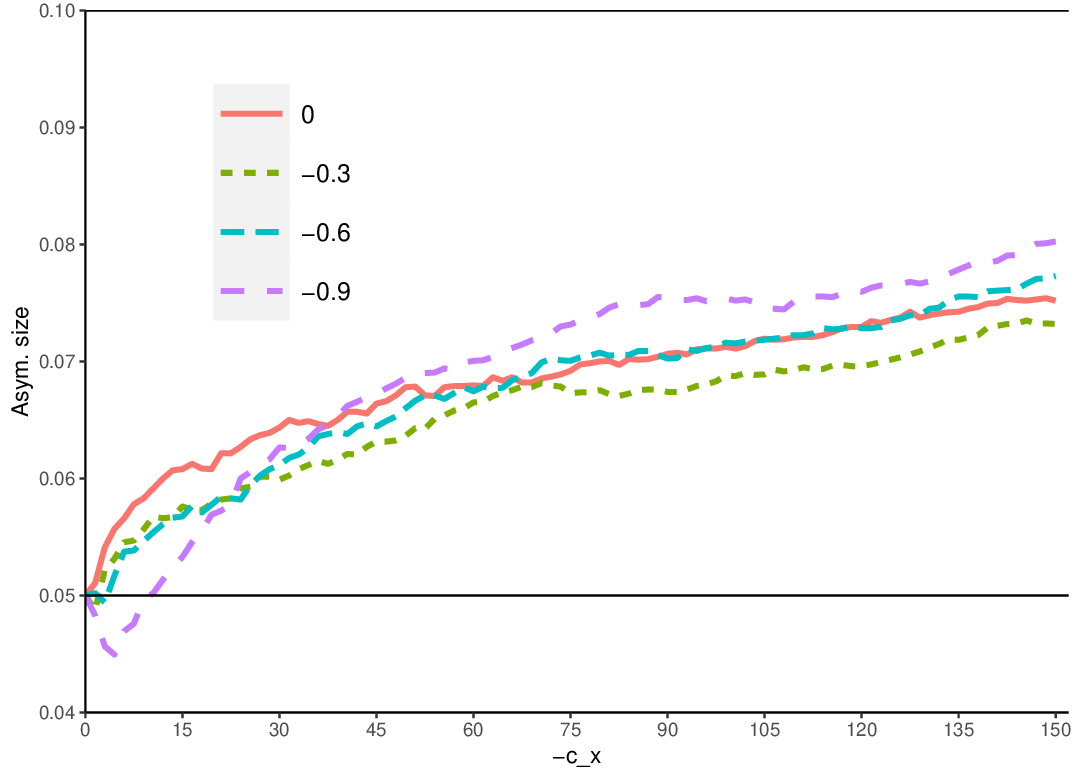}
		\caption{Asymptotic sizes under the 0.05 nominal level ($\mathrm{Corr}(\varepsilon_{x,t}, \eta_t)=-0.3$)}
		\label{fig:size03_asym_prod}
	\end{subfigure}
	\caption{Asymptotic behavior of the Product test}
    \label{fig:prod_asym}
\end{sidewaysfigure}

The analysis for the Sum and Product tests is somewhat complicated because they depend on both $\mathrm{Corr}(\varepsilon_{x,t},\varepsilon_{y,t})$ and $\mathrm{Corr}(\varepsilon_{x,t},\eta_t)$. Moreover, they also depend on $\mathrm{Corr}(\varepsilon_{y,t},\eta_t)$. To gain some insight, we consider the case where $\varepsilon_{y,t}$ is symmetric, which is satisfied when, e.g., $\varepsilon_{y,t}$ is normally distributed. This yields $\mathrm{Corr}(\varepsilon_{y,t},\eta_t)=0$, and thus we can focus on the effect of $\mathrm{Corr}(\varepsilon_{x,t},\varepsilon_{y,t})$ and $\mathrm{Corr}(\varepsilon_{x,t},\eta_t)$ on the asymptotic size. We consider $\mathrm{Corr}(\varepsilon_{x,t},\varepsilon_{y,t})=0,-0.3,-0.6,-0.9$ and $\mathrm{Corr}(\varepsilon_{x,t},\eta_t)=0,-0.3$.\footnote{We do not consider the cases with $\mathrm{Corr}(\varepsilon_{x,t},\eta_t)=-0.6,-0.9$ because these values can cause the variance-covariance matrix of $(\varepsilon_{x,t}, \varepsilon_{y,t}, \eta_t)'$ to be an indefinite matrix.} In Figures \ref{fig:sum_asym} and \ref{fig:prod_asym}, we plot the asymptotic quantiles and size of the Sum and Product tests as functions of $-c_x$ and $\mathrm{Corr}(\varepsilon_{x,t},\varepsilon_{y,t})=0,-0.3,-0.6,-0.9$, fixing the value of $\mathrm{Corr}(\varepsilon_{x,t},\eta_t)$.

In Figure \ref{fig:quantiles0_asym_sum}, we plot the asymptotic 0.95 quantiles of the Sum test statistic when $\mathrm{Corr}(\varepsilon_{x,t},\eta_t)=0$. It attains the minimum value when $-c_x=0$ and increases as $-c_x$ gets larger, although in a non-monotonic way. This leads to a slight over-rejection (see Figure \ref{fig:size0_asym_sum}). However, we suspect that this result is mostly due to the approximation error for the asymptotic null distribution of the Sum test statistic, as in the case of the Wald test. Indeed, we found that the type 1 error is closer to or below the nominal level if we increase the steps for the approximation of $(W_x,J_x,W_y,W_\eta)$ and the number of replications. When $\mathrm{Corr}(\varepsilon_{x,t},\eta_t)=-0.3$ (Figure \ref{fig:quantiles03_asym_sum}), the asymptotic 0.95 quantile is decreasing in $-c_x$ in general and attains the maximum at $-c_x=0$. Therefore, the asymptotic size is equal to or below the nominal 0.05 level.

Figure \ref{fig:quantiles0_asym_prod} plots the asymptotic 0.95 quantiles of the Product statistic when $\mathrm{Corr}(\varepsilon_{x,t},\eta_t)=0$. It attains the minimum value when $-c_x=0$ and increases as $-c_x$ gets larger, and thus the Product test is asymptotically oversized (see Figure \ref{fig:size0_asym_prod}). When $\mathrm{Corr}(\varepsilon_{x,t},\eta_t)=-0.3$, the same results are obtained (Figures \ref{fig:quantiles03_asym_prod} and \ref{fig:size03_asym_prod}). Therefore, the Product test should not be based on subsampling, even in the special case (where $\varepsilon_{y,t}$ is symmetric).

\begin{rem}
	The hybrid approach proposed by \citet{andrewsHybridSizeCorrectedSubsampling2009} is not directly applicable to the Sum and Product tests. This is because the asymptotic null distributions of these test statistics depend on the correlation between the LM and Wald tests, even when $c_x= -\infty$, which leads to violation of Assumption K of \citet{andrewsHybridSizeCorrectedSubsampling2009}. Size correction methods for these tests will be a topic for future works.
\end{rem}

The finite-sample size and power of the subsampling LM, Wald, and Sum tests will be investigated in Section 6.

\section{Conditionally Heteroskedastic Case}

In this section, we extend our discussions thus far to the conditionally heteroskedastic case, which is more plausible in many applications. Incorporating the conditional heteroskedasticity requires us to modify the tests considered in the preceding sections.

\subsection{LM test}

The heteroskedasticity-robust version of the LM test is derived by \citet{hansenTestingParameterInstability1992}, which is of the form
\begin{align}
\mathrm{LM}_T^*	\coloneqq \frac{1}{T\sum_{t=1}^{T}x_{t-1}^2z_t^2(\hat{\beta})}\sum_{i=1}^{T}\Bigl(\sum_{t=1}^{i}x_{t-1}z_t(\hat{\beta})\Bigr)^2.
\end{align}
Subsampling counterparts are defined similarly.

\subsection{Wald test}

For the Wald test, the usual modification involving the use of HC variance estimator is insufficient in our setting. The problem is that, when predictive regression error $\varepsilon_{y,t}$ is conditionally heteroskedastic, then $\varepsilon_{y,t}^2$ is serially correlated, which induces serial correlation in $\eta_t = \varepsilon_{y,t}^2 - \sigma_y^2$ and violates Assumption \ref{asm:stationary_case_wald}(iv).

To solve this problem, we follow \citet{nishiStochasticLocalModerate2024} to modify the Wald test. To facilitate the discussion, the following regularity condition is imposed on $\varepsilon_{x,t}$, $v_t$ and $\eta_t$.

\begin{asm}
    \begin{itemize}
        \item[(i)] $(\varepsilon_{x,t},v_t)'$ is stationary and $\alpha$-mixing with mixing coefficients $\alpha(m) = O(m^{-\gamma})$ for some $\gamma>6$. Moreover, $E[|\varepsilon_{x,t}|^{6}] +E[|v_t|^6] + E[|\eta_t|^6] < \infty$.

        \item[(ii)] $(\varepsilon_{x,t},v_t)'$ is an m.d.s. with respect to some filtration $\mathcal{F}_t$ to which $(\varepsilon_{x,t},v_t)'$ is adapted.
        
        \item [(iii)]  The long-run variance-covariance matrix of $(\varepsilon_{x,t}, \eta_t)'$ is positive definite:
        \begin{align}
	   \begin{pmatrix}
		  \sigma_{x}^2 & \sigma_{x\eta, lr} \\
		  \sigma_{x\eta, lr} & \sigma_{\eta,lr}^2
	   \end{pmatrix}
            > 0,
        \end{align}
        where $\sigma_x^2=\mathrm{E}[\varepsilon_{x,t}^2]$, $\sigma_{\eta,lr}^2\coloneqq E[\eta_1^2]+2\sum_{t=1}^{\infty}E[\eta_1\eta_{t+1}]$, and $\sigma_{x\eta,lr}\coloneqq E[\varepsilon_{x,1}\eta_1]+\sum_{t=1}^{\infty}E[\varepsilon_{x,1}\eta_{t+1}] + \sum_{t=1}^{\infty}E[\eta_1\varepsilon_{x,t+1}]$.
    \end{itemize}
	\label{asm:serial_corr}
\end{asm}

Note that $\varepsilon_{y,t}$ and $\eta_t$ are stationary and $\alpha$-mixing with the same mixing coefficients as $(\varepsilon_{x,t},v_t)'$ under Assumption \ref{asm:serial_corr}(i). Assumption \ref{asm:serial_corr}(i) allows for weak serial correlation in $\eta_t$, while we keep the no serial correlation assumption for $\varepsilon_{x,t}$ and $v_t$ (and thus $\varepsilon_{y,t}$) in (ii) (cf. Assumption \ref{asm:stationary_case_wald}(iv)). Under this condition, conditional heteroskedasticity in $\varepsilon_{y,t}$ may be permitted. In Assumption \ref{asm:serial_corr}(iii), we assume that $\varepsilon_{x,t}$ and $\eta_t$ have a finite 6th moment. This condition is a technical one to guarantee that Theorem 3.1 of \citet{liangWeakConvergenceStochastic2016} holds. This assumption might be weakened if a more sophisticated asymptotic theory is available.

Under Assumption \ref{asm:serial_corr}, we have $\Bigl(T^{-1/2}\sigma_{x}^{-1}\sum_{t=1}^{\lfloor T\cdot \rfloor}\varepsilon_{x,t}, T^{-1/2}\sigma_{\eta,lr}^{-1}\sum_{t=1}^{\lfloor T\cdot \rfloor}\eta_t\Bigr)' \Rightarrow \Bigl(W_x(\cdot), W_{\eta,lr}(\cdot)\Bigr)'$, where $(W_x, W_{\eta,lr})'$ is a vector Brownian motion. Furthermore, the one-sided long-run covariance between $\varepsilon_{x,t}$ and $\eta_t$, $\Lambda_{x\eta} \coloneqq \sum_{t=1}^{\infty}E[\varepsilon_{x,1}\eta_{t+1}]$, exists.

Under Assumption \ref{asm:serial_corr}, we can show that the asymptotic null distribution of the Wald test statistic becomes
\begin{align}
	\mathrm{W}_T(\hat{\beta}) &\Rightarrow \sigma_\eta^{-2}
	\begin{pmatrix}
		\sigma_{\eta,lr}\int_{0}^{1}\tilde{J}_{x,1}(r)dW_{\eta,lr}(r) + \Lambda_{x\eta}/\sigma_x \\ \sigma_{\eta,lr}\int_{0}^{1}\tilde{J}_{x,2}(r)dW_{\eta,lr}(r) + 2(\Lambda_{x\eta}/\sigma_x)\int_{0}^{1}J_x(r)dr
	\end{pmatrix}' \\
	&\times\begin{pmatrix}
		\int_{0}^{1}\tilde{J}_{x,1}(r)^2dr & \int_{0}^{1}\tilde{J}_{x,1}(r)\tilde{J}_{x,2}(r)dr \\ \int_{0}^{1}\tilde{J}_{x,2}(r)\tilde{J}_{x,1}(r)dr & \int_{0}^{1}\tilde{J}_{x,2}(r)^2dr
	\end{pmatrix}^{-1} 
	\begin{pmatrix}
		\sigma_{\eta,lr}\int_{0}^{1}\tilde{J}_{x,1}(r)dW_{\eta,lr}(r) + \Lambda_{x\eta}/\sigma_x \\ \sigma_{\eta,lr}\int_{0}^{1}\tilde{J}_{x,2}(r)dW_{\eta,lr}(r) + 2(\Lambda_{x\eta}/\sigma_x)\int_{0}^{1}J_x(r)dr
	\end{pmatrix};
 \label{dist:wald_serial_corr}
\end{align} 
see Appendix A for the proof. In view of this expression, we modify $\mathrm{W}_T(\hat{\beta})$ in the following way:
\begin{align}
	\mathrm{W}_T^*(\hat{\beta}) \coloneqq \hat{\sigma}_{\tilde{\xi},lr}^{-2}(\hat{\beta})\Biggl(\hat{\theta} - \begin{pmatrix}
		X'X
	\end{pmatrix}^{-1}
	\begin{pmatrix}
		T\hat{\Lambda}_{x\eta,T} \\
		2\hat{\Lambda}_{x\eta,T}\sum_{t=1}^{T}x_{t-1}
	\end{pmatrix}\Biggr)'\begin{pmatrix}
		X'X
	\end{pmatrix} \Biggl(\hat{\theta} - \begin{pmatrix}
	X'X
	\end{pmatrix}^{-1}
	\begin{pmatrix}
	T\hat{\Lambda}_{x\eta,T} \\
	2\hat{\Lambda}_{x\eta,T}\sum_{t=1}^{T}x_{t-1}
	\end{pmatrix}\Biggr),
\end{align}
where $\hat{\sigma}_{\tilde{\xi},lr}^2(\hat{\beta})$ and $\hat{\Lambda}_{x\eta,T}$ are consistent estimators of $\sigma_{\eta,lr}^2$ and $\Lambda_{x\eta}$, respectively. In this article, the long-run (co)variance estimators are constructed nonparametrically using the Parzen kernel, i.e.,
\begin{align}
	\hat{\sigma}_{\tilde{\xi},lr}^2(\hat{\beta}) &= T^{-1}\sum_{t=1}^{T}\hat{\tilde{\xi}}_t^2 + \sum_{i=1}^{l}w\Bigl(\frac{i}{l+1}\Bigr)\Bigl(2T^{-1}\sum_{t=i+1}^{T}\hat{\tilde{\xi}}_{t-i}\hat{\tilde{\xi}}_{t}\Bigr), \notag \\
	\hat{\Lambda}_{x\eta,T} &= \sum_{i=1}^{l}w\Bigl(\frac{i}{l+1}\Bigr)\Bigl(T^{-1}\sum_{t=i+1}^{T}\Delta x_{t-i}\hat{\tilde{\xi}}_t\Bigr), \label{def:lambda_x_eta}
\end{align}
where $\hat{\tilde{\xi}}_t$ is the OLS residual from \eqref{eqn:linearized_demean}, and
\begin{align}
	w(x) = \begin{cases}
		1-6x^2+6x^3 & \mathrm{for} \ x \in [0,1/2] \\
		2(1-x)^3 & \mathrm{for} \ x \in (1/2,1] \\
		0 & \mathrm{otherwise}.
	\end{cases}
\end{align}
Here, $l$ denotes the truncation number, for which we set $l=\lfloor 4(T/100)^{1/3}\rfloor$ in this article.

\begin{rem}
    Another way to construct $\hat{\Lambda}_{x\eta,T}$ would be to replace $\Delta x_{t-i}$ with $\hat{\varepsilon}_{x,t-i}$, the residual obtained from the regression of $x_t$ on $(1,x_{t-1})$. This choice is valid in our setting, where $x_t$ is modeled as (near) $\mathrm{I}(1)$. However, if $x_{t}$ is $\mathrm{I}(0)$, the correction involving $\hat{\Lambda}_{x\eta,T}$ should be asymptotically negligible. In such a case, we need $\hat{\Lambda}_{x\eta,T}=o_p(T^{-1/2})$. As shown in Theorem 4.1 and Lemma 8.1 of \citet{phillipsFullyModifiedLeast1995}, this is achieved by using \eqref{def:lambda_x_eta} and setting $l = cT^{\kappa}$ for some $c>0$ and $\kappa \in(1/4,2/3)$.
\end{rem}

Under the null $\mathrm{H}_0:\omega_\beta=0$, a standard argument will show that these estimators are consistent because $\hat{\beta}-\beta=O_p(T^{-1})$. It will be tedious and much involved to establish the consistency of these estimators under the alternative hypothesis. However, we expect that they are consistent under the sequences of local alternatives considered in this article because the convergence rates of those local alternatives are sufficiently fast that the presence of random coefficient does not affect the consistency of the short-run variance estimator $\hat{\sigma}_{\tilde{\xi}}^2(\hat{\beta})$ (see Lemmas \ref{lem:wald_sta_components} and \ref{lem:wald_nonsta_components}).

The following theorem establishes that $W_T^{*}(\hat{\beta})$ has the same asymptotic null distribution as $W_T(\hat{\beta})$ has under Assumption \ref{asm:stationary_case_wald} (up to the difference between $W_{\eta}$ and $W_{\eta,lr}$).

\begin{thm}
    Suppose Assumption \ref{asm:serial_corr} and the null $\mathrm{H}_0:\omega_{\beta}=0$ hold. Then, we have
	\begin{align}
			W_T^{*}(\hat{\beta}) \Rightarrow &\begin{pmatrix}
				\int_{0}^{1}\tilde{J}_{x,1}(r)dW_{\eta,lr}(r) \\ \int_{0}^{1}\tilde{J}_{x,2}(r)dW_{\eta,lr}(r) 
			\end{pmatrix}' \begin{pmatrix}
				\int_{0}^{1}\tilde{J}_{x,1}(r)^2dr & \int_{0}^{1}\tilde{J}_{x,1}(r)\tilde{J}_{x,2}(r)dr \\ \int_{0}^{1}\tilde{J}_{x,2}(r)\tilde{J}_{x,1}(r)dr & \int_{0}^{1}\tilde{J}_{x,2}(r)^2dr
			\end{pmatrix}^{-1}\begin{pmatrix}
				\int_{0}^{1}\tilde{J}_{x,1}(r)dW_{\eta,lr}(r) \\ \int_{0}^{1}\tilde{J}_{x,2}(r)dW_{\eta,lr}(r) 
			\end{pmatrix}
		 \label{dist:mod_wald_serial}.
	\end{align}
    \label{thm:long_run_wald_mod}
\end{thm}

\begin{rem}
	When $x_t$ is $\mathrm{I}(0)$, it is necessary that $E[x_{t-1}\eta_t] = E[x^2_{t-1}\eta_t]=0$ holds to identify the parameters in \eqref{eqn:linearized}, from which the Wald-based tests are derived. For example, these conditions do not hold when $\varepsilon_{x,t}$ is conditionally heteroskedastic and $\gamma \neq 0$, where $\gamma$ is the correlation coefficient between $\varepsilon_{x,t}$ and $\varepsilon_{y,t}$.
	\label{rem:i0_regressor_garch}
\end{rem}

\begin{rem}
    The above modification made to the Wald test does not allow for the unconditional heteroskedasticity in $\varepsilon_{y,t}$. To see this, recall that the Wald test is based on regression \eqref{eqn:linearized}. If $\varepsilon_{y,t}$ is unconditionally heteroskedastic and $E[\varepsilon_{y,t}^2] = \sigma_{y,t}^2$, say, then the appropriate regression model becomes
    \begin{align}
        z_t^2(\beta) = \sigma_{y,t}^2 + \delta x_{t-1} + \omega_\beta^2x_{t-1}^2 + \xi_t.
    \end{align}
    Because the Wald test is based on the OLS estimator from \eqref{eqn:linearized}, the time-varying intercept, $\sigma_{y,t}^2$, cannot be eliminated by the usual demeaning approach as in \eqref{eqn:linearized_demean}. To accommodate the unconditional heteroskedasticity in $\varepsilon_{y,t}$, a different construction will be required for the Wald test. Note that the heteroskedasticity-robust LM test, $\mathrm{LM}_T^*$, allows for the unconditional heteroskedasticity (see \citet{georgievTestingParameterInstability2018}). 
\end{rem}

\section{Monte Carlo Simulation}

In this section, Monte Carlo simulations are conducted to investigate the finite-sample performance of the tests. The test statistics considered are $\mathrm{LM}_T^*$, $\mathrm{W}_T^*(\hat{\beta})$ and their sum and product. These tests are based on subsampling with the choice of subsample size explained in Section 4.2. To see whether the heteroskedasticity-robust subsampling tests behave well, we use the data generating process given by \eqref{model:pr_sta} along with
\begin{align}
	\begin{split}
		s_{\beta,t} &= s_{\beta,t}^*/\sqrt{\mathrm{Var}(s_{\beta,t}^*)}, \\
		s_{\beta,t}^* &= \rho_\beta s_{\beta,t-1}^* + \varepsilon_{\beta,t}, \\
		\varepsilon_{y,t} &= \gamma \varepsilon_{x,t} + v_t.
	\end{split}
\end{align}
Here, we set $(\alpha_y,\beta,\alpha_x,s_{x,0},s_{\beta,0})=(0,0,0,0,0)$, $\rho_x=0.95$ and $\gamma=-0.8$.\footnote{We calculate the Wald statistic using demeaned $y_t$ and $x_t$, allowing for nonzero location parameters (see Remark \ref{rem:demeaning}). We also do this when conducting an empirical analysis in Section 7.} For the persistence of $s_{\beta,t}$, we consider $\rho_\beta \in \{0.6,0.8,0.9,0.98\}$. The coefficient variance $\omega_\beta^2$ is taken as $\omega_\beta^2 \in \{0.01,0.05,0.1,0.5\}$. For the driver process, we specify that $v_t$ is GARCH(1,1):
\begin{align}
	\begin{pmatrix}
		\varepsilon_{x,t} \\ v_t \\ \varepsilon_{\beta,t}
	\end{pmatrix} = \begin{pmatrix}
		1 & 0 & 0 \\ 0 & \sigma_{v,t} & 0 \\ 0 & 0 & 1
	\end{pmatrix} \begin{pmatrix}
		u_{x,t} \\ u_{v,t} \\ u_{\beta,t}
	\end{pmatrix},
\end{align}
where $(u_{x,t},u_{v,t},u_{\beta,t})' \sim \mathrm{i.i.d.} \ N(0,\mathrm{diag}(1,0.36,1))$, and $\sigma_{v,t}^2=c_0+c_1\sigma_{v,t-1}^2+c_2v_{t-1}^2$. For the GARCH parameters, we consider the following configurations:
\begin{itemize}
	\item[]DGP1 (weak case): $(c_0,c_1,c_2) = (0.6,0.2,0.2)$.
	\item[]DGP2 (moderate case): $(c_0,c_1,c_2) = (0.4,0.3,0.3)$.
	\item[]DGP3 (strong case): $(c_0,c_1,c_2) = (0.2,0.4,0.4)$.
\end{itemize}

All the tests are implemented under 5\% significance level.

\subsection{Size}

Empirical sizes are displayed in Table \ref{tab:size}. Because we have found that the Product test is oversized if based on subsampling (see Section 4.3.2), we do not consider the Product test here. For DGP1, the sizes of all the tests are around 0.02-0.04, which indicates that these tests are valid in that their sizes are below the nominal level. The sizes for DGP2 and DGP3 are similar, so the same comment applies.

As noted in Remark \ref{ref:subsample_size_condition}, our approach does not necessarily imply $b^{\mathrm{sel}}/T\to 0$ as $T\to \infty$, which is a necessary condition for subsampling to work. Therefore, we report in Table \ref{tab:b_sel} the medians of $b^{\mathrm{sel}}/T$ to verify whether the condition holds. Since the median of $b^{\mathrm{sel}}/T$ moves toward zero as $T$ increases for all the tests and DGPs, the condition $b^{\mathrm{sel}}/T\to 0$ seems to hold.

In Appendix C, we conduct further experiments where the source of GARCH effects for $\varepsilon_{y,t}$ is those for $\varepsilon_{x,t}$. The GARCH parameters are chosen such that the simulated variables mimic the real data used in Section 7. The result is similar to that for DGPs 1-3.

\subsection{Power comparison}

To investigate the power properties of the tests, we check the size-adjusted power and the power of subsampling-based tests, under $T=100$ and DGP1. The size-adjusted powers are displayed in Table \ref{tab:size_adjusted_power}. For the case of $\rho_\beta=0.6$, $s_{\beta,t}$ is a stationary AR(1) process, and the Wald test outperforms the LM test, as expected. Furthermore, the Sum and Product tests are as powerful as the Wald test, with the Sum test being the best. When $\rho_\beta=0.8$, $s_{\beta,t}$ is more persistent, and hence the LM test can perform better than the Wald test (along with the Sum and Product tests) for small $\omega_\beta^2$. However, for moderate to large $\omega_\beta^2$ ($= 0.05,0.1,0.5$), the LM test is dominated by the others. The best-performing test in these cases is the Product test. A similar pattern in power properties is observed in the case of more persistent $s_{\beta,t}$ with $\rho_\beta=0.9$, but the performance of the LM test gets better. When $\rho_\beta=0.98$, $s_{\beta,t}$ is almost a unit root, and the LM test performs the best and the Wald test worst for all $\omega_\beta^2$ considered, as expected from our theoretical analysis. As for the combination tests, the powers of the Sum test are quite similar to those of the Wald test, while the Product test behaves similarly to the LM test, ranking second or first.

Table \ref{tab:subsampling_power} shows the powers when tests are based on subsampling. The product test, which should not be based on subsampling, is not considered here. The pattern of the power of the tests is the same as that of the size-adjusted power: the Wald and Sum tests perform well when $s_{\beta,t}$ is stationary, with the Sum test attaining a higher power, and the LM test is the best test when $s_{\beta,t}$ is strongly persistent. Note that the Sum test is at least the second best test in all cases (because the Product test is not considered here).

\section{Empirical Application}

This section illustrates how the tests considered thus far can be applied to real data analysis. Our analytical framework is exactly the same as that employed by \citet{devpuraStockReturnPredictability2018}, who investigate whether 14 financial variables have a time-varying predictive ability for the U.S. stock market excess returns. They do this by first estimating predictive coefficient at each time point by recursive-window estimation, calculating p-values associated with the predictive coefficient estimates, and then concluding predictability is time-varying if 25-90\% of the p-values are less than a nominal level (0.05, say). A problem of this approach, however, is that it conducts statistical tests sequentially without any adjustment for multiple testing, resulting in uncontrolled size. Moreover, the ``25-90\%" criterion is arbitrary. To get around these problems, we use the tests considered in this article (except the Product test). Note that these tests, if based on subsampling, generally do not suffer from the issues considered by \citet{devpuraStockReturnPredictability2018}, namely, persistence, long-run endogeneity and conditional heteroskedasticity of predictors.

Our working regression is as in Section 6 with the U.S stock market excess returns used as dependent variable, and each of the 14 variables as a predictor. Specifically, the predictors are book-to-market ratio (BM), dividend payout ratio (DE), default yield spread (DFY), default return spread (DFR), dividend-price ratio (DP), dividend yield (DY), earnings-to-price (EP), inflation (INFL), long-term bond return (LTR), long-term bond yield (LTY), net equity expansion (NTIS), stock variance (SVAR), T-bill rate (TBL) and term spread (TMS); see \citet{devpuraStockReturnPredictability2018} for the detailed description of the variables. All the data are monthly, spanning 1927:1-2014:12 ($T=1055$).\footnote{We use data from 1927:1 to 2014:11 for predictors and from 1927:2 to 2014:12 for the dependent variable.} The data are taken from Amit Goyal's web page (\url{https://sites.google.com/view/agoyal145}). The estimated GARCH(1,1) effects for predictors are displayed in Table \ref{tab:garch_effects_rho}, and the testing results in Table \ref{tab:empirical}.

Before discussing the testing results, it should be recalled that the Wald-based tests may suffer from the identification problem when the predictor is a stationary GARCH process (see Remark \ref{rem:i0_regressor_garch}). Therefore, we should not conduct the Wald-based tests when such a predictor is used. According to Table \ref{tab:garch_effects_rho}, DFR, INFL, LTR and SVAR are such predictors, so we do not perform the Wald and Sum tests when these variables are used as the predictor. We note here that \citeauthor{chenTestingSmoothStructural2012}'s (\citeyear{chenTestingSmoothStructural2012}) tests for parameter instability, which are constructed under the stationarity of $x_t$, may be applied to these predictors (as long as the assumptions on which their tests rely are satisfied).

\citet{devpuraStockReturnPredictability2018} determine that 7 predictors (EP, DFY, LTR, LTY, NTIS, SVAR and TBL) have time-varying predictive ability. According to our results, the LM test does not reject the null of constant predictability for any predictors at 5\% level.\footnote{For DFR, INFL, LTR and SVAR, we also performed the standard LM test using the critical value from Table 1 of \citet{hansenTestingParameterInstability1992}, which is calculated under the assumption of $\mathrm{I}(0)$ regressor. However, the standard LM test also failed to reject the null for these predictors.} In sharp contrast, the Wald and Sum tests reject the null for 5 predictors (BM, DE, DP, DY, DFY). Neither of the tests reject the null for EP, LTY, NTIS, TBL and TMS, 4 of which are judged to have time-varying predictive ability by \citet{devpuraStockReturnPredictability2018}.

\section{Conclusion}

In this article, we proposed a Wald-type test for coefficient constancy that is designed to detect stationary random coefficient. We analyzed power properties of the Wald test and \citeauthor{nyblomTestingConstancyParameters1989}'s (\citeyear{nyblomTestingConstancyParameters1989}) LM test under the sequence of local alternatives that takes different forms depending on whether the random coefficient is stationary or integrated and which test is considered. We found that the Wald test outperforms the LM test when the coefficient process is stationary, while the LM test is superior to the Wald test when the coefficient process is a (near) unit root. We also considered two combination tests of the Wald and LM, which take their sum and product. Our simulation study revealed the LM test is the best when random coefficient is highly persistent, the Wald and Sum tests are the most powerful tests when coefficient process is stationary, and the Product test has good power properties irrespective of the persistence of random coefficient, with the best performance for moderate persistence. The message from this result is that the best test for coefficient randomness differs from context to context, and the persistence of the random coefficient determines which test is the best one.

To deal with the nuisance-parameter problem associated with the above tests, we proposed to base them on subsampling. However, there may be a better way (in terms of performance and/or computation) to tackle the nuisance-parameter problem than the subsampling approach. A promising approach will be the IVX method proposed by \citet{phillipsEconometricInferenceVicinity2009}. It has been shown in the literature \citep{kostakisRobustEconometricInference2015,phillipsRobustEconometricInference2016} that the IVX approach leads to valid inference in predictive regressions, irrespective of regressors' persistence and long-run endogeneity. The use of the IVX method would also make Wald-based tests feasible when the predictor is GARCH and has an autoregressive root distant from unity.

We applied the tests to the U.S. stock returns data. The result obtained from this exercise mostly contradicted the conclusion of an earlier research that sequentially applies hypothesis tests to investigate the coefficient constancy. This implies that earlier findings based only on sequential estimation of predictability might be misleading.

\bibliographystyle{standard}
\bibliography{random_coefficient_PR}

@article{andrewsHybridSizeCorrectedSubsampling2009,
  title = {Hybrid and {{Size-Corrected Subsampling Methods}}},
  author = {Andrews, Donald W. K. and Guggenberger, Patrik},
  year = {2009},
  journal = {Econometrica},
  volume = {77},
  number = {3},
  eprint = {40263842},
  eprinttype = {jstor},
  pages = {721--762},
  publisher = {{[Wiley, The Econometric Society]}},
  issn = {0012-9682},
  urldate = {2023-02-01}
}

@article{andrewsTestsParameterInstability1993,
  title = {Tests for {{Parameter Instability}} and {{Structural Change With Unknown Change Point}}},
  author = {Andrews, Donald W. K.},
  year = {1993},
  journal = {Econometrica},
  volume = {61},
  number = {4},
  eprint = {2951764},
  eprinttype = {jstor},
  pages = {821--856},
  publisher = {{[Wiley, Econometric Society]}},
  issn = {0012-9682},
  doi = {10.2307/2951764},
  urldate = {2023-10-14}
}

@article{campbellEfficientTestsStock2006,
  title = {Efficient Tests of Stock Return Predictability},
  author = {Campbell, John Y. and Yogo, Motohiro},
  year = {2006},
  month = jul,
  journal = {Journal of Financial Economics},
  volume = {81},
  number = {1},
  pages = {27--60},
  issn = {0304-405X},
  urldate = {2022-03-26},
  langid = {english}
}

@article{chenTestingSmoothStructural2012,
  title = {Testing for {{Smooth Structural Changes}} in {{Time Series Models}} via {{Nonparametric Regression}}},
  author = {Chen, Bin and Hong, Yongmiao},
  year = {2012},
  journal = {Econometrica},
  volume = {80},
  number = {3},
  pages = {1157--1183},
  issn = {1468-0262},
  doi = {10.3982/ECTA7990},
  urldate = {2022-07-20},
  langid = {english}
}

@article{chiangForecastingTreasuryBill1991,
  title = {Forecasting the {{Treasury Bill Rate}}: {{A Time-Varying Coefficient Approach}}},
  shorttitle = {Forecasting the {{Treasury Bill Rate}}},
  author = {Chiang, Thomas C. and Kahl, Douglas R.},
  year = {1991},
  journal = {Journal of Financial Research},
  volume = {14},
  number = {4},
  pages = {327--336},
  issn = {1475-6803},
  doi = {10.1111/j.1475-6803.1991.tb00670.x},
  urldate = {2023-12-25},
  copyright = {{\textcopyright} The Southern Finance Association and the Southwestern Finance Association},
  langid = {english}
}

@article{choiSubsamplingVectorAutoregressive2005,
  title = {Subsampling Vector Autoregressive Tests of Linear Constraints},
  author = {Choi, In},
  year = {2005},
  month = jan,
  journal = {Journal of Econometrics},
  volume = {124},
  number = {1},
  pages = {55--89},
  issn = {0304-4076},
  doi = {10.1016/j.jeconom.2003.12.013},
  urldate = {2023-05-22},
  langid = {english}
}

@article{devpuraStockReturnPredictability2018,
  title = {Is Stock Return Predictability Time-Varying?},
  author = {Devpura, Neluka and Narayan, Paresh Kumar and Sharma, Susan Sunila},
  year = {2018},
  month = jan,
  journal = {Journal of International Financial Markets, Institutions and Money},
  volume = {52},
  pages = {152--172},
  issn = {1042-4431},
  doi = {10.1016/j.intfin.2017.06.001},
  urldate = {2022-03-26},
  langid = {english}
}

@article{fuSpecificationTestsTimevarying2023,
  title = {Specification Tests for Time-Varying Coefficient Models},
  author = {Fu, Zhonghao and Hong, Yongmiao and Su, Liangjun and Wang, Xia},
  year = {2023},
  month = aug,
  journal = {Journal of Econometrics},
  volume = {235},
  number = {2},
  pages = {720--744},
  issn = {0304-4076},
  doi = {10.1016/j.jeconom.2022.08.001},
  urldate = {2024-01-01}
}

@article{georgievTestingParameterInstability2018,
  title = {Testing for Parameter Instability in Predictive Regression Models},
  author = {Georgiev, Iliyan and Harvey, David I. and Leybourne, Stephen J. and Taylor, A. M. Robert},
  year = {2018},
  month = may,
  journal = {Journal of Econometrics},
  volume = {204},
  number = {1},
  pages = {101--118},
  issn = {0304-4076},
  doi = {10.1016/j.jeconom.2018.01.005},
  urldate = {2022-07-21},
  langid = {english}
}

@article{guidolinTimeVaryingStock2013,
  title = {Time Varying Stock Return Predictability: {{Evidence}} from {{US}} Sectors},
  shorttitle = {Time Varying Stock Return Predictability},
  author = {Guidolin, Massimo and McMillan, David G. and Wohar, Mark E.},
  year = {2013},
  month = mar,
  journal = {Finance Research Letters},
  volume = {10},
  number = {1},
  pages = {34--40},
  issn = {1544-6123},
  doi = {10.1016/j.frl.2012.07.002},
  urldate = {2022-05-28},
  langid = {english}
}

@article{hammamiUnderstandingTimevaryingShorthorizon2020,
  title = {Understanding Time-Varying Short-Horizon Predictability},
  author = {Hammami, Yacine and Zhu, Jie},
  year = {2020},
  month = jan,
  journal = {Finance Research Letters},
  volume = {32},
  pages = {101097},
  issn = {1544-6123},
  doi = {10.1016/j.frl.2019.01.009},
  urldate = {2022-05-30},
  langid = {english}
}

@article{hansenConvergenceStochasticIntegrals1992,
  title = {Convergence to Stochastic Integrals for Dependent Heterogeneous Processes},
  author = {Hansen, Bruce E.},
  year = {1992},
  journal = {Econometric Theory},
  volume = {8},
  number = {4},
  pages = {489--500},
  publisher = {{Cambridge University Press}},
  urldate = {2021-11-03},
  langid = {english}
}

@article{hansenTestingParameterInstability1992,
  title = {Testing for Parameter Instability in Linear Models},
  author = {Hansen, Bruce E.},
  year = {1992},
  month = aug,
  journal = {Journal of Policy Modeling},
  volume = {14},
  number = {4},
  pages = {517--533},
  issn = {0161-8938},
  doi = {10.1016/0161-8938(92)90019-9},
  urldate = {2023-03-25},
  langid = {english}
}

@article{hansenTestsParameterInstability2002,
  title = {Tests for {{Parameter Instability}} in {{Regressions}} with {{I}}(1) {{Processes}}},
  author = {Hansen, Bruce E.},
  year = {2002},
  journal = {Journal of Business \& Economic Statistics},
  volume = {20},
  number = {1},
  eprint = {1392149},
  eprinttype = {jstor},
  pages = {45--59},
  publisher = {{[American Statistical Association, Taylor \& Francis, Ltd.]}},
  issn = {0735-0015},
  urldate = {2024-01-01}
}

@article{jachSubsamplingInferenceMean2012,
  title = {Subsampling Inference for the Mean of Heavy-Tailed Long-Memory Time Series},
  author = {Jach, Agnieszka and McElroy, Tucker and Politis, Dimitris N.},
  year = {2012},
  journal = {Journal of Time Series Analysis},
  volume = {33},
  number = {1},
  pages = {96--111},
  issn = {1467-9892},
  doi = {10.1111/j.1467-9892.2011.00742.x},
  urldate = {2023-04-02},
  langid = {english}
}

@article{kostakisRobustEconometricInference2015,
  title = {Robust {{Econometric Inference}} for {{Stock Return Predictability}}},
  author = {Kostakis, Alexandros and Magdalinos, Tassos and Stamatogiannis, Michalis P.},
  year = {2015},
  journal = {The Review of Financial Studies},
  volume = {28},
  number = {5},
  pages = {1506--1553},
  publisher = {{[Oxford University Press, The Society for Financial Studies]}},
  issn = {0893-9454},
  urldate = {2022-03-31}
}

@article{leeRevisitingDemandMoney2013,
  title = {Revisiting the Demand for Money Function: Evidence from the Random Coefficients Approach},
  shorttitle = {Revisiting the Demand for Money Function},
  author = {Lee, Chien-Chiang and Chang, An-Hsing},
  year = {2013},
  month = sep,
  journal = {Quantitative Finance},
  volume = {13},
  number = {9},
  pages = {1491--1502},
  publisher = {{Routledge}},
  issn = {1469-7688},
  doi = {10.1080/14697688.2011.653386},
  urldate = {2023-12-25}
}

@article{liangWeakConvergenceStochastic2016,
  title = {Weak {{Convergence}} to {{Stochastic Integrals}} for {{Econometric Applications}}},
  author = {Liang, Hanying and Phillips, Peter C. B. and Wang, Hanchao and Wang, Qiying},
  year = {2016},
  month = dec,
  journal = {Econometric Theory},
  volume = {32},
  number = {6},
  pages = {1349--1375},
  publisher = {{Cambridge University Press}},
  urldate = {2021-03-25},
  langid = {english}
}

@article{linTestingParameterConstancy1999,
  title = {Testing Parameter Constancy in Linear Models against Stochastic Stationary Parameters},
  author = {Lin, Chien-Fu Jeff and Ter{\"a}svirta, Timo},
  year = {1999},
  month = jun,
  journal = {Journal of Econometrics},
  volume = {90},
  number = {2},
  pages = {193--213},
  issn = {0304-4076},
  doi = {10.1016/S0304-4076(98)00041-4},
  urldate = {2023-08-20}
}

@article{nishiStochasticLocalModerate2024,
  title = {Stochastic Local and Moderate Departures from a Unit Root and Its Application to Unit Root Testing},
  author = {Nishi, Mikihito and Kurozumi, Eiji},
  year = {2024},
  journal = {Journal of Time Series Analysis},
  volume = {45},
  number = {1},
  pages = {133--157},
  issn = {1467-9892},
  doi = {10.1111/jtsa.12691},
  urldate = {2024-01-01},
  copyright = {{\textcopyright} 2023 The Authors. Journal of Time Series Analysis published by John Wiley \& Sons Ltd.},
  langid = {english}
}

@misc{nishiTestingCoefficientRandomness2023,
  title = {Testing for {{Coefficient Randomness}} in {{Local-to-Unity Autoregressions}}},
  author = {Nishi, Mikihito},
  year = {2023},
  number = {arXiv:2301.04853},
  eprint = {2301.04853},
  primaryclass = {econ, stat},
  publisher = {{arXiv}},
  url = {https://arxiv.org/abs/2301.04853} ,
  urldate = {2023-01-13},
  archiveprefix = {arxiv}
}

@article{nyblomTestingConstancyParameters1989,
  title = {Testing for the {{Constancy}} of {{Parameters Over Time}}},
  author = {Nyblom, Jukka},
  year = {1989},
  journal = {Journal of the American Statistical Association},
  volume = {84},
  number = {405},
  pages = {223--230},
  publisher = {{[American Statistical Association, Taylor \& Francis, Ltd.]}},
  issn = {0162-1459},
  doi = {10.2307/2289867},
  urldate = {2022-08-28}
}

@article{perronUsefulnessLackThereof2016,
  title = {On the {{Usefulness}} or {{Lack Thereof}} of {{Optimality Criteria}} for {{Structural Change Tests}}},
  author = {Perron, Pierre and Yamamoto, Yohei},
  year = {2016},
  month = may,
  journal = {Econometric Reviews},
  volume = {35},
  number = {5},
  pages = {782--844},
  publisher = {{Taylor \& Francis}},
  issn = {0747-4938},
  doi = {10.1080/07474938.2014.977621},
  urldate = {2023-11-06}
}

@article{phillipsEconometricInferenceVicinity2009,
  title = {Econometric {{Inference}} in the {{Vicinity}} of {{Unity}}},
  author = {Phillips, Peter C. B. and Magdalinos, Tassos},
  year = {2009},
  month = apr,
  journal = {Working Papers},
  number = {CoFie-06-2009},
  publisher = {{Singapore Management University, Sim Kee Boon Institute for Financial Economics}},
  urldate = {2023-07-23},
  langid = {english}
}

@article{phillipsFullyModifiedLeast1995,
  title = {Fully {{Modified Least Squares}} and {{Vector Autoregression}}},
  author = {Phillips, Peter C. B.},
  year = {1995},
  journal = {Econometrica},
  volume = {63},
  number = {5},
  eprint = {2171721},
  eprinttype = {jstor},
  pages = {1023--1078},
  publisher = {{[Wiley, Econometric Society]}},
  issn = {0012-9682},
  doi = {10.2307/2171721},
  urldate = {2023-05-22}
}

@article{phillipsRobustEconometricInference2016,
  title = {Robust Econometric Inference with Mixed Integrated and Mildly Explosive Regressors},
  author = {Phillips, Peter C. B. and Lee, Ji Hyung},
  year = {2016},
  month = jun,
  journal = {Journal of Econometrics},
  series = {Innovations in {{Multiple Time Series Analysis}}},
  volume = {192},
  number = {2},
  pages = {433--450},
  issn = {0304-4076},
  urldate = {2022-09-30},
  langid = {english}
}

@article{rahmanPredictivePowerDividend2019,
  title = {Predictive Power of Dividend Yields and Interest Rates for Stock Returns in {{South Asia}}: {{Evidence}} from a Bias-Corrected Estimator},
  shorttitle = {Predictive Power of Dividend Yields and Interest Rates for Stock Returns in {{South Asia}}},
  author = {Rahman, Md Lutfur and Shamsuddin, Abul and Lee, Doowon},
  year = {2019},
  month = jul,
  journal = {International Review of Economics \& Finance},
  volume = {62},
  pages = {267--286},
  issn = {1059-0560},
  doi = {10.1016/j.iref.2019.04.010},
  urldate = {2022-05-30},
  langid = {english}
}

@article{romanoSubsamplingIntervalsAutoregressive2001,
  title = {Subsampling {{Intervals}} in {{Autoregressive Models}} with {{Linear Time Trend}}},
  author = {Romano, Joseph P. and Wolf, Michael},
  year = {2001},
  journal = {Econometrica},
  volume = {69},
  number = {5},
  eprint = {2692222},
  eprinttype = {jstor},
  pages = {1283--1314},
  publisher = {{[Wiley, Econometric Society]}},
  issn = {0012-9682},
  urldate = {2023-02-23}
}

@article{swamyLinearPredictionEstimation1980,
  title = {Linear Prediction and Estimation Methods for Regression Models with Stationary Stochastic Coefficients},
  author = {Swamy, P. A. V. B. and Tinsley, P. A.},
  year = {1980},
  month = feb,
  journal = {Journal of Econometrics},
  volume = {12},
  number = {2},
  pages = {103--142},
  issn = {0304-4076},
  doi = {10.1016/0304-4076(80)90001-9},
  urldate = {2023-12-25}
}

@book{whiteAsymptoticTheoryEconometricians2000,
  title = {Asymptotic {{Theory}} for {{Econometricians}}},
  author = {White, Halbert},
  year = {2000},
  month = oct,
  edition = {Revised edition},
  publisher = {{Emerald Group Pub Ltd}},
  address = {{San Diego}},
  isbn = {978-0-12-746652-1},
  langid = {english}
}

\clearpage
\begin{table} \caption{Convergence rate of $\omega_\beta$ under which tests have nontrivial asymptotic power}
	\centering
	\begin{threeparttable}
		\renewcommand{\arraystretch}{2.5}	\begin{tabular}{ccc}
			\hline &$\mathrm{I}(0)$ $s_{\beta,t}$&$\mathrm{I}(1)$ $s_{\beta,t}$ \\ 
			\hline Wald&$T^{-3/4}$&$T^{-5/4}$ \\
			\hline LM&$T^{-1/2}$&$T^{-3/2}$ \\
			\hline 
		\end{tabular}
	\end{threeparttable}	\label{tab:convergence_rates}
\end{table} 

\clearpage
\begin{table} \caption{Empirical sizes of the tests}
	\centering
	\begin{threeparttable}
		\renewcommand{\arraystretch}{1.8}	\begin{tabular}{c@{\hskip 14pt}c@{\hskip 14pt}@{\hskip 13pt}c@{\hskip 13pt}@{\hskip 13pt}c@{\hskip 13pt}}
			\hline &LM&Wald&Sum \\
			\hline \multicolumn{4}{c}{\textbf{DGP1}} \\
			\hline $T=100$&0.021&0.037&0.037 \\
			 $T=200$&0.018&0.038&0.036 \\
			 $T=500$&0.029&0.028&0.028 \\
			\hline \multicolumn{4}{c}{\textbf{DGP2}} \\
			\hline $T=100$&0.020&0.037&0.037 \\
			 $T=200$&0.018&0.035&0.035 \\
			 $T=500$&0.027&0.029&0.028 \\
			\hline \multicolumn{4}{c}{\textbf{DGP3}} \\
			\hline $T=100$&0.021&0.041&0.040 \\
			 $T=200$&0.018&0.037&0.035 \\
			 $T=500$&0.029&0.026&0.028 \\
			\hline 
		\end{tabular}
	\begin{tablenotes}
		\footnotesize
		\item[] Entries are based on 5000 replications with significance level 0.05.
	\end{tablenotes}
	\end{threeparttable} \label{tab:size}
\end{table} 

\clearpage
\begin{table} \caption{Medians of $b^{\mathrm{sel}}/T$}
	\centering
	\begin{threeparttable}
		\renewcommand{\arraystretch}{1.8}	\begin{tabular}{c@{\hskip 22pt}c@{\hskip 13pt}@{\hskip 13pt}c@{\hskip 13pt}@{\hskip 13pt}c@{\hskip 13pt}}
			\hline &LM&Wald&Sum \\
			\hline \multicolumn{4}{c}{\textbf{DGP1}} \\
			\hline $T=100$&0.12&0.26&0.26 \\
			 $T=200$&0.06&0.17&0.17 \\
			 $T=500$&0.03&0.06&0.06 \\
			\hline \multicolumn{4}{c}{\textbf{DGP2}} \\
			\hline $T=100$&0.12&0.26&0.26 \\
			 $T=200$&0.06&0.17&0.17 \\
			 $T=500$&0.03&0.06&0.06 \\
			\hline \multicolumn{4}{c}{\textbf{DGP3}} \\
			\hline $T=100$&0.12&0.26&0.26 \\
			 $T=200$&0.06&0.17&0.17 \\
			\ $T=500$&0.03&0.06&0.06 \\
			\hline 
		\end{tabular}
		\begin{tablenotes}
			\footnotesize
			\item[] Entries are based on 5000 replications with significance level 0.05.
		\end{tablenotes}
	\end{threeparttable} \label{tab:b_sel}
\end{table} 

\clearpage
\begin{table} \caption{Size adjusted power}
	\centering
	\begin{threeparttable}
		\renewcommand{\arraystretch}{1.8}	\begin{tabular}{cc@{\hskip 22pt}c@{\hskip 13pt}@{\hskip 13pt}c@{\hskip 13pt}@{\hskip 13pt}c@{\hskip 13pt}@{\hskip 13pt}c@{\hskip 13pt}}
			\hline $\rho_\beta$&$\omega_\beta^2$&LM&Wald&Sum&Prod \\
			\hline \multirow{4}{*}{0.6}&0.01&0.078&\textit{0.099}&\textbf{0.100}&0.097 \\
			 &0.05&0.157&\textit{0.399}&\textbf{0.403}&0.397 \\
			 &0.1&0.201&\textit{0.628}&\textbf{0.633}&0.614 \\
			 &0.5&0.269&\textit{0.920}&\textbf{0.924}&0.903 \\
			\hline \multirow{4}{*}{0.8}&0.01&\textbf{0.133}&0.083&0.085&\textit{0.123} \\
			 &0.05&0.309&0.325&\textit{0.335}&\textbf{0.439} \\
			 &0.1&0.395&0.512&\textit{0.525}&\textbf{0.643} \\
			 &0.5&0.522&0.837&\textit{0.850}&\textbf{0.926} \\
			\hline \multirow{4}{*}{0.9}&0.01&\textbf{0.184}&0.073&0.077&\textit{0.143} \\
			 &0.05&\textit{0.435}&0.248&0.266&\textbf{0.441} \\
			 &0.1&\textit{0.541}&0.410&0.433&\textbf{0.624} \\
			 &0.5&0.706&0.744&\textit{0.775}&\textbf{0.921} \\
			\hline \multirow{4}{*}{0.98}&0.01&\textbf{0.170}&0.056&0.058&\textit{0.116} \\
			 &0.05&\textbf{0.429}&0.110&0.123&\textit{0.314} \\
			 &0.1&\textbf{0.563}&0.189&0.213&\textit{0.453} \\
			 &0.5&\textit{0.808}&0.510&0.557&\textbf{0.809} \\
			\hline 
		\end{tabular}
		\begin{tablenotes}
			\footnotesize
			\item[(i)] Entries are based on 5000 replications with significance level 0.05, $T=100$ and DGP1.
			\item[(ii)] Entries in \textbf{bold} signify the highest power for corresponding $(\rho_\beta,\omega_\beta^2)$. Entries in \textit{italics} signify the second highest power for corresponding $(\rho_\beta,\omega_\beta^2)$.
		\end{tablenotes}
	\end{threeparttable} \label{tab:size_adjusted_power}
\end{table} 

\clearpage
\begin{table} \caption{Power of subsampling tests}
	\centering
	\begin{threeparttable}
		\renewcommand{\arraystretch}{1.8}	\begin{tabular}{cc@{\hskip 22pt}c@{\hskip 13pt}@{\hskip 13pt}c@{\hskip 13pt}@{\hskip 13pt}c@{\hskip 13pt}}
			\hline $\rho_\beta$&$\omega_\beta^2$&LM&Wald&Sum \\
			\hline \multirow{4}{*}{0.6}&0.01&0.037&\textbf{0.070}&\textit{0.068} \\
			&0.05&0.078&\textit{0.261}&\textbf{0.262} \\
			&0.1&0.099&\textit{0.403}&\textbf{0.404} \\
			&0.5&0.128&\textbf{0.503}&\textit{0.502} \\
			\hline \multirow{4}{*}{0.8}&0.01&0.060&\textit{0.062}&\textbf{0.065} \\
			&0.05&0.169&\textit{0.228}&\textbf{0.232} \\
			&0.1&0.217&\textit{0.347}&\textbf{0.353} \\
			&0.5&0.286&\textit{0.446}&\textbf{0.448} \\
			\hline \multirow{4}{*}{0.9}&0.01&\textbf{0.098}&0.053&\textit{0.055} \\
			&0.05&\textbf{0.273}&0.188&\textit{0.196} \\
			&0.1&\textbf{0.354}&0.303&\textit{0.311} \\
			&0.5&\textbf{0.461}&0.427&\textit{0.437} \\
			\hline \multirow{4}{*}{0.98}&0.01&\textbf{0.099}&\textit{0.043}&\textit{0.043} \\
			&0.05&\textbf{0.299}&0.084&\textit{0.096} \\
			&0.1&\textbf{0.423}&0.153&\textit{0.175} \\
			&0.5&\textbf{0.649}&0.382&\textit{0.412} \\
			\hline 
		\end{tabular}
		\begin{tablenotes}
			\footnotesize
			\item[a] Entries are based on 5000 replications with significance level 0.05, $T=100$ and DGP1.
			\item[b] Entries in \textbf{bold} signify the highest power for corresponding $(\rho_\beta,\omega_\beta^2)$. Entries in \textit{italics} signify the second highest power for corresponding $(\rho_\beta,\omega_\beta^2)$.
		\end{tablenotes}
	\end{threeparttable} \label{tab:subsampling_power}
\end{table} 

\clearpage
\begin{table} \caption{GARCH effects and persistence of predictors}
	\centering
	\begin{threeparttable} \renewcommand{\arraystretch}{1.8}
		\begin{tabular}{@{\hskip 10pt}c@{\hskip 10pt}@{\hskip 18pt}c@{\hskip 18pt}@{\hskip 18pt}c@{\hskip 18pt}@{\hskip 18pt}c@{\hskip 18pt}@{\hskip 18pt}c@{\hskip 18pt}}
			\hline Predictor&$c_0'$&$c_1'$&$c_2'$&$\rho_x$ \\
			\hline BM&$2.82\times10^{-5}$&0.829&0.159&0.993 \\
			 DE&$6.62\times10^{-5}$&$5.56\times10^{-7}$&0.999&1.000 \\
			 DP&$5.83\times10^{-5}$&0.860&0.125&0.999 \\
			 DY&$5.82\times10^{-5}$&0.862&0.123&0.999 \\
			 EP&$8.38\times10^{-5}$&0.800&0.199&0.994 \\
			 DFR&$4.44\times10^{-6}$&0.780&0.219&-0.175 \\
			 DFY&$6.57\times10^{-9}$&0.823&0.166&0.968 \\
			 INFL&$2.85\times10^{-7}$&0.838&0.160&0.459 \\
			 LTR&$3.06\times10^{-6}$&0.814&0.185&0.068 \\
			 LTY&$1.58\times10^{-8}$&0.797&0.201&0.998 \\
			 NTIS&$5.95\times10^{-7}$&0.900&0.067&0.960 \\
			 SVAR&$3.39\times10^{-8}$&0.970&0.020&0.650 \\
			 TBL&$2.32\times10^{-8}$&0.788&0.209&0.999 \\
			 TMS&$3.87\times10^{-8}$&0.830&0.169&0.991 \\
			\hline 
		\end{tabular}
		\begin{tablenotes}
			\footnotesize
			\item[] We estimate the following model: $x_{t}=\alpha_x + \rho_x x_{t-1} + \varepsilon_{x,t}$ with $\varepsilon_{x,t}=\sigma_{x,t}u_{x,t}$, $\sigma_{x,t}^2 = c_0'+c_1'\sigma_{x,t-1}^2 + c_2'\varepsilon_{x,t-1}^2$, and $u_{x,t}$ being normally distributed.
		\end{tablenotes} 
	\end{threeparttable}
	\label{tab:garch_effects_rho}
\end{table} 

\clearpage
\begin{table} \caption{Results of coefficient constancy tests}
	\centering
	\begin{threeparttable}
		\renewcommand{\arraystretch}{1.8}	\begin{tabular}{@{\hskip 10pt}c@{\hskip 10pt}@{\hskip 14pt}c@{\hskip 14pt}@{\hskip 13pt}c@{\hskip 13pt}@{\hskip 13pt}c@{\hskip 13pt}}
			\hline Predictor&LM&Wald&Sum \\
			\hline BM&-&Yes&Yes \\
			 DE&-&Yes&Yes \\
			 DP&-&Yes&Yes \\
			 DY&-&Yes&Yes \\
			 EP&-&-&- \\
			 DFR&-&NA&NA \\
			 DFY&-&Yes&Yes \\
			 INFL&-&NA&NA \\
			 LTR&-&NA&NA \\
			 LTY&-&-&- \\
			 NTIS&-&-&- \\
			 SVAR&-&NA&NA \\
			 TBL&-&-&- \\
			 TMS&-&-&- \\
			\hline 
		\end{tabular}
		\begin{tablenotes}
			\footnotesize
			\item[] For entries in the second to fifth columns, ``Yes" (``-") signifies the rejection (nonrejection) of the null $H_0:\omega_\beta^2=0$ by the corresponding test with 5\% significance level. ``NA'' means that the corresponding test is not conducted.
		\end{tablenotes}
	\end{threeparttable} \label{tab:empirical}
\end{table} 

\clearpage

\clearpage

\clearpage

\clearpage

\clearpage

\titleformat*{\section}{\Large\centering}
\section*{Appendix to ``Testing for Stationary or Persistent Coefficient Randomness in Predictive Regressions" by M.Nishi}

\titleformat*{\section}{\large\bfseries}

\section*{Appendix A: Proofs of the Main Results}
\setcounter{equation}{0}
\renewcommand{\theequation}{A.\arabic{equation}}
\begin{lemapp}
	\begin{itemize}
		\item[(i)] Suppose the condition of Theorem \ref{thm:asym_dists_sta}(i) holds. Then, we have, under model \eqref{model:pr_sta_set0},
		\begin{align}
			T(\hat{\beta}-\beta) \Rightarrow \frac{\sigma_y \int_{0}^{1}J_x(r)dW_y(r)}{\sigma_x \int_{0}^{1}J_x(r)^2dr}.
		\end{align}
		\item[(ii)] Suppose the condition of Theorem \ref{thm:asym_dists_sta}(ii) holds. Then, we have, under model \eqref{model:pr_sta_set0},
		\begin{align}
			T(\hat{\beta}-\beta) \Rightarrow \frac{\sigma_y \int_{0}^{1}J_x(r)dW_y(r)}{\sigma_x \int_{0}^{1}J_x(r)^2dr} + g\sigma_{\beta,lr}\frac{\int_{0}^{1}J_x(r)^2dW_{\beta,lr}(r) + 2(\Lambda_{x\beta}/\sigma_x\sigma_{\beta,lr})\int_{0}^{1}J_x(r)dr}{\int_{0}^{1}J_x(r)^2dr}.
		\end{align}
		\item[(iii)] Suppose the condition of Theorem \ref{thm:wald_nonsta} holds. Then, we have, under model \eqref{model:pr_nonsta_set0},
		\begin{align}
			T^{3/4}(\hat{\beta}-\beta) \Rightarrow \frac{g\int_{0}^{1}J_x(r)^2J_\beta(r)dr}{\int_{0}^{1}J_x(r)^2dr}.
		\end{align}
	\end{itemize}
	\label{lem:beta_hat}
\end{lemapp}

\noindent \begin{proof}
	\begin{itemize}
		\item[(i)] Since
		\begin{align}
			\hat{\beta} &= \beta + \frac{gT^{-3/4}\sum_{t=1}^{T}x_{t-1}^2s_{\beta,t} + \sum_{t=1}^{T}x_{t-1}\varepsilon_{y,t}}{\sum_{t=1}^{T}x_{t-1}^2}, \\
			\intertext{we have}
			T(\hat{\beta} - \beta) &= \frac{T^{-1/4}g\sigma_x^2\sigma_{\beta,lr}\int_{0}^{1}J_{x,T}(r)^2dW_{\beta,lr,T}(r) + \sigma_x\sigma_y\int_{0}^{1}J_{x,T}(r)dW_{y,T}(r)}{\sigma_x^2\int_{0}^{1}J_{x,T}(r)^2dr} \\
			&\Rightarrow \frac{\sigma_y \int_{0}^{1}J_x(r)dW_y(r)}{\sigma_x \int_{0}^{1}J_x(r)^2dr}.
		\end{align}
		because the first term in the numerator is $O_p(T^{-1/4})$ and the second term weakly converges to $\sigma_x\sigma_y \int_{0}^{1}J_x(r)dW_y(r)$ by the application of Theorem 2.1 of \citet{hansenConvergenceStochasticIntegrals1992}.
		\item[(ii)] As in part (i), we have
		\begin{align}
			\hat{\beta} &= \beta + \frac{gT^{-1/2}\sum_{t=1}^{T}x_{t-1}^2s_{\beta,t} + \sum_{t=1}^{T}x_{t-1}\varepsilon_{y,t}}{\sum_{t=1}^{T}x_{t-1}^2}, \\
			\intertext{from which we get}
			T(\hat{\beta}-\beta) &= \frac{ g\sigma_x^2\sigma_{\beta,lr}\int_{0}^{1}J_{x,T}(r)^2dW_{\beta,lr,T}(r) + \sigma_x\sigma_y\int_{0}^{1}J_{x,T}(r)dW_{y,T}(r)}{\sigma_x^2\int_{0}^{1}J_{x,T}(r)^2dr}.
		\end{align}
		As for the first term in the numerator, because $\varepsilon_{x,t}$ and $s_{\beta,t}$ are stationary and $\alpha$-mixing processes satisfying Assumption \ref{asm:stationary_case_lm}, we can apply Theorem 3.1 of \citetapp{liangWeakConvergenceStochastic2016} to obtain
		\begin{align}
			\int_{0}^{1}J_{x,T}(r)^2dW_{\beta,lr,T}(r) \Rightarrow \int_{0}^{1}J_x(r)^2dW_{\beta,lr}(r) + 2\frac{\Lambda_{x\beta}}{\sigma_x\sigma_{\beta,lr}}\int_{0}^{1}J_x(r)dr.
		\end{align}
		Therefore, an application of the continuous mapping theorem (CMT) gives the desired result. 
		\item[(iii)] Since
		\begin{align}
			\hat{\beta} &= \beta + \frac{gT^{-5/4}\sum_{t=1}^{T}x_{t-1}^2s_{\beta,t} + \sum_{t=1}^{T}x_{t-1}\varepsilon_{y,t}}{\sum_{t=1}^{T}x_{t-1}^2}, \\
			\intertext{it follows that, by letting $J_{\beta,T}(r) \coloneqq T^{-1/2}s_{\beta,\lfloor Tr\rfloor}$,}
			T^{3/4}(\hat{\beta}-\beta) &= \frac{ g\sigma_x^2\int_{0}^{1}J_{x,T}(r)^2J_{\beta,T}(r)dr + T^{-1/4}\sigma_x\sigma_y\int_{0}^{1}J_{x,T}(r)dW_{y,T}(r)}{\sigma_x^2\int_{0}^{1}J_{x,T}(r)^2dr} \\
			&\Rightarrow \frac{g\int_{0}^{1}J_x(r)^2J_\beta(r)dr}{\int_{0}^{1}J_x(r)^2dr},
		\end{align}
		by the CMT.
	\end{itemize}	
\end{proof}

The next lemma collects the asymptotic behavior of the components of $\mathrm{W}_T(\hat{\beta})$ when $s_{\beta,t}$ is $\mathrm{I}(0)$.

\begin{lemapp}
	Suppose the condition of Theorem \ref{thm:asym_dists_sta}(i) holds. Then, under model \eqref{model:pr_sta_set0}, we have the following results.
	\begin{itemize}
		\item[(i)] 
		\begin{align}
			T^{-1}\sum_{t=1}^{T}\tilde{x}_{1,t-1}z_t^2(\hat{\beta}) \Rightarrow
			\sigma_x\sigma_{\eta}\int_{0}^{1}\tilde{J}_{x,1}(r)dW_\eta(r) + g^2\sigma_x^3\int_{0}^{1}\tilde{J}_{x,1}(r)\tilde{J}_{x,2}(r)dr+2gm\sigma_x^2\sigma_y\int_{0}^{1}\tilde{J}_{x,1}(r)^2dr.
		\end{align}
		\item[(ii)]
		\begin{align}
			T^{-3/2}\sum_{t=1}^{T}\tilde{x}_{2,t-1}z_t^2(\hat{\beta}) \Rightarrow
			\sigma_x^2\sigma_{\eta}\int_{0}^{1}\tilde{J}_{x,2}(r)dW_\eta(r) + g^2\sigma_x^4\int_{0}^{1}\tilde{J}_{x,2}(r)^2dr+2gm\sigma_x^3\sigma_y\int_{0}^{1}\tilde{J}_{x,1}(r)\tilde{J}_{x,2}(r)dr.
		\end{align}
		\item[(iii)] $\hat{\sigma}_{\tilde{\xi}}^2(\hat{\beta}) \stackrel{p}{\to}\sigma_{\eta}^2.$
	\end{itemize}
	\label{lem:wald_sta_components}
\end{lemapp}

\noindent \begin{proof}
	\begin{itemize}
		\item[(i)] From equation \eqref{model:pr_sta_set0}, we have
		\begin{align}
			T^{-1}\sum_{t=1}^{T}\tilde{x}_{1,t-1}z_t^2(\hat{\beta}) &= T^{-1}\sum_{t=1}^{T}\tilde{x}_{1,t-1}(y_t - \hat{\beta}x_{t-1})^2 \\
			&= T^{-1}\sum_{t=1}^{T}\tilde{x}_{1,t-1}\{\omega_\beta s_{\beta,t}x_{t-1}+\varepsilon_{y,t} - (\hat{\beta}- \beta)x_{t-1}\}^2 \\
			&= T^{-1}\sum_{t=1}^{T}\tilde{x}_{1,t-1}\{\sigma_{y}^2 + (\varepsilon_{y,t}^2 - \sigma_y^2)+\omega_\beta^2x_{t-1}^2 + \omega_\beta^2x_{t-1}^2(s_{\beta,t}^2-1) + (\hat{\beta}-\beta)^2x_{t-1}^2 \\
			& \quad \ +2\omega_\beta\sigma_{y\beta}x_{t-1} + 2\omega_\beta x_{t-1}(s_{\beta,t}\varepsilon_{y,t}- \sigma_{y\beta}) - 2\omega_\beta(\hat{\beta}-\beta)x_{t-1}^2s_{\beta,t} - 2(\hat{\beta}-\beta)x_{t-1}\varepsilon_{y,t}\} \\
			&=T^{-1}\sum_{t=1}^{T}\tilde{x}_{1,t-1}\eta_t + g^2T^{-5/2}\sum_{t=1}^{T}\tilde{x}_{1,t-1}\tilde{x}_{2,t-1} + 2gm\sigma_yT^{-2}\sum_{t=1}^{T}(\tilde{x}_{1,t-1})^2 + A_{1,T} + A_{2,T}, \label{eqn:sum_x_z2_sta} \\
			\intertext{where}
			A_{1,T} &\coloneqq g^2T^{-5/2}\sum_{t=1}^{T}\tilde{x}_{1,t-1}x_{t-1}^2(s_{\beta,t}^2-1) + 2gT^{-7/4}\sum_{t=1}^{T}\tilde{x}_{1,t-1}x_{t-1}(s_{\beta,t}\varepsilon_{y,t}-\sigma_{y\beta}) \\
			& \quad -2T(\hat{\beta}-\beta)T^{-2}\sum_{t=1}^{T}\tilde{x}_{1,t-1}x_{t-1}\varepsilon_{y,t} - 2gT(\hat{\beta}-\beta)T^{-11/4}\sum_{t=1}^{T}\tilde{x}_{1,t-1}x_{t-1}^2s_{\beta,t} \\
			\intertext{and}
			A_{2,T} &\coloneqq T^2(\hat{\beta}-\beta)^2T^{-3}\sum_{t=1}^{T}\tilde{x}_{1,t-1}\tilde{x}_{2,t-1}.
		\end{align}
	We show $A_{1,T},A_{2,T}=o_p(1)$. The first term of $A_{1,T}$ can be rewritten as
	\begin{align}
		g^2T^{-5/2}\sum_{t=1}^{T}\tilde{x}_{1,t-1}x_{t-1}^2(s_{\beta,t}^2-1) = T^{-1/2}g^2\sigma_x^3\int_{0}^{1}\tilde{J}_{x,1,T}(r)J_x(r)^2dW_{1,T}(r),
	\end{align}
	where $W_{1,T}(r) \coloneqq T^{-1/2}\sum_{t=1}^{\lfloor Tr \rfloor}(s_{\beta,t}^2-1)$. Under Assumption \ref{asm:stationary_case_wald}, $\{s_{\beta,t}^2-1\}$ satisfies the FCLT for mixing processes \citepapp[Theorem 5.20 of][]{whiteAsymptoticTheoryEconometricians2000}, and hence $W_{1,T}(r)$ is $O_p(1)$ uniformly in $r$. Therefore, the first term of $A_{1,T}$ is $O_p(T^{-1/2})$. Similarly, we can show that the other terms of $A_{1,T}$ are $O_p(T^{-1/4})$. $A_{2,T}$ is $o_p(1)$, because
	\begin{align}
		A_{2,T} = T^2(\hat{\beta}-\beta)^2\times T^{-1/2}\sigma_x^3\int_{0}^{1}\tilde{J}_{x,1,T}(r)\tilde{J}_{x,2,T}(r)dr = O_p(T^{-1/2}),
	\end{align}
	by the CMT. Substituting these results into \eqref{eqn:sum_x_z2_sta} yields
	\begin{align}
		T^{-1}\sum_{t=1}^{T}\tilde{x}_{1,t-1}z_t^2(\hat{\beta}) &= \sigma_x\sigma_{\eta}\int_{0}^{1}\tilde{J}_{x,1,T}(r)dW_{\eta,T}(r) + g^2\sigma_x^3\int_{0}^{1}\tilde{J}_{x,1,T}(r)\tilde{J}_{x,2,T}(r)dr \\ &+ 2gm\sigma_x^2\sigma_y\int_{0}^{1}\tilde{J}_{x,1,T}(r)^2dr + o_p(1) \\
		&\Rightarrow \sigma_x\sigma_{\eta}\int_{0}^{1}\tilde{J}_{x,1}(r)dW_\eta(r) + g^2\sigma_x^3\int_{0}^{1}\tilde{J}_{x,1}(r)\tilde{J}_{x,2}(r)dr+2gm\sigma_x^2\sigma_y\int_{0}^{1}\tilde{J}_{x,1}(r)^2dr,
	\end{align}
	by virtue of Theorem 2.1 of \citetapp{hansenConvergenceStochasticIntegrals1992} and the CMT.
	\item[(ii)] The proof is essentially the same as that of part (i) and thus is omitted.
	\item[(iii)] For $\hat{\sigma}_{\tilde{\xi}}^2(\hat{\beta})$, we have
	\begin{align}
		\hat{\sigma}_{\tilde{\xi}}^2(\hat{\beta}) &= T^{-1}Z_2(\hat{\beta})'M_XZ_2(\hat{\beta}) \\
		&=T^{-1}\sum_{t=1}^{T}\{\widetilde{z_t^2}(\hat{\beta})\}^2 - T^{-1} \begin{pmatrix}
			\sum_{t=1}^{T}\tilde{x}_{1,t-1}z_t^2(\hat{\beta}) \\
			\sum_{t=1}^{T}\tilde{x}_{2,t-1}z_t^2(\hat{\beta})
		\end{pmatrix}' \\
		&\qquad \qquad \qquad \qquad \qquad \times \begin{pmatrix}
			\sum_{t=1}^{T}(\tilde{x}_{1,t-1})^2 & \sum_{t=1}^{T}\tilde{x}_{1,t-1}\tilde{x}_{2,t-1} \\
			\sum_{t=1}^{T}\tilde{x}_{2,t-1}\tilde{x}_{1,t-1} & \sum_{t=1}^{T}(\tilde{x}_{2,t-1})^2
		\end{pmatrix}^{-1} 
		\begin{pmatrix}
			\sum_{t=1}^{T}\tilde{x}_{1,t-1}z_t^2(\hat{\beta}) \\
			\sum_{t=1}^{T}\tilde{x}_{2,t-1}z_t^2(\hat{\beta})
		\end{pmatrix} \\
		&\eqqcolon B_{1,T} - B_{2,T}. \label{eqn:sigma_hat_xi_expand}
	\end{align}
	$B_{1,T}$ becomes
	\begin{align}
		B_{1,T} &= T^{-1}\sum_{t=1}^{T}\biggl\{z_t^2(\hat{\beta}) - T^{-1}\sum_{t=1}^{T}z_t^2(\hat{\beta})\biggr\}^2 \\
		&=T^{-1}\sum_{t=1}^{T}z_t^4(\hat{\beta}) - \biggl\{T^{-1}\sum_{t=1}^{T}z_t^2(\hat{\beta})\biggr\}^2 \\
		&=T^{-1}\sum_{t=1}^{T}\{\varepsilon_{y,t}+gT^{-3/4}x_{t-1}s_{\beta,t} - (\hat{\beta}-\beta)x_{t-1}\}^4 \\
		& \qquad -\Bigl[T^{-1}\sum_{t=1}^{T}\{\varepsilon_{y,t}+gT^{-3/4}x_{t-1}s_{\beta,t} - (\hat{\beta}-\beta)x_{t-1}\}^2\Bigr]^2 \\
		&=T^{-1}\sum_{t=1}^{T}\varepsilon_{y,t}^4 + B_{1,1,T} - \Bigl[T^{-1}\sum_{t=1}^{T}\varepsilon_{y,t}^2+B_{1,2,T}\Bigr]^2, \label{eqn:b_1T_expand}
	\end{align}
	where
	\begin{align}
		B_{1,1,T} &\coloneqq 4gT^{-7/4}\sum_{t=1}^{T}x_{t-1}\varepsilon_{y,t}^3s_{\beta,t} + 6g^2T^{-5/2}\sum_{t=1}^{T}x_{t-1}^2\varepsilon_{y,t}^2s_{\beta,t}^2 + 4g^3T^{-13/4}\sum_{t=1}^{T}x_{t-1}^3\varepsilon_{y,t}s_{\beta,t}^3 \\
		&+g^4T^{-4}\sum_{t=1}^{T}x_{t-1}^4s_{\beta,t}^4 - 4(\hat{\beta}-\beta)T^{-1}\sum_{t=1}^{T}(\varepsilon_{y,t}+gT^{-3/4}x_{t-1}s_{\beta,t})^3x_{t-1} \\
		&+6(\hat{\beta}-\beta)^2T^{-1}\sum_{t=1}^{T}(\varepsilon_{y,t}+gT^{-3/4}x_{t-1}s_{\beta,t})^2x_{t-1}^2 - 4(\hat{\beta}-\beta)^3T^{-1}\sum_{t=1}^{T}(\varepsilon_{y,t}+gT^{-3/4}x_{t-1}s_{\beta,t})x_{t-1}^3 \\
		&+(\hat{\beta}-\beta)^4T^{-1}\sum_{t=1}^{T}x_{t-1}^4,
	\end{align}
	and
	\begin{align}
		B_{1,2,T} &\coloneqq 2gT^{-7/4}\sum_{t=1}^{T}x_{t-1}\varepsilon_{y,t}s_{\beta,t} + g^2T^{-5/2}\sum_{t=1}^{T}x_{t-1}^2s_{\beta,t}^2 \\
		& \qquad -2(\hat{\beta}-\beta)T^{-1}\sum_{t=1}^{T}(\varepsilon_{y,t}+gT^{-3/4}x_{t-1}s_{\beta,t})x_{t-1} + (\hat{\beta}-\beta)^2T^{-1}\sum_{t=1}^{T}x_{t-1}^2.
	\end{align}
	We show $B_{1,1,T}$ and $B_{1,2,T}$ are $o_p(1)$. For $B_{1,1,T}$, the first term satisfies
	\begin{align}
		|4gT^{-7/4}\sum_{t=1}^{T}x_{t-1}\varepsilon_{y,t}^3s_{\beta,t}| &\leq 4g\sup_{0\leq r\leq1}|T^{-1/2}x_{\lfloor Tr \rfloor}|T^{-5/4}\sum_{t=1}^{T}|\varepsilon_{y,t}^3s_{\beta,t}| \\
		&=O_p(T^{-1/4}),
	\end{align}
	because $\sup_{0\leq r\leq1}|T^{-1/2}x_{\lfloor r \rfloor}|\Rightarrow \sup_{0\leq r\leq1}|J_x(r)|$ and $T^{-1}\sum_{t=1}^{T}|\varepsilon_{y,t}^3x_{\beta,t}|\stackrel{p}{\to}E[|\varepsilon_{y,t}^3s_{\beta,t}|]$ by the law of large numbers (LLN) for mixing processes \citepapp[Corollary 3.48 of][]{whiteAsymptoticTheoryEconometricians2000} under Assumption \ref{asm:stationary_case_wald}. The other terms of $B_{1,1,T}$ and $B_{1,2,T}$ can be shown to be $o_p(1)$ in a similar fashion. Therefore, we can proceed from \eqref{eqn:b_1T_expand} to
	\begin{align}
		B_{1,T} &= T^{-1}\sum_{t=1}^{T}\varepsilon_{y,t}^4 - \Bigl\{T^{-1}\sum_{t=1}^{T}\varepsilon_{y,t}^2\Bigr\}^2 + o_p(1) \stackrel{p}{\to}E[\varepsilon_{y,t}^4] - \sigma_y^4 = \sigma_{\eta}^2.
	\end{align}
	As for $B_{2,T}$, an application of the CMT yields $B_{2,T}=O_p(T^{-1})$, given that $T^{-1}\sum_{t=1}^{T}\tilde{x}_{1,t-1}z_t^2(\hat{\beta})$ and $T^{-3/2}\sum_{t=1}^{T}\tilde{x}_{2,t-1}z_t^2(\hat{\beta})$ are $O_p(1)$ from parts (i) and (ii), and $T^{-2}\sum_{t=1}^{T}(\tilde{x}_{1,t-1})^2$, $T^{-5/2}\sum_{t=1}^{T}\tilde{x}_{1,t-1}\tilde{x}_{2,t-1}$ and $T^{-3}\sum_{t=1}^{T}(\tilde{x}_{2,t-1})^2$ are $O_p(1)$. Hence, we conclude $\hat{\sigma}_{\tilde{\xi}}^2(\hat{\beta}) \stackrel{p}{\to}\sigma_{\eta}^2$.
	\end{itemize}
\end{proof}

We also collect the asymptotic behavior of the components of $\mathrm{LM}_T$ in the next lemma.

\begin{lemapp}
	Suppose the condition of Theorem \ref{thm:asym_dists_sta}(ii) holds. Then, under model \eqref{model:pr_sta_set0}, we have the following results.
	\begin{itemize}
		\item[(i)] $T^{-1/2}\sum_{t=1}^{\lfloor Tr \rfloor}z_t(\hat{\beta}) \Rightarrow
			\sigma_yA(r) + g\sigma_x\sigma_{\beta,lr}B(r)$.
		\item[(ii)] $\hat{\sigma}_T^2 \Rightarrow \sigma_y^2 + g^2 \sigma_x^2\int_{0}^{1}J_x(r)^2dr$.
	\end{itemize}
	\label{lem:lm_sta_components}
\end{lemapp}

\noindent \begin{proof}
	\begin{itemize}
		\item[(i)] \begin{align}
			T^{-1/2}\sum_{t=1}^{\lfloor Tr \rfloor}z_t(\hat{\beta})&=T^{-1/2}\sum_{t=1}^{\lfloor Tr \rfloor}\{\varepsilon_{y,t}+gT^{-1/2}x_{t-1}s_{\beta,t} - (\hat{\beta}-\beta)x_{t-1}\} \\
			&=\sigma_yW_{y,T}(r) + g\sigma_x\sigma_{\beta,lr}\int_{0}^{r}J_{x,T}(s)dW_{\beta,lr,T}(s) -T(\hat{\beta}-\beta)\sigma_x\int_{0}^{r}J_{x,T}(s)ds \\
			&\Rightarrow \sigma_yW_y(r) + g\sigma_x\sigma_{\beta,lr}\Bigl(\int_{0}^{r}J_x(s)dW_{\beta,lr}(s) + r\frac{\Lambda_{x\beta}}{\sigma_x\sigma_{\beta,lr}}\Bigr) \\
			&\quad - \Bigl(\frac{\sigma_y\int_{0}^{1}J_x(r)dW_y(r)}{\sigma_x\int_{0}^{1}J_x(r)^2dr} + g\sigma_{\beta,lr}\frac{\int_{0}^{1}J_x(r)^2dW_{\beta,lr}(r) + 2(\Lambda_{x\beta}/\sigma_x\sigma_{\beta,lr})\int_{0}^{1}J_x(r)dr}{\int_{0}^{1}J_x(r)^2dr}\Bigr) \\ 
			&\quad \times \sigma_x\int_{0}^{r}J_x(s)ds \\
			&=\sigma_yA(r) + g\sigma_x\sigma_{\beta,lr}B(r),
		\end{align}
		by Lemma \ref{lem:beta_hat}(ii), Theorem 3.1 of \citetapp{liangWeakConvergenceStochastic2016} and the CMT.
		\item[(ii)] \begin{align}
			\hat{\sigma}_T^2 &= T^{-1}\sum_{t=1}^{T}z_t^2(\hat{\beta}) \\
			&= T^{-1}\sum_{t=1}^{T}\{\varepsilon_{y,t}+gT^{-1/2}s_{\beta,t}x_{t-1}-(\hat{\beta}-\beta)x_{t-1}\}^2 \\
			&=T^{-1}\sum_{t=1}^{T}\varepsilon_{y,t}^2 + g^2T^{-2}\sum_{t=1}^{T}x_{t-1}^2+g^2T^{-2}\sum_{t=1}^{T}x_{t-1}^2(s_{\beta,t}^2-1) + T^2(\hat{\beta}-\beta)^2T^{-3}\sum_{t=1}^{T}x_{t-1}^2 \\
			& \quad +2gm\sigma_yT^{-7/4}\sum_{t=1}^{T}x_{t-1} + 2gT^{-3/2}\sum_{t=1}^{T}x_{t-1}(s_{\beta,t}\varepsilon_{y,t}-\sigma_{y\beta}) \\
			& \quad -2T(\hat{\beta}-\beta)T^{-2}\sum_{t=1}^{T}x_{t-1}\varepsilon_{y,t} - 2gT(\hat{\beta}-\beta)T^{-5/2}\sum_{t=1}^{T}x_{t-1}^2s_{\beta,t} \\
			&\eqqcolon T^{-1}\sum_{t=1}^{T}\varepsilon_{y,t}^2 + g^2T^{-2}\sum_{t=1}^{T}x_{t-1}^2 + C_T.
		\end{align}
	As for $C_T$, the first term can be written as $g^2T^{-1/2}\int_{0}^{1}J_{x,T}(r)^2dW_{s_\beta^2-1,T}(r)$, say, because $\{s_{\beta,t}^2-1\}$ satisfies the FCLT under Assumption \ref{asm:stationary_case_lm}. Therefore, the first term of $C_T$ is $O_p(T^{-1/2})$. Similarly, it can be shown that the other terms of $C_T$ are $O_p(T^{-1/4})$. Hence, we deduce
	\begin{align}
		\hat{\sigma}_T^2 = T^{-1}\sum_{t=1}^{T}\varepsilon_{y,t}^2 + g^2\sigma_x^2\int_{0}^{1}J_{x,T}(r)^2dr + o_p(1) \Rightarrow \sigma_y^2 + g^2\sigma_x^2\int_{0}^{1}J_x(r)^2dr.
	\end{align}
	\end{itemize}
\end{proof}

Finally, we establish the limiting behavior of the components of the Wald test statistic when $s_{\beta,t}$ is $\mathrm{I}(1)$.

\begin{lemapp}
	Suppose the condition of Theorem \ref{thm:wald_nonsta} holds. Then, under model \eqref{model:pr_nonsta_set0}, we have the following results.
	\begin{itemize}
	\item[(i)] $T^{-1}\sum_{t=1}^{T}\tilde{x}_{1,t-1}z_t^2(\hat{\beta}) \Rightarrow\sigma_x\sigma_\eta\int_{0}^{1}\tilde{J}_{x,1}(r)dW_\eta(r) + g^2\sigma_x^3\int_{0}^{1}\tilde{J}_{x,1}(r)J_x(r)^2Q(r)^2dr$.
	\item[(ii)] $T^{-3/2}\sum_{t=1}^{T}\tilde{x}_{2,t-1}z_t^2(\hat{\beta}) \Rightarrow \sigma_x^2\sigma_\eta\int_{0}^{1}\tilde{J}_{x,2}(r)dW_\eta(r) + g^2\sigma_x^4\int_{0}^{1}\tilde{J}_{x,2}(r)J_x(r)^2Q(r)^2dr$.
	\item[(ii)] $\hat{\sigma}_{\tilde{\xi}}^2(\hat{\beta}) \stackrel{p}{\to}\sigma_{\eta}^2$.
	\end{itemize}
	\label{lem:wald_nonsta_components}
\end{lemapp}

\noindent \begin{proof}
	\begin{itemize}
		\item[(i)] Let $\rho_\beta \coloneqq 1+c_\beta T^{-1}$ and $\sigma_{y\beta} \coloneqq \mathrm{Cov}(\varepsilon_{y,t},\varepsilon_{\beta,t})$, with an abuse of notation. Then, we have
		\begin{align}
			T^{-1}\sum_{t=1}^{T}\tilde{x}_{1,t-1}z_t^2(\hat{\beta}) &= T^{-1}\sum_{t=1}^{T}\tilde{x}_{1,t-1}[\varepsilon_{y,t}+gT^{-5/4}x_{t-1}\varepsilon_{\beta,t}+T^{-3/4}x_{t-1}\{g\rho_\beta T^{-1/2}s_{\beta,t-1}-T^{3/4}(\hat{\beta}-\beta)\}]^2 \\
			&=T^{-1}\sum_{t=1}^{T}\tilde{x}_{1,t-1}(\varepsilon_{y,t}+gT^{-5/4}x_{t-1}\varepsilon_{\beta,t})^2 \\
			& \quad + 2T^{-7/4}\sum_{t=1}^{T}\tilde{x}_{1,t-1}x_{t-1}\{g\rho_\beta T^{-1/2}s_{\beta,t-1}-T^{3/4}(\hat{\beta}-\beta)\}(\varepsilon_{y,t}+gT^{-5/4}x_{t-1}\varepsilon_{\beta,t}) \\
			& \quad + T^{-5/2}\sum_{t=1}^{T}\tilde{x}_{1,t-1}x_{t-1}^2\{g\rho_\beta T^{-1/2}s_{\beta,t-1}-T^{3/4}(\hat{\beta}-\beta)\}^2 \\
			&= \sigma_x\sigma_\eta\int_{0}^{1}\tilde{J}_{x,1,T}(r)dW_{\eta,T}(r) + 2g\sigma_x^2\sigma_{y\beta}T^{-1/4}\int_{0}^{1}\tilde{J}_{x,1,T}(r)^2dr(1+o_p(1)) \\
			& \quad +g^2\sigma_x^3T^{-1}\int_{0}^{1}\tilde{J}_{x,1,T}(r)\tilde{J}_{x,2,T}(r)dr(1+o_p(1)) \\
			& \quad + 2\sigma_x^2\int_{0}^{1}\tilde{J}_{x,1,T}(r)J_{x,T}(r)\{g\rho_\beta J_{\beta,T}(r)-T^{3/4}(\hat{\beta}-\beta)\}\{\sigma_yT^{-1/4}dW_{y,T}(r) \\
			&\qquad \qquad \qquad \qquad \qquad \qquad \qquad  +g\sigma_xT^{-1}J_{x,T}(r)dW_{\beta,T}(r)\} \\
			& \quad + \sigma_x^3\int_{0}^{1}\tilde{J}_{x,1,T}(r)J_{x,T}(r)^2\{g\rho_\beta J_{\beta,T}(r)-T^{3/4}(\hat{\beta}-\beta)\}^2dr.
		\end{align}
	Applying Lemma \ref{lem:beta_hat}(iii) and the CMT gives the desired result.
	\item[(ii)] The proof is almost identical to that of part (i) and thus is omitted.
	\item[(iii)] Recall that $\hat{\sigma}_{\tilde{\xi}}^2(\hat{\beta})$ can be written as \eqref{eqn:sigma_hat_xi_expand}. Let $Q_T(\frac{t-1}{T}) \coloneqq g\rho_\beta T^{-1/2}s_{\beta,t-1} - T^{3/4}(\hat{\beta}-\beta)$. Then, $B_{1,T}$ becomes
	\begin{align}
		B_{1,T} &= T^{-1}\sum_{t=1}^{T}z_t^4(\hat{\beta}) - \Bigl\{T^{-1}\sum_{t=1}^{T}z_t^2(\hat{\beta})\Bigr\}^2 \\
		&= T^{-1}\sum_{t=1}^{T}\Bigl\{\varepsilon_{y,t}+ gT^{-5/4}x_{t-1}\varepsilon_{\beta,t} +  T^{-3/4}x_{t-1}Q_T\Bigl(\frac{t-1}{T}\Bigr)\Bigr\}^4 \\
		& \quad - \Bigl[T^{-1}\sum_{t=1}^{T}\Bigl\{\varepsilon_{y,t}+ gT^{-5/4}x_{t-1}\varepsilon_{\beta,t} +  T^{-3/4}x_{t-1}Q_T\Bigl(\frac{t-1}{T}\Bigr)\Bigr\}^2\Bigr]^2 \\
		&=T^{-1}\sum_{t=1}^{T}(\varepsilon_{y,t}+gT^{-5/4}x_{t-1}\varepsilon_{\beta,t})^4 + 4T^{-7/4}\sum_{t=1}^{T}(\varepsilon_{y,t}+gT^{-5/4}x_{t-1}\varepsilon_{\beta,t})^3x_{t-1}Q_T\Bigl(\frac{t-1}{T}\Bigr) \\
		&\quad +6T^{-5/2}\sum_{t=1}^{T}(\varepsilon_{y,t}+gT^{-5/4}x_{t-1}\varepsilon_{\beta,t})^2x_{t-1}^2Q_T\Bigl(\frac{t-1}{T}\Bigr)^2 \\
		&\quad +4T^{-13/4}\sum_{t=1}^{T}(\varepsilon_{y,t}+gT^{-5/4}x_{t-1}\varepsilon_{\beta,t})x_{t-1}^3Q_T\Bigl(\frac{t-1}{T}\Bigr)^3 +T^{-4}\sum_{t=1}^{T}x_{t-1}^4Q_T\Bigl(\frac{t-1}{T}\Bigr)^4 \\
		&\quad - \Bigl[T^{-1}\sum_{t=1}^{T}(\varepsilon_{y,t}+gT^{-5/4}x_{t-1}\varepsilon_{\beta,t})^2 + 2T^{-7/4}\sum_{t=1}^{T}(\varepsilon_{y,t}+gT^{-5/4}x_{t-1}\varepsilon_{\beta,t})x_{t-1}Q_T\Bigl(\frac{t-1}{T}\Bigr) \\
		&\quad + T^{-5/2}\sum_{t=1}^{T}x_{t-1}^2Q_T\Bigl(\frac{t-1}{T}\Bigr)^2\Bigr]^2 \\
		&=T^{-1}\sum_{t=1}^{T}\varepsilon_{y,t}^4 - \Bigl[T^{-1}\sum_{t=1}^{T}\varepsilon_{y,t}^2\Bigr]^2 + O_p(T^{-1/4}),
	\end{align}
	because $Q_T(\frac{t}{T})=O_p(1)$ uniformly in $t$. It can also be shown that $B_{2,T}=O_p(T^{-1})$ by parts (i) and (ii) and the CMT. Therefore, $\hat{\sigma}_{\tilde{\xi}}^2(\hat{\beta})=T^{-1}\sum_{t=1}^{T}\varepsilon_{y,t}^4 - \Bigl[T^{-1}\sum_{t=1}^{T}\varepsilon_{y,t}^2\Bigr]^2+o_p(1) \stackrel{p}{\to}\sigma_\eta^2$.
	\end{itemize}
\end{proof}

\noindent \begin{proof}[Proof of Theorem \ref{thm:asym_dists_sta}]
	\begin{itemize}
		\item[(i)] Letting $D_T \coloneqq \mathrm{diag}(T,T^{3/2})$, $\mathrm{W}_T(\hat{\beta})$ can be written as
		\begin{align}
			\mathrm{W}_T(\hat{\beta}) &= \hat{\sigma}_{\tilde{\xi}}^{-2}(\hat{\beta})(X'Z_2(\hat{\beta}))'(X'X)^{-1}(X'Z_2(\hat{\beta})) \\
			&=\hat{\sigma}_{\tilde{\xi}}^{-2}(\hat{\beta})(D_T^{-1}X'Z_2(\hat{\beta}))'(D_T^{-1}X'XD_T^{-1})^{-1}(D_T^{-1}X'Z_2(\hat{\beta})) \\
			&=\hat{\sigma}_{\tilde{\xi}}^{-2}(\hat{\beta}) \begin{pmatrix}
				T^{-1}\sum_{t=1}^{T}\tilde{x}_{1,t-1}z_t^2(\hat{\beta}) \\
				T^{-3/2}\sum_{t=1}^{T}\tilde{x}_{2,t-1}z_t^2(\hat{\beta})
			\end{pmatrix}' \begin{pmatrix}
				T^{-2}\sum_{t=1}^{T}(\tilde{x}_{1,t-1})^2 & T^{-5/2}\sum_{t=1}^{T}\tilde{x}_{1,t-1}\tilde{x}_{2,t-1} \\
				T^{-5/2}\sum_{t=1}^{T}\tilde{x}_{2,t-1}\tilde{x}_{1,t-1} & T^{-3}\sum_{t=1}^{T}(\tilde{x}_{2,t-1})^2
			\end{pmatrix}^{-1} \\
			& \qquad \qquad \qquad \qquad \qquad \qquad \times \begin{pmatrix}
				T^{-1}\sum_{t=1}^{T}\tilde{x}_{1,t-1}z_t^2(\hat{\beta}) \\
				T^{-3/2}\sum_{t=1}^{T}\tilde{x}_{2,t-1}z_t^2(\hat{\beta})
			\end{pmatrix}.
		\end{align}
		Applying Lemma \ref{lem:wald_sta_components} and the CMT, we obtain the stated result.
		
		\item[(ii)] To prove part (ii), we follow the proof of Theorem 1 of \citetapp{georgievTestingParameterInstability2018}. Note that $\mathrm{LN}_T$ can be written as
		\begin{align}
			\mathrm{LM}_T &= \frac{1}{\hat{\sigma}_T^2T^{-2}\sum_{t=1}^{T}x_{t-1}^2}\sum_{i=1}^{T}\Bigl(T^{-1}\sum_{t=1}^{i}x_{t-1}z_t(\hat{\beta})\Bigr)^2\frac{1}{T} \\
			&=\frac{1}{\hat{\sigma}_T^2\int_{0}^{1}J^2_{x,T}(r)dr}\int_{0}^{1}\Bigl(\int_{0}^{r}J_{x,T}(s)dZ_T(s)\Bigr)^2dr,
		\end{align}
	where $Z_T(r) \coloneqq T^{-1/2}\sum_{t=1}^{\lfloor Tr \rfloor}z_t(\hat{\beta})$. Then, the desired result directly follows from Lemma \ref{lem:lm_sta_components} and the CMT.
	\end{itemize}
\end{proof}

\noindent \begin{proof}[Proof of Theorem \ref{thm:wald_nonsta}]
	Using Lemma \ref{lem:wald_nonsta_components} and the CMT, the proof proceeds in the same way as that of Theorem \ref{thm:asym_dists_sta}(i), and hence the details are omitted.
\end{proof}

\begin{lemapp}
     Under Assumption \ref{asm:serial_corr} and under $\mathrm{H}_0:\omega_{\beta}=0$, the following results hold:
	\begin{itemize}
        
		\item[(i)] 
		\begin{align}
			T^{-3/2}\sum_{t=1}^{T}\tilde{x}_{2,t-1}z_t^2(\hat{\beta}) \Rightarrow \sigma_{\eta,lr}\sigma_x^2\int_{0}^{1}\tilde{J}_{x,2}(r)dW_{\eta,lr}(r) + 2\sigma_x\Lambda_{x\eta}\int_0^1J_x(r)dr,
		\end{align}
		
		\item[(ii)] \begin{align}
			T^{-1}\sum_{t=1}^{T}\tilde{x}_{1,t-1}z_t^2(\hat{\beta}) \Rightarrow \sigma_{\eta,lr}\sigma_x\int_{0}^{1}\tilde{J}_{x,1}(r)dW_{\eta,lr}(r) + \Lambda_{x\eta},
		\end{align}

        \item[(iii)] $\hat{\sigma}_{\tilde{\xi}}^2(\hat{\beta}) \stackrel{p}{\to} \sigma_\eta^2$.
	\end{itemize} \label{lemapp:wc_serial_corr}
\end{lemapp}

\noindent \begin{proof}
    (i) Noting that $z_t^2(\hat{\beta}) = (y_t - \hat{\beta}x_{t-1})^2 = \{\varepsilon_{y,t} - (\hat{\beta}-\beta)x_{t-1}\}^2$, we have
    \begin{align}
        T^{-3/2}\sum_{t=1}^{T}\tilde{x}_{2,t-1}z_t^2(\hat{\beta}) &= T^{-3/2}\sum_{t=1}^T\tilde{x}_{2,t-1}\{\sigma_y^2+(\varepsilon_{y,t}^2-\sigma_y^2) - 2(\hat{\beta}-\beta)x_{t-1}\varepsilon_{y,t}+(\hat{\beta}-\beta)^2x_{t-1}^2\} \\
        &= \sigma_x^2\sigma_{\eta,lr}T^{-3/2}\sum_{t=1}^T\Bigl(\frac{x_{t-1}^2}{\sigma_x^2} - T^{-1}\sum_{s=1}^T\frac{x_{s-1}^2}{\sigma_x^2}\Bigr)\frac{\eta_t}{\sigma_{\eta,lr}} \\
        &\quad -2T(\hat{\beta}-\beta)T^{-5/2}\sum_{t=1}\tilde{x}_{2,t-1}x_{t-1}\varepsilon_{y,t} + T^{2}(\hat{\beta}-\beta)^2T^{-7/2}\sum_{t=1}^T(\tilde{x}_{2,t-1})^2 \\
        &=\sigma_x^2\sigma_{\eta,lr}\Biggl\{\sum_{t=1}^T\Bigl(\frac{x_{t-1}}{\sigma_x\sqrt{T}}\Bigr)^2\frac{\eta_t}{\sigma_{\eta,lr}\sqrt{T}} - \sum_{s=1}^T\Bigl(\frac{x_{s-1}}{\sigma_x\sqrt{T}}\Bigr)^2\frac{1}{T}\times \sum_{t=1}^T\frac{\eta_t}{\sigma_{\eta,lr}\sqrt{T}}\Biggr\} \\
        &\quad -2T(\hat{\beta}-\beta)T^{-5/2}\sum_{t=1}\tilde{x}_{2,t-1}x_{t-1}\varepsilon_{y,t} + T^{2}(\hat{\beta}-\beta)^2T^{-7/2}\sum_{t=1}^T(\tilde{x}_{2,t-1})^2.
        \label{eqn:sum_tildex2_z2}
    \end{align}
    The second term in the braces of \eqref{eqn:sum_tildex2_z2} weakly converges to $\int_{0}^1J_x^2(r)drW_{\eta,lr}(1)$ by the FCLT and the CMT. For the first term, we can apply Theorem 3.1 of \citetapp{liangWeakConvergenceStochastic2016} to obtain
    \begin{align}
        \sum_{t=1}^T\Bigl(\frac{x_{t-1}}{\sigma_x\sqrt{T}}\Bigr)^2\frac{\eta_t}{\sigma_{\eta,lr}\sqrt{T}} \Rightarrow \int_0^1J_x^2(r)dW_{\eta,lr}(r) + 2\frac{\Lambda_{x\eta}}{\sigma_x\sigma_{\eta,lr}}\int_0^1J_x(r)dr.
    \end{align}
    The second term of \eqref{eqn:sum_tildex2_z2} is $O_p(T^{-1/2})$ since $\hat{\beta}-\beta=O_p(T^{-1})$ (by the fact that we consider $\mathrm{H}_0:\omega_\beta=0$ and $\varepsilon_{y,t}$ is a stationary $\alpha$-mixing m.d.s. with a finite 6th moment) and $\sum_{t=1}^T\tilde{x}_{2,t-1}x_{t-1}\varepsilon_{y,t}$ is $O_p(T^2)$. Similarly, the third term of \eqref{eqn:sum_tildex2_z2} is $O_p(T^{-1/2})$. Combining the above results yields
    \begin{align}
        T^{-3/2}\sum_{t=1}^{T}\tilde{x}_{2,t-1}z_t^2(\hat{\beta}) &\Rightarrow \sigma_x^2\sigma_{\eta,lr}\Biggl(\int_0^1J_x^2(r)dW_{\eta,lr}(r) + 2\frac{\Lambda_{x\eta}}{\sigma_x\sigma_{\eta,lr}}\int_0^1J_x(r)dr - \int_{0}^1J_x^2(r)drW_{\eta,lr}(1)\Biggr) \\
        & = \sigma_x^2\sigma_{\eta,lr}\int_0^1\tilde{J}_{x,2}(r)dW_{\eta,lr}(r) + 2\sigma_x\Lambda_{x\eta}\int_0^1J_x(r)dr.
    \end{align}

    (ii) Similarly to part (i), we have
    \begin{align}
        T^{-1}\sum_{t=1}^{T}\tilde{x}_{1,t-1}z_t^2(\hat{\beta}) &= \sigma_x\sigma_{\eta,lr}\Biggl\{\sum_{t=1}^T\frac{x_{t-1}}{\sigma_x\sqrt{T}}\frac{\eta_t}{\sigma_{\eta,lr}\sqrt{T}} - \sum_{s=1}^T\frac{x_{s-1}}{\sigma_x\sqrt{T}}\frac{1}{T}\times \sum_{t=1}^T\frac{\eta_t}{\sigma_{\eta,lr}\sqrt{T}}\Biggr\} \\
        & \quad -2T(\hat{\beta}-\beta)T^{-2}\sum_{t=1}^T\tilde{x}_{1,t-1}x_{t-1}\varepsilon_{y,t} + T^{2}(\hat{\beta}-\beta)^2T^{-3}\sum\tilde{x}_{1,t-1}x_{t-1}^2,
    \end{align}
    where the last two terms are $O_p(T^{-1/2})$. Applying Theorem 3.1 of \citetapp{liangWeakConvergenceStochastic2016} and the CMT completes the proof.

    (iii) As in part (iii) of Lemma \ref{lem:wald_sta_components}, we have
    \begin{align}
	   \hat{\sigma}_{\tilde{\xi}}^2(\hat{\beta}) &= T^{-1}Z_2(\hat{\beta})'M_XZ_2(\hat{\beta}) =     B_{1,T} - B_{2,T},
    \end{align} 
    where $B_{1,T}=T^{-1}\sum_{t=1}^{T}\{\widetilde{z_t^2}(\hat{\beta})\}^2$ and
    \begin{align}
        B_{2,T} = T^{-1} \begin{pmatrix}
			\sum_{t=1}^{T}\tilde{x}_{1,t-1}z_t^2(\hat{\beta}) \\
			\sum_{t=1}^{T}\tilde{x}_{2,t-1}z_t^2(\hat{\beta})
		\end{pmatrix}'\begin{pmatrix}
			\sum_{t=1}^{T}(\tilde{x}_{1,t-1})^2 & \sum_{t=1}^{T}\tilde{x}_{1,t-1}\tilde{x}_{2,t-1} \\
			\sum_{t=1}^{T}\tilde{x}_{2,t-1}\tilde{x}_{1,t-1} & \sum_{t=1}^{T}(\tilde{x}_{2,t-1})^2
		\end{pmatrix}^{-1} 
		\begin{pmatrix}
			\sum_{t=1}^{T}\tilde{x}_{1,t-1}z_t^2(\hat{\beta}) \\
			\sum_{t=1}^{T}\tilde{x}_{2,t-1}z_t^2(\hat{\beta})
		\end{pmatrix}.
    \end{align}
	$B_{1,T}$ becomes
	\begin{align}
		B_{1,T} &= T^{-1}\sum_{t=1}^{T}\biggl\{z_t^2(\hat{\beta}) - T^{-1}\sum_{t=1}^{T}z_t^2(\hat{\beta})\biggr\}^2 \\
		&=T^{-1}\sum_{t=1}^{T}z_t^4(\hat{\beta}) - \biggl\{T^{-1}\sum_{t=1}^{T}z_t^2(\hat{\beta})\biggr\}^2 \\
		&=T^{-1}\sum_{t=1}^{T}\{\varepsilon_{y,t} - (\hat{\beta}-\beta)x_{t-1}\}^4 -\Bigl[T^{-1}\sum_{t=1}^{T}\{\varepsilon_{y,t} - (\hat{\beta}-\beta)x_{t-1}\}^2\Bigr]^2 \\
		&=T^{-1}\sum_{t=1}^{T}\varepsilon_{y,t}^4 + B_{1,1,T} - \Bigl[T^{-1}\sum_{t=1}^{T}\varepsilon_{y,t}^2+B_{1,2,T}\Bigr]^2,
	\end{align}
	where
	\begin{align}
		B_{1,1,T} &= - 4(\hat{\beta}-\beta)T^{-1}\sum_{t=1}^{T}\varepsilon_{y,t}^3x_{t-1}+6(\hat{\beta}-\beta)^2T^{-1}\sum_{t=1}^{T}\varepsilon_{y,t}^2x_{t-1}^2 \\
        &- 4(\hat{\beta}-\beta)^3T^{-1}\sum_{t=1}^{T}\varepsilon_{y,t}x_{t-1}^3+(\hat{\beta}-\beta)^4T^{-1}\sum_{t=1}^{T}x_{t-1}^4,
	\end{align}
	and
	\begin{align}
		B_{1,2,T} &= -2(\hat{\beta}-\beta)T^{-1}\sum_{t=1}^{T}\varepsilon_{y,t}x_{t-1} + (\hat{\beta}-\beta)^2T^{-1}\sum_{t=1}^{T}x_{t-1}^2.
	\end{align}
    Using the fact that $\hat{\beta}-\beta=O_p(T^{-1})$ and $\sup_{0\leq r\leq1}|T^{-1/2}x_{\lfloor Tr \rfloor}| = O_p(1)$, it can be readily shown that $B_{1,1,T}$ and $B_{1,2,T}$ are $o_p(1)$. This implies
	\begin{align}
		B_{1,T} &= T^{-1}\sum_{t=1}^{T}\varepsilon_{y,t}^4 - \Bigl\{T^{-1}\sum_{t=1}^{T}\varepsilon_{y,t}^2\Bigr\}^2 + o_p(1) \stackrel{p}{\to}E[\varepsilon_{y,t}^4] - \sigma_y^4 = \sigma_{\eta}^2.
	\end{align}
	For $B_{2,T}$, an application of the CMT yields $B_{2,T}=O_p(T^{-1})$, given that $T^{-1}\sum_{t=1}^{T}\tilde{x}_{1,t-1}z_t^2(\hat{\beta})$ and $T^{-3/2}\sum_{t=1}^{T}\tilde{x}_{2,t-1}z_t^2(\hat{\beta})$ are $O_p(1)$ from parts (i) and (ii), and $T^{-2}\sum_{t=1}^{T}(\tilde{x}_{1,t-1})^2$, $T^{-5/2}\sum_{t=1}^{T}\tilde{x}_{1,t-1}\tilde{x}_{2,t-1}$ and $T^{-3}\sum_{t=1}^{T}(\tilde{x}_{2,t-1})^2$ are $O_p(1)$. Hence, we conclude $\hat{\sigma}_{\tilde{\xi}}^2(\hat{\beta}) \stackrel{p}{\to}\sigma_{\eta}^2$.
\end{proof}

\noindent \begin{proof}[Proof of \eqref{dist:wald_serial_corr}]
    As in the proof of Theorem \ref{thm:asym_dists_sta}, we have
    \begin{align}
        W_T(\hat{\beta}) 
		&=\hat{\sigma}_{\tilde{\xi}}^{-2}(\hat{\beta})\begin{pmatrix} T^{-1}\sum_{t=1}^{T}\tilde{x}_{1,t-1}z_t^2(\hat{\beta}) \\ T^{-3/2}\sum_{t=1}^{T}\tilde{x}_{2,t-1}z_t^2(\hat{\beta}) \\
		\end{pmatrix}' \begin{pmatrix}
			T^{-2}\sum_{t=1}^{T}(\tilde{x}_{1,t-1})^2 & T^{-5/2}\sum_{t=1}^{T}\tilde{x}_{1,t-1}\tilde{x}_{2,t-1} \\
			T^{-5/2}\sum_{t=1}^{T}\tilde{x}_{2,t-1}\tilde{x}_{1,t-1} & T^{-3}\sum_{t=1}^{T}(\tilde{x}_{2,t-1})^2 \\ 
		\end{pmatrix}^{-1} \\
		& \ \ \ \ \ \ \ \ \ \ \ \times \begin{pmatrix} T^{-1}\sum_{t=1}^{T}\tilde{x}_{1,t-1}z_t^2(\hat{\beta}) \\ T^{-3/2}\sum_{t=1}^{T}\tilde{x}_{2,t-1}z_t^2(\hat{\beta}) \\
		\end{pmatrix}.
    \end{align}
    Given the above equation, \eqref{dist:wald_serial_corr} is an immediate consequence of Lemma \ref{lemapp:wc_serial_corr} and the CMT.
    
\end{proof}

\noindent \begin{proof}[Proof of Theorem \ref{thm:long_run_wald_mod}]
    First note that
    \begin{align}
    \hat{\theta} - \begin{pmatrix}
		X'X
	\end{pmatrix}^{-1}
	\begin{pmatrix}
		T\hat{\Lambda}_{x\eta,T} \\
		2\hat{\Lambda}_{x\eta,T}\sum_{t=1}^{T}x_{t-1}
	\end{pmatrix} &= (X'X)^{-1}\Biggl\{\begin{pmatrix}
        \sum_{t=1}^T\tilde{x}_{1,t-1}z_t^2(\hat{\beta}) \\ \sum_{t=1}^T\tilde{x}_{2,t-1}z_t^2(\hat{\beta})
    \end{pmatrix} - \begin{pmatrix}
            T\hat{\Lambda}_{x\eta,T} \\ 2\hat{\Lambda}_{x\eta,T}\sum_{t=1}^Tx_{t-1}
        \end{pmatrix}\Biggr\} \\
            &\eqqcolon (X'X)^{-1}F_T.
    \end{align}
    Therefore, using Lemma \ref{lemapp:wc_serial_corr}, $W_T^{*}(\hat{\beta})$ satisfies
    \begin{align}
        W_T^{*}(\hat{\beta}) &= \frac{F_T'(X'X)^{-1}F_T}{\hat{\sigma}^2_{\tilde{\xi},lr}(\hat{\beta})} \\
            &= \bigl\{\hat{\sigma}^2_{\tilde{\xi},lr}(\hat{\beta})\bigr\}^{-1} \Biggl\{\begin{pmatrix}
        T^{-1}\sum_{t=1}^T\tilde{x}_{1,t-1}z_t^2(\hat{\beta}) \\ T^{-3/2}\sum_{t=1}^T\tilde{x}_{2,t-1}z_t^2(\hat{\beta})
    \end{pmatrix} - \begin{pmatrix}
            \hat{\Lambda}_{x\eta,T} \\ 2\hat{\Lambda}_{x\eta,T}T^{-3/2}\sum_{t=1}^Tx_{t-1}
        \end{pmatrix} \Biggr\}' \\
            &\times \begin{pmatrix}
                T^{-2}\sum_{t=1}^T\tilde{x}_{1,t-1}^2 & T^{-5/2}\sum_{t=1}^T\tilde{x}_{1,t-1}\tilde{x}_{2,t-1} \\
                T^{-5/2}\sum_{t=1}^T\tilde{x}_{2,t-1}\tilde{x}_{1,t-1} & T^{-3}\sum_{t=1}^T\tilde{x}_{2,t-1}^2
            \end{pmatrix}^{-1} \\
            &\times \Biggl\{\begin{pmatrix}
        T^{-1}\sum_{t=1}^T\tilde{x}_{1,t-1}z_t^2(\hat{\beta}) \\ T^{-3/2}\sum_{t=1}^T\tilde{x}_{2,t-1}z_t^2(\hat{\beta})
    \end{pmatrix} - \begin{pmatrix}
            \hat{\Lambda}_{x\eta,T} \\ 2\hat{\Lambda}_{x\eta,T}T^{-3/2}\sum_{t=1}^Tx_{t-1}
        \end{pmatrix}\Biggr\} \\
            &\Rightarrow \sigma_{\eta,lr}^{-2}\Biggl\{\begin{pmatrix}
                \sigma_{\eta,lr}\sigma_x\int_0^1\tilde{J}_{x,1}(r)dW_{\eta,lr}(r) \\
                \sigma_{\eta,lr}\sigma_x^2\int_0^1\tilde{J}_{x,2}(r)dW_{\eta,lr}(r)
            \end{pmatrix}\Biggr\}'\begin{pmatrix}
                \sigma_x^2\int_0^1\tilde{J}_{x,1}^2(r)dr & \sigma_x^3\int_0^1\tilde{J}_{x,1}(r)\tilde{J}_{x,2}(r)dr \\
                \sigma_x^3\int_0^1\tilde{J}_{x,2}(r)\tilde{J}_{x,1}(r)dr & \sigma_x^4\int_0^1\tilde{J}_{x,2}^2(r)dr
            \end{pmatrix}^{-1} \\
            &\times \Biggl\{\begin{pmatrix}
                \sigma_{\eta,lr}\sigma_x\int_0^1\tilde{J}_{x,1}(r)dW_{\eta,lr}(r) \\
                \sigma_{\eta,lr}\sigma_x^2\int_0^1\tilde{J}_{x,2}(r)dW_{\eta,lr}(r)
            \end{pmatrix}\Biggr\} \\
            &= \begin{pmatrix}
			\int_{0}^{1}\tilde{J}_{x,1}(r)dW_{\eta,lr}(r) \\ \int_{0}^{1}\tilde{J}_{x,2}(r)dW_{\eta,lr}(r) \end{pmatrix}'\begin{pmatrix}
			\int_{0}^{1}\bigl(\tilde{J}_{x,1}\bigr)^2(r)dr & \int_{0}^{1}\tilde{J}_{x,1}(r)\tilde{J}_{x,2}(r)dr \\ \int_{0}^{1}\tilde{J}_{x,2}(r)\tilde{J}_{x,1}(r)dr & \int_{0}^{1}\bigl(\tilde{J}_{x,2}\bigr)^2(r)dr
		\end{pmatrix}^{-1}\begin{pmatrix}
			\int_{0}^{1}\tilde{J}_{x,1}(r)dW_{\eta,lr}(r) \\ \int_{0}^{1}\tilde{J}_{x,2}(r)dW_{\eta,lr}(r) \end{pmatrix}.
    \end{align}
\end{proof}

\section*{Appendix B: Asymptotics for the LM test of \citetapp{linTestingParameterConstancy1999}}
\setcounter{equation}{0}
\renewcommand{\theequation}{B.\arabic{equation}}

\citetapp{linTestingParameterConstancy1999} (hereafter LT) derive the LM test for coefficient constancy in linear models under the alternative hypothesis of stationary AR random coefficient. Their test statistic in our setting is given by
\begin{align}
	\mathrm{LM}_T^{lt} \coloneqq \frac{\{\sum_{t=1}^{T}(x_{t-1}^2 - T^{-1}\sum_{t=1}^{T}x_{t-1}^2)z_t^2(\hat{\beta})\}^2}{2\hat{\sigma}_T^4\sum_{t=1}^{T}(x_{t-1}^2 - T^{-1}\sum_{t=1}^{T}x_{t-1}^2)^2}. \label{def:lt_lm}
\end{align}
In fact, this is the Wald statistic from linearized model \eqref{eqn:linearized} with the restriction of $\delta=0$ and normality assumption on $\varepsilon_{y,t}$. In other words, \eqref{def:lt_lm} is the Wald statistic, or equivalently the squared t statistic, from the restricted model
\begin{align}
	z_t^2(\beta) = \sigma_y^2 + \omega_\beta^2x_{t-1}^2 + \xi_t, \label{eqn:linearized_restricted}
\end{align}
with $\varepsilon_{y,t} \sim N(0,\sigma_y^2)$. Because $\mathrm{Var}(\xi_t) = \mathrm{Var}(\eta_t) = 2\sigma_y^4$ under the null and normality of $\varepsilon_{y,t}$, the squared t statistic from \eqref{eqn:linearized_restricted} with the variance estimator calculated under the null coincides with $\mathrm{LM}_T^{lt}$. Using Lemmas \ref{lem:wald_sta_components} and \ref{lem:wald_nonsta_components} in Appendix A, it can be shown that $\mathrm{LM}_T^{lt}$ has the asymptotic alternative distribution when $\omega_\beta = gT^{-3/4}$ and $s_{\beta,t}$ is $\mathrm{I}(0)$, and when $\omega_\beta = gT^{-5/4}$ and $s_{\beta,t}$ is $\mathrm{I}(1)$, as our Wald test does.\footnote{LT in fact propose to use the sum of $\mathrm{LM}_T^{lt}$ and another statistic that reflects the AR structure of the random coefficient, to detect coefficient randomness more efficiently. However, the test statistic cannot be calculated without the knowledge about the number of the AR lags of the random coefficient. Moreover, the misspecification of the AR structure of the random coefficient can decrease the power of the test, as shown by their simulation (pp.203-204).}

The difference between LT's test and our Wald test lies in the treatment of parameter $\delta$ in \eqref{eqn:linearized}. Because $\delta=0$ is implied by $\sigma_{y\beta} = \mathrm{E}[\varepsilon_{y,t}s_{\beta,t}]=0$, LT's test essentially assumes that the regression disturbance and random coefficient are uncorrelated. Therefore, LT's test will be more efficient than our Wald test if this condition holds or $\sigma_{y\beta}$ is near zero, but the Wald test will outperform LT's test otherwise, as pointed out by \citetapp{nishiTestingCoefficientRandomness2023} in the context of autoregressions.

\bibliographystyleapp{standard}
\bibliographyapp{random_coefficient_PR}

\clearpage
\section*{Appendix C: Additional Monte Carlo Experiments}
\setcounter{equation}{0}
\renewcommand{\theequation}{C.\arabic{equation}}

In addition to the Monte Carlo simulation in Section 6, we conduct experiments where $\varepsilon_{x,t}$ (and thus $\varepsilon_{y,t}$) is a GARCH(1,1) process. The DGPs used here are identical to those in Section 6, except that the innovation process is specified as follows:
\begin{align}
	\begin{pmatrix}
		\varepsilon_{x,t} \\ v_t \\ \varepsilon_{\beta,t}
	\end{pmatrix} = \begin{pmatrix}
		\sigma_{x,t} & 0 & 0 \\ 0 & 1 & 0 \\ 0 & 0 & 1
	\end{pmatrix} \begin{pmatrix}
		u_{x,t} \\ u_{v,t} \\ u_{\beta,t}
	\end{pmatrix},
\end{align}
where $(u_{x,t},u_{v,t},u_{\beta,t})' \stackrel{\mathrm{i.i.d.}}{\sim} N(0,\mathrm{diag}(1,1,1))$, and $\sigma_{x,t}^2=c_0'+c_1'\sigma_{x,t-1}^2+c_2'\varepsilon_{x,t-1}^2$. We choose GARCH parameters similar to the estimates obtained from the real data analysis in Section 7, which are
\begin{itemize}
	\item[]DGP4 : $(c_0',c_1',c_2') = (3\times 10^{-8},0.9,0.05)$.
	\item[]DGP5 : $(c_0',c_1',c_2') = (3\times 10^{-5},0.8,0.15)$.
	\item[]DGP6 : $(c_0',c_1',c_2') = (6\times 10^{-5},5\times 10^{-7},0.99)$.
\end{itemize}
The results are shown in Tables \ref{tab_b:size} and \ref{tab_b:b_sel}.

\begin{table} \renewcommand\thetable{C.1} \caption{Empirical sizes of the tests}
	\centering
	\begin{threeparttable}
		\renewcommand{\arraystretch}{1.8}	\begin{tabular}{c@{\hskip 22pt}c@{\hskip 13pt}@{\hskip 13pt}c@{\hskip 13pt}@{\hskip 13pt}c@{\hskip 13pt}}
			\hline &LM&Wald&Sum \\
			\hline \multicolumn{4}{c}{\textbf{DGP4}} \\
			\hline $T=100$&0.016&0.040&0.041 \\
			 $T=200$&0.022&0.038&0.036 \\
			 $T=500$&0.033&0.028&0.028 \\
			\hline \multicolumn{4}{c}{\textbf{DGP5}} \\
			\hline $T=100$&0.016&0.042&0.042 \\
			 $T=200$&0.021&0.037&0.037 \\
			 $T=500$&0.030&0.028&0.028 \\
			\hline \multicolumn{4}{c}{\textbf{DGP6}} \\
			\hline $T=100$&0.018&0.044&0.045 \\
			 $T=200$&0.023&0.042&0.040 \\
			 $T=500$&0.030&0.031&0.029 \\
			\hline 
		\end{tabular}
		\begin{tablenotes}
			\footnotesize
			\item[] Entries are based on 5000 replications with significance level 0.05.
		\end{tablenotes}
	\end{threeparttable} \label{tab_b:size}
\end{table} 

\begin{table} \renewcommand\thetable{C.2} \caption{Medians of $b^{\mathrm{sel}}/T$}
	\centering
	\begin{threeparttable}
		\renewcommand{\arraystretch}{1.8}	\begin{tabular}{c@{\hskip 22pt}c@{\hskip 13pt}@{\hskip 13pt}c@{\hskip 13pt}@{\hskip 13pt}c@{\hskip 13pt}}
			\hline &LM&Wald&Sum \\
			\hline \multicolumn{4}{c}{\textbf{DGP4}} \\
			\hline $T=100$&0.12&0.26&0.25 \\
			 $T=200$&0.06&0.17&0.17 \\
			 $T=500$&0.03&0.07&0.06 \\
			\hline \multicolumn{4}{c}{\textbf{DGP5}} \\
			\hline $T=100$&0.12&0.25&0.25 \\
			 $T=200$&0.06&0.17&0.17 \\
			 $T=500$&0.03&0.06&0.07 \\
			\hline \multicolumn{4}{c}{\textbf{DGP6}} \\
			\hline $T=100$&0.12&0.26&0.26 \\
			 $T=200$&0.06&0.17&0.17 \\
			 $T=500$&0.03&0.07&0.07 \\
			\hline 
		\end{tabular}
		\begin{tablenotes}
			\footnotesize
			\item[] Entries are based on 5000 replications with significance level 0.05.
		\end{tablenotes}
	\end{threeparttable} \label{tab_b:b_sel}
\end{table}

\end{document}